\documentclass{book}
\usepackage{amssymb}
\usepackage{amsfonts}
\usepackage{amsmath}

\newtheorem{theorem}{Theorem}

\newtheorem{corollary}[theorem]{Corollary}

\newtheorem{example}[theorem]{Example}

\newtheorem{lemma}[theorem]{Lemma}

\newenvironment{proof}[1][Proof]{\noindent\textbf{#1.} }{\ \rule{0.5em}{0.5em}}
\input{tcilatex}

\begin{document}

\frontmatter
\title{MATHEMATICAL ASPECTS OF COMPUTER ENGINEERING\\
\textit{Advances in Science and Technology in the USSR}\\
\textbf{Design of Optimal Dynamic Analyzers: Mathematical Aspects of Wave
Pattern Recognition}}
\author{\textbf{V P Belavkin and V P Maslov} \\
Moscow Institute of Electronics and Mathematics \\
109028 Moscow USSR}
\date{Mir Publisher 1988}
\maketitle
\tableofcontents

\chapter*{Preface}

\markboth{PREFACE}{PREFACE}We give a review of the most important results on
optimal tomography as mathematical wave-pattern recognition theory emerged
in the 70's in connection with the problems\ of optimal estimation and
hypothesis testing in quantum theory. The key problems in this theory is
mathematical analysis and synthesis of the optimal dynamic analyzer
discriminating between a given, discrete or continuous, family of pure or
mixed \textit{a priori} unknown wave patterns. Classical pattern recognition
theory as a part of classical mathematical statistics cannot tackle such
problems since it operates with given sample data and is not concerned how
these date should be obtained from the physical wave states in an optimal
way. In quantum theory this problem is sometimes referred as the problem of
optimal measurement of an unknown quantum state, and is the main problem of
the emerging mathematical theory of quantum statistics.

We develop the results of optimal quantum measurement theory, most of which
belong to VPB \cite{17}--\cite{34}, further into the direction of wave,
rather than particle statistical estimation and hypothesis testing theory,
with the aim to include not only quantum matter waves but also classical
wave patterns like optical and acoustic waves. We apply the developed
methods of this new pattern recognition theory to the problems of
mathematical design of optimal wave analyzer discriminating the visual and
sound patterns. We conclude that Hilbert space and operator methods
developed in quantum theory are equally useful in the classical wave theory,
as soon as the possible observations are restricted to only intensity
distributions of waves, i.e. when the wave states are not the allowed
observables, as they are not the observables of individual particles in the
quantum theory. We will show that all characteristic attributes of quantum
theory such as complementarity, entanglement or Heisenberg uncertainty
relations are also attributes of the generalized wave pattern recognition
theory.

\chapter{Introduction}

The problem of automatic recognition of coherent wave patterns is coming to
the forefront in connection with one of the most important problems in
building fifth generation computational systems, that are either of
classical wave input and output or of quantum nature. The problem of
efficiency for such wave pattern analyzers contains a number of mathematical
subproblems, the most important of which have been studied in the
seventeen's in connection with the quantum estimation and hypothesis testing
theory. Without going into details of this and related quantum measurement
theory these problems can be explained as the problems of (mathematical)
design of an optimal analyzers of classical optical or acoustic waves, the
devices occupying the center of the stage in the problem of automatic
recognition of visual or sound patterns. An example of such a device is
given in \cite{1}: a receiver of acoustic waves $v(x-ct)$ whose idealized
model is a point-like resonator (or cavity) capable of measuring the
intensities of the vibrational modes excited by the waves. The modes are the
natural vibrations of one or several standards placed at point $x=0$ and are
described by an orthonormal set of functions $\chi _{k}(t)$ on a given
interval of observation $[0,T]$. A typical example of such a resonator is
the spectrum analyzer, a device that measures the intensity distribution
over the discrete frequencies $f_{k}=k/T$, $k\in N$, and can be represented
by a selective filter of harmonic waves 
\begin{equation*}
v_{k}(x-ct)=2\func{Re}\phi ^{k}\exp \{2\pi \mathrm{j}k(x-ct)/cT\},\quad 
\mathrm{j}=\sqrt{-1},
\end{equation*}%
in the output of which one can measure the positive numbers $\nu ^{k}=|\phi
^{k}|^{2}$ determined by the complex-valued amplitudes $\phi ^{k}\in \mathbb{%
C}^{1}$. The spectrum selector described by the harmonic functions%
\begin{equation*}
\chi _{k}(t)=\exp \{-2\pi \mathrm{j}kt/T\}\text{, }k=0,1,\ldots 
\end{equation*}%
which form an orthonormal set $\{\chi _{k}\}$ with respect to the scalar
product 
\begin{equation*}
(\chi _{i}|\chi _{k})=T^{-1}\int_{0}^{T}\chi _{i}(t)^{\ast }\chi _{k}(t)\,%
\mathrm{d}t,
\end{equation*}%
is ideally suited for the discrimination of pure tones with multiple
frequencies $\{f_{i}\}$, tones described by disjoint complex-valued
amplitudes $\phi _{i}^{k}=0$ at $i\neq k$ corresponding to the harmonic waves%
\begin{equation*}
v_{i}(x-ct)=2\func{Re}\varphi _{i}(t-x/c),
\end{equation*}%
where $\varphi _{i}(t)=\sum_{k=0}^{\infty }\phi _{i}^{k}\chi _{k}(t)$. To
establish which of the tones in $\{\varphi _{i}\}$ with different
frequencies and nonzero intensities $\nu _{i}=|\phi _{i}^{k}|^{2}\neq 0$ at $%
i=k$ is actually detected by such a receiver, it is sufficient to find the
number $i$ of the excited standard tuned to one of the harmonic modes in $%
\{\chi _{i}\}$ corresponding to the set $\{\varphi _{i}\}:\chi
_{i}(t)=\varphi _{i}(t)/\left\Vert \varphi _{i}\right\Vert $. The
vibrational energy of such a standard will coincide with the intensity $%
|(\chi _{i}\mid \varphi _{i})|^{2}=\left\Vert \varphi _{i}\right\Vert ^{2}$
of the detected signal $\varphi _{i}$, while the other standards remain
unexcited: $(\chi _{k}\mid \varphi _{i})=0$ at $k\neq i$. A segment of human
speech of duration $T$ containing a finished sentence consists, however, not
of a single pure tone but, generally, of an infinitude of pure tones of
different amplitudes and frequencies (the frequencies may be assumed to be
multiples of $1/T$ at a fixed interval $T$ of a single reception act).
Unharmonic signals, described by spectral amplitudes $\varphi _{i}=[\phi
_{i}^{k}]_{k=0}^{\infty }$ or, in the temporal representation, by the
analytic signals 
\begin{equation*}
\varphi _{i}(t)=\sum_{k=0}^{\infty }\phi _{i}^{k}\exp \{-2\pi \mathrm{j}%
kt/T\},
\end{equation*}%
may be indistinguishable in spectral measurements, even if they are
orthogonal. For example, if the $\varphi _{i}(t)$, $i=1,\ldots ,m$ are
disjoint pulses obtained through the shift by $T$ of the pulsed signal $%
\varphi _{0}(t)$ of length $\Delta t=T/m$, these pulses have corresponding
to them the orthogonal spectral amplitudes $\varphi _{i}^{k}=\phi
_{0}^{k}\exp \{2\pi \mathrm{j}ik/m\}$ with the same intensity distributions $%
\nu _{i}^{k}=|\phi _{i}^{k}|^{2}=|\phi _{0}^{k}|^{2}$, $i=1,\ldots ,m$, $%
k\in N$.

Orthogonal sound signals $\{\varphi _{i}\}$ on $[0,T]$ may be identified in
a similar manner by selective filters matched with the signal modes $\chi
_{i}(t)=\varphi _{i}(t)/\left\Vert \varphi _{i}\right\Vert $ that measure
the intensity distribution $\nu ^{k}=(\chi _{k}\mid \varphi _{i})|^{2}$ in
the modes $\{\chi _{i}\}$, a distribution that has a different form for
different values of $i$, namely, $\nu _{i}^{k}=\left\Vert \varphi
_{i}\right\Vert ^{2}\delta _{i}^{k}$. However, different sentences in human
speech correspond ordinarily to nonorthogonal sound signals $\varphi _{i}(t)$%
, which from the viewpoint of their meaning are identified if they are
collinear, that is, differ only in the total energy $\left\Vert \varphi
_{i}\right\Vert ^{2}=\int |\varphi _{i}(t)|^{2}\mathrm{d}t$. For the
recognition of nonorthogonal signals $\{\varphi _{i}\}_{i=1}^{m}$ one cannot
employ matched filtration since the filters described by the nonorthogonal
modes $\chi _{i}=\varphi _{i}/\left\Vert \varphi _{i}\right\Vert $ are
noncommutative and, hence, cannot be matched in a single selector;
otherwise, the total measured intensity $\sum_{k=1}^{m}|(\chi _{k}\mid \chi
_{i})|^{2}$, would exceed the total energy $\left\Vert \varphi
_{i}\right\Vert ^{2}$ of the received signal $\varphi _{i}$ if $(\varphi
_{i}\mid \varphi _{k})\neq 0$ at least for one $k\neq i$. Thus, we have an
indefinite situation, formally similar to the incompatibility of
noncommutative quantum mechanical observables, a situation arising from
Bohr's complementarity principle \cite{2} and Heisenberg's uncertainty
relation \cite{3}. Typical examples of noncommutative filters are the
frequency and temporal filters, which are incompatible, just as position and
momentum measurements are incompatible in a quantum mechanical system. So
which of the disjoint selectors described by orthonormal sets of modes $%
\{\chi _{k}\}$ must we employ to discern the nonorthogonal sound signals
from a given set $\{\varphi _{i}\}$? It is natural to look for the answer to
this nontrivial question in the form of a solution to an optimization
problem by selecting a criterion of discernment quality such that the
optimal selector does not depend on the gauge transformation $\varphi
_{i}\rightarrow \lambda \varphi _{i}$ for every complex-valued $\lambda $.
The latter condition is satisfied by the criterion of the maximum of the
total intensity $\sum_{i=1}^{m}|(\chi _{i}\mid \varphi _{i})|^{2}$ of the
true received amplitudes or by the criterion of the minimum of the lost
intensity $\sum_{i\neq k}|(\chi _{i}\mid \varphi _{i})|^{2}$. The
corresponding extremal problem for arbitrary nonorthogonal amplitudes $%
\{\varphi _{k}\}$ describing quantum mechanical states normalized to prior
probabilities was first studied in the general form in \cite{4}, \cite{5}, 
\cite{6}. Particular solution of this problem for the case of two
nonorthogonal amplitudes $\{\varphi _{0},\varphi _{1}\}$ were obtained in 
\cite{7}, \cite{8}, while the case of several linearly independent
amplitudes $\{\varphi _{i}\}$ was also considered in \cite{9}, \cite{10}.

The above-noted analogy between optimal recognition of sound signals and
discernment of quantum mechanical states suggested the possibility of
constructing a wave theory of noncommutative measurements within the scope
of which one could solve more general problems of testing wave hypotheses
for estimating the wave parameters. From the formal viewpoint this theory
generalizes the quantum theory of optimal measurements, hypothesis testing,
and estimation of parameters \cite{11}, \cite{12}, while actually it carries
the mathematical methods and ideas developed in \cite{17}-\cite{34} into the
new, practically more realistic, field of applications. A short author's
review of optimal processing of quantum signals is given in \cite{30}.
(Additional literature on the quantum theory of detection, hypothesis
testing, and parameter estimation can be found in the references cited in 
\cite{11}, \cite{12}.)

In the present text we give a systematic description of the wave theory of
representation and measurement based on analogies with quantum mechanics.
This theory is then employed to solve the problems of detection,
discrimination, identification, and estimation of the parameters of sound
signals and visual patterns within the framework of the noncommutative
theory of testing wave hypotheses developed here. The idea of employing the
methods of quantum mechanics for discerning wave patterns emerged at the
beginning of the 1970s, when a seminar on quantum mechanics and pattern
recognition was organized in the Physics Department of Moscow State
University. The seminar was directed by V.P. Maslov, along with Yu.P.
Pyt'ev, and the author, V.P. Belavkin, attended it. Interest in wave
tomography as reconstruction of the complete wave field rather than only the
energy illumination of the image in a certain plane was stimulated by the
rapid development of holography, which was invented by Gabor \cite{36}, and
then underwent a revival \cite{37} when coherent sources of light, or
lasers, were created. The emerging optimization problems of discerning wave
fronts are similar to the problems of discerning quantum mechanical states
and cannot be solved by classical methods \cite{38} of pattern recognition
since it is impossible to register directly and exactly the phase and
amplitude of a wave field by measuring the energy parameters. A detailed
study of these problems at the time showed \cite{39} that the then existing
quasiclassical methods of solving quantum mechanical problems were also
inadequate, and the solution had to be postponed until a consistent
noncommutative theory of measurements was developed in the then rapidly
advancing field of quantum theory.

Some particular problems of optimal processing optical wave signals, such as
those of optical localization \cite{40}, detection, and discrimination of
two signals from closely positioned sources of coherent radiation \cite{41}, 
\cite{42}, have been already well studied by methods of quantum statistical
and nonlinear optics \cite{43}, \cite{44}, \cite{45}.

In this book, in addition to discussing the noncommutative theory of
measurements common for quantum states and classical sound signals and
coherent optical waves, we provide solutions to a number of classical wave
recognition problems from the quantum optimal measurement theory (obtained
by the first author, V.P. Belavkin, in the 70's, when solving similar
problems for quantum signals). Content and commentaries to the list of
literature are presented in brief summaries at the beginning of each
section. Similar summaries are given at the beginning of each subsection.
The subsections are written as a series of articles of increasing complexity
so that each can be read independently, although the best way to understand
the material is to carefully read the articles in the order given.

\mainmatter

\chapter{Representation and Measurement of Waves}

In this section we discuss the mathematical apparatus of the wave theory of
representation and measurement of sound and visual patterns. Along with pure
wave patterns, which are described by coherent signals and fields, we also
consider the representation of mixed patterns, which are described by
partially coherent and incoherent signals and fields. In addition to the
spatial-frequency (coordinate) and wave-temporal (momentum) representations
we introduce the joint canonical representation, in which the pure and mixed
patterns are described by entire functions and kernels in a phase complex
space. We develop the mathematical theory of idea filters and quasifilters,
disjoint selectors, and quasiselectors, which describe ordinary, successive,
and indirect measurements of wave-pattern intensity distributions. Using
this theory as a basis, we analyze coordinate and momenta measurements as
well as quasimeasurements of joint coordinate-momentum distributions. The
mathematical tools used here are in many respects similar to those used in
the quantum theory of representations and measurements \cite{3}, which
recently received a new impetus in connection with problems of quantum
states recognition \cite{4}, \cite{5}, \cite{6}, \cite{7}, \cite{8}, \cite{9}%
, \cite{10}; however, we give a wave rather than a statistical
interpretation of the apparatus in accordance with the application
considered here.

\section{Mathematical Description of Wave Patterns}

In this section we describe three basic types of representation of pure and
mixed wave patterns; the coordinate, or spatial-frequency, the momentum, or
wave-temporal, and the canonical, in which the wave patterns are represented
by holomorphic amplitudes in the complex coordinate-momentum plane. The
third representation, which emerged in quantum optics \cite{43}, proves
useful in an analysis of the frequency-temporal structure of sound and
visual patterns and in holography in an analysis of the spatial-temporal
structure of such patterns.

\subsection{Wave Sound and Visual Patterns}

The sound and visual patterns considered here are commonly described by wave
amplitudes $v(t,\mathbf{q})$ that are real-valued function of time $t$ and
coordinates $\mathbf{q}$ in a spatial-temporal region accessible for
measurement. Although the simplest wave equations describing the behaviour
of such physical fields are linear in amplitudes $v(t,\mathbf{q})$, for
purposes of measurement of sound and visual patterns the most interesting
are functions quadratic in $v$ that describe the distribution of sound on
the standards of the dynamic analyzer or of light on photodetectors, a
distribution that in spatial-temporal measurements is described by the
intensity function $v^{2}(t,\mathbf{q})/2$. More useful information is
proved not by the intensities of a sound or visual pattern at points $(t,%
\mathbf{q})$ but by the distribution of the sound intensity at frequency $f$
(a spatial-frequency distribution), as is common in an analysis of colour
patterns. Such a distribution is determined by the intensity function 
\begin{equation}
\iota (f,\mathbf{q})=|\varphi (f,\mathbf{q})|^{2},\quad f\geq 0,
\label{1 1 1}
\end{equation}%
with $\varphi $ the complex-valued spectral amplitudes, 
\begin{equation*}
\varphi (f,\mathbf{q})=\int_{-\infty }^{\infty }v(t,\mathbf{q})\mathrm{e}%
^{2\pi \mathrm{j}ft}\,\mathrm{d}t,
\end{equation*}%
used to represent the wave field $v(t,\mathbf{q})$ in the form of a linear
combination of harmonic oscillations at each point $\mathbf{q}$: 
\begin{equation*}
v(t,\mathbf{q})=2\func{Re}\int_{0}^{\infty }\varphi (f,\mathbf{q})\mathrm{e}%
^{-2\pi \mathrm{j}ft}\,\mathrm{d}f.
\end{equation*}%
Bearing in mind the well-known advantage of employing complex-valued
amplitudes, in what follows we consider complex-valued signals $\varphi
(x)=\varphi (f,\mathbf{q})\equiv \varphi (q)$, with $x=f(f,\mathbf{q})\equiv
q$, assuming that in a given spatial-frequency region of measurements $%
\Omega $ they possess a finite total intensity 
\begin{equation}
I(\varphi )=\int_{\Omega }|\varphi (q)|^{2}\,\mathrm{d}q.  \label{1 1 2}
\end{equation}%
When registering sound signals for which the spatial region of measurements
is much smaller than the characteristic length of the sound wave, we can
take for $\Omega $ a one-dimensional region, which is usually determined by
a positive-frequency pass band $\Phi \in \mathbb{R}_{+}$ of the dynamic
analyzer, assuming that $x=f$ at the point of its localization $\mathbf{q}=0$%
; recognizing visual patterns usually requires only a three-dimensional
region $\Omega =\Phi \times S$, where $\Phi $ is the optical frequency band
and $S$ the surface on which the pattern is localized; for static patterns $%
x=\mathbf{q}$\quad $(t=0)$.

From the standpoint of physics the admissible amplitudes are smooth
amplitudes $\varphi (q)$, $q\in \Omega $, with compact supports inside a $%
(d+1)$-dimensional region $\Omega \subseteq \mathbb{R}^{d+1}$ or, if $\Omega 
$ is noncompact, amplitudes that fall off rapidly at infinity together with
all their derivatives. Such amplitudes generate a Hilbert space $\mathcal{H}%
=L^{2}(\Omega )$ of amplitudes $\chi (q)$ of finite intensity $\left\Vert
\chi \right\Vert ^{2}<\infty $: 
\begin{equation}
(\varphi \mid \chi )=\int_{\Omega }\varphi (q)^{\ast }\chi (q)\,\mathrm{d}q.
\label{1 1 3}
\end{equation}%
Generally the set $\mathcal{D}$ of basic amplitudes $\varphi $ can form an
arbitrary complex-values space with a positive Hermitian form $I(\varphi
)=(\varphi \mid \varphi )$ that is invariant with respect to complex
conjugation $\varphi ^{\ast }\left( q\right) =\varphi (q)^{\ast }$. This
form defines a finite intensity, $I(\varphi )\neq 0$, for any nonzero $%
\varphi $. We denote the completion of this space in norm $\left\Vert
\varphi \right\Vert =I^{1/2}(\varphi )$ by $\mathcal{H}$ and consider it to
be a Hilbert space equipped with an isometric involution $\chi \mapsto \chi
^{\ast }$ with respect to the scalar produce (\ref{1 1 3}), which is linear
in the second argument $\varphi \in \mathcal{H}$.

The use of complex-valued amplitudes not only considerably simplifies the
formulas for calculating the observed distributions of the fields but also
makes it possible to employ analogies from quantum theory. Specifically, as
in quantum theory, complex-valued amplitudes differing in a phase factor
must be assumed ``equal'' since they lead to the same intensities defined by
Hermitian forms of $\varphi $. Note that in the quantum description of
optical and sound signals we usually take the mean number of the
corresponding quanta (photons and phonons) as the intensity functions for
light and sound, respectively. These quantities are determined by the same
Hermitian forms of the complex-valued amplitudes $\varphi $ as in the
classical mode of description, provided that the quantum mechanical states
are coherent \cite{43}, \cite{44}, that is, are described by Poisson
probability amplitudes $|\varphi \rangle $. Thus, restricting ourselves to
intensity measurements, we postulate that only distributions of quanta are
observable, while the only characteristics of signals $\varphi $ of interest
to physics are those obtained as a result of measurement of such
distributions.

\subsection{Momentum Representation}

In problems dealing with the recognition of moving patterns what may be of
interest is not the spatial-frequency intensity distribution (\ref{1 1 1})
but the momentum-temporal distribution described by the function 
\begin{equation}
\widetilde{\iota }(t,\mathbf{p})=|\widetilde{\varphi }(t,\mathbf{p}%
)|^{2},\quad (t,\mathbf{p})\in \mathbb{R}^{d+1},  \label{1 1 4}
\end{equation}%
where $\widetilde{\varphi }(t,\mathbf{p})$ is the involution Fourier
transform, 
\begin{equation}
\widetilde{\varphi }(t,\mathbf{p})=\iint \varphi (f,\mathbf{q})^{\ast }%
\mathrm{e}^{2\pi \mathrm{j}(tf+\mathbf{p\cdot q})}\,\mathrm{d}f\,\mathrm{d}%
\mathbf{q}.  \label{1 1 5}
\end{equation}%
We introduce the notation $x=(t,\mathbf{p})\equiv p$. The representation of
amplitudes $\varphi $ in terms of the functions $\widetilde{\varphi }(x)=%
\widetilde{\varphi }(t,\mathbf{p})\equiv \widetilde{\varphi }(p)$ is called
the momentum representation.\footnote{%
Thanks to the introduction of the involution $\varphi \mapsto \varphi ^{\ast
}$ in transformation (\ref{1 1 5}), the inverse transformation to the
coordinate representation is carried out by the same formula (\ref{1 1 5} ), 
$\varphi =\widetilde{\widetilde{\varphi }}$, which can be extended onto
generalized amplitudes $\chi $ in the standard manner.} Note that this
representation differs from the common one by complex conjugation, $%
\widetilde{\varphi }^{\ast }=\widetilde{\varphi }^{\ast }$, but this
difference is \textquotedblleft unobservable\textquotedblright\ from the
viewpoint of measuring intensities described by Hermitian forms (\ref{1 1 1}%
) and (\ref{1 1 4}), which are invariant under such conjugation.

Allowing for Plancherel's equality 
\begin{equation}
\int |\widetilde{\varphi }(p)|^{2}\mathrm{d}p=\int |\varphi (q)|^{2}\mathrm{d%
}q,  \label{1 1 6}
\end{equation}%
we find that the total intensity described by the distribution function (\ref%
{1 1 4}) coincides with the total intensity (\ref{1 1 2}) for any amplitude $%
\varphi $ with support in $\Omega $: 
\begin{equation}
\int \widetilde{\iota }(p)\,\mathrm{d}p=I(\varphi )=\int \iota (q)\,\mathrm{d%
}q.  \label{1 1 7}
\end{equation}

In what follows we consider the values of $\iota (q)$ and $\widetilde{\iota }%
(p)$ in (\ref{1 1 1}) and (\ref{1 1 4}) as being functions of $\varphi $,
denoting by $\widetilde{\iota }(\varphi ,q)$ and $\widetilde{\iota }(\varphi
,p)$ the functionals connected by involutions $\widetilde{\iota }(\varphi
,p)=\iota (\widetilde{\varphi },p)$ and $\widetilde{\widetilde{\iota }}%
(\varphi ,q)=\iota (\varphi ,q)$, respectively. Note that the Fourier
transformation $\iota (q)\mapsto \widetilde{\iota }(p)$ of the Hermitian
functional $\iota (\varphi ,q)=|\varphi (q)|^{2}$ cannot be reduced to the
Fourier transformation of its value $\iota (\varphi )$ as a function of $q$.
More than that, measuring the spatial distribution $\widetilde{\iota }(q)$
for a single value of $\varphi $ does not generally make it possible in any
way to calculate the corresponding distribution $\widetilde{\iota }(p)$, and
vice versa. Nevertheless, these distributions satisfy certain relationships,
the simplest of which are (\ref{1 1 7}) and the inequality 
\begin{equation}
\int (p-\overline{p})^{2}\widetilde{\iota }(p)\,\mathrm{d}p\int (q-\overline{%
q})^{2}\iota (q)\,\mathrm{d}q\geq \tfrac{1}{(4\pi )^{2}}I^{2}(\varphi )
\label{1 1 8}
\end{equation}%
(where $\overline{p}=\int p\widetilde{\iota }(p)\,\mathrm{d}p/I(\varphi )$
and $\overline{q}=\int q\iota (q)\,\mathrm{d}q/I(\varphi )$ for each of the
components $p=p_{k}$ and $q=q_{k}$, $k=0,\ldots ,d$), which is known as the
uncertainty relation. Using the commutation relations 
\begin{equation}
\widehat{p}\widehat{q}-\widehat{q}\widehat{p}=(2\pi \mathrm{j})^{-1}\widehat{%
1}  \label{1 1 9}
\end{equation}%
For each pair of operators $\widehat{q},\widehat{p}$ in the $p$%
-representation, with $\widehat{p}=p-\overline{p}$ and $\widehat{q}=\partial
/\partial (2\pi \mathrm{j}p)=\overline{q}$, we can easily arrive at (\ref{1
1 8}) as a corollary of Schwarz's inequality 
\begin{equation}
\left\Vert \widehat{p}\widetilde{\varphi }\right\Vert \,\left\Vert \widehat{q%
}\widetilde{\varphi }\right\Vert \geq \left\vert (\widehat{p}\widetilde{%
\varphi }\mid \widehat{q}\widetilde{\varphi })\right\vert \geq \left\vert 
\func{Im}(\widehat{p}\widetilde{\varphi }\mid \widehat{q}\widetilde{\varphi }%
)\right\vert =\tfrac{1}{2\pi }\left\Vert \widetilde{\varphi }\right\Vert
^{2}.  \label{1 1 10}
\end{equation}%
Indeed, according to definition (\ref{1 1 4}) we have 
\begin{equation*}
\int (p-\overline{p})^{2}\widetilde{\iota }(p)\,\mathrm{d}p=\int \left\vert
(p-\overline{p})\widetilde{\varphi }(p)\right\vert ^{2}\,\mathrm{d}%
p=\left\Vert \widehat{p}\widetilde{\varphi }\right\Vert ^{2}
\end{equation*}%
and, similarly, allowing for (\ref{1 1 6}), for the Fourier transform $%
\widehat{p}\widetilde{\varphi }$ of the function $(q-\overline{q})\varphi
(q) $ we obtain from (\ref{1 1 1}) 
\begin{equation*}
\int (q-\overline{q})^{2}\iota (q)\,\mathrm{d}q=\int \left\vert (q-\overline{%
q})\varphi (q)\right\vert ^{2}\,\mathrm{d}q=\left\Vert \widehat{q}\widetilde{%
\varphi }\right\Vert ^{2}.
\end{equation*}%
Thus, inequality (\ref{1 1 8}) is equivalent to (\ref{1 1 10}), where $%
\left\Vert \widetilde{\varphi }\right\Vert ^{2}=I(\varphi )$. For nonzero
amplitudes $\varphi $ this inequality is usually written as 
\begin{equation}
\sigma _{p}\sigma _{q}\geq 1/4\pi ,  \label{1 1 11}
\end{equation}%
where $\sigma _{p}$ and $\sigma _{q}$ are the standard deviations, 
\begin{eqnarray}
\sigma _{p}^{2} &=&\int (p-\overline{p})^{2}\widetilde{\iota }(p)\,\mathrm{d}%
p/I(\varphi ),  \notag \\
\sigma _{q}^{2} &=&\int (q-\overline{q})^{2}\iota (q)\,\mathrm{d}q/I(\varphi
),  \label{1 1 12}
\end{eqnarray}%
of momentum $p$ and coordinate $q$ in the wave packet $\varphi $ from their
mean values $\overline{p}$ and $\overline{q}$. In this form (\ref{1 1 11})
is similar to the quantum mechanical Heisenberg uncertainty relation;
however, here the standard deviations (\ref{1 1 12}) have no statistical
meaning but characterize the extent to which the intensity distributions are
localized in the coordinate and momentum spaces. The lower bound $1/(4\pi )$
in this relation is achieved only in the case of an unbounded region $\Omega
=\mathbb{R}^{d+1}$ for the Gaussian amplitudes 
\begin{equation}
\varphi (q)=C_{q}\exp \{2\pi \mathrm{j}(q-\frac{\overline{q}}{2})\overline{q}%
-\frac{\left\vert q-\overline{q}\right\vert ^{2}}{4\sigma _{q}^{2}}\}.
\label{1 1 13}
\end{equation}%
These amplitudes, with the normalization constants $C_{q}=1/(2\pi \sigma
_{p}^{2})^{(d+1)/4}$, have a similar form in the $p$-representation: 
\begin{equation*}
\widetilde{\varphi }(p)=C_{p}\exp \{2\pi \mathrm{j}(p-\frac{\overline{p}}{2})%
\overline{q}-\frac{\left\vert p-\overline{p}\right\vert ^{2}}{4\sigma
_{p}^{2}}\},
\end{equation*}%
with $C_{p}=1/(2\pi \sigma _{p}^{2})^{(d+1)/4}$ and $\sigma _{p}\sigma
_{q}=1/(4\pi )$, are called standard canonical (Poisson) amplitudes and are
denoted by $\psi _{\alpha }=|\alpha )$, with%
\begin{equation*}
\alpha =\frac{1}{2}(\frac{\overline{q}}{\sigma _{q}}+\mathrm{j}\frac{%
\overline{p}}{\sigma _{p}})
\end{equation*}%
if $\sigma _{q}$ and $\sigma _{p}$ are fixed. Note that for $\alpha \neq
\alpha ^{\prime }$ such amplitudes are nonorthogonal: 
\begin{equation}
\left( \alpha \mid \alpha ^{\prime }\right) =\exp \{-\frac{1}{2}\left\vert
\alpha ^{\prime }\right\vert ^{2}+\alpha ^{\prime }\alpha ^{\dagger }-\frac{1%
}{2}\left\vert \alpha \right\vert ^{2}\},  \label{1 1 14}
\end{equation}%
with $\alpha ^{\prime }\alpha ^{\dagger }$ defined as the scalar product $%
\sum_{i=0}^{d}\alpha _{i}^{\ast }\alpha _{i}^{\prime }$.

\subsection{Mixed Signals}

Due to limits in present-day technology, only a fraction of the information
on the intensity distribution of amplitude $\varphi $ in this or another
region can usually be obtained when analyzing sound and visual patterns. For
instance, in sound pattern recognition the common method is to use only the
frequency or temporal distribution obtain through integration 
\begin{equation}
\iota (f)=\int \left\vert \varphi (f,\mathbf{q})\right\vert ^{2}\,\mathrm{d}%
\mathbf{q},\;\;\,\widetilde{\iota }(t)=\int \left\vert \widetilde{\varphi }%
(t,\mathbf{p})\right\vert ^{2}\,\mathrm{d}\mathbf{p}  \label{1 1 15}
\end{equation}%
of distributions (\ref{1 1 1}) and (\ref{1 1 4}) over the spatial or wave
region of measurement. In visual pattern recognition often only black and
white patterns are considered. These are obtained as the result of mixing 
\begin{equation}
\iota (\mathbf{q})=\int \left\vert \varphi (f,\mathbf{q})\right\vert ^{2}%
\mathrm{d}f,\,\;\;\widetilde{\iota }(\mathbf{p})=\int \left\vert \widetilde{%
\varphi }(t,\mathbf{p})\right\vert ^{2}\,\mathrm{d}t  \label{1 1 16}
\end{equation}%
of the appropriate colour patterns in the spatial or wave region of
measurement. To obtain such incomplete distributions there is no need to
provide a total description of the signal by amplitude $\varphi (f,\mathbf{q}%
)$. For instance, in describing sound it is sufficient to specify only the
Hermitian kernel 
\begin{equation*}
S(f^{\prime },f)=\int \varphi (f^{\prime },\mathbf{q})\varphi ^{\ast }(f,%
\mathbf{q})\,\mathrm{d}q,
\end{equation*}%
for which $\iota (f)=S(f,f)$ and $\widetilde{\iota }(t)=\widetilde{S}(t,t)$,
where 
\begin{equation*}
\widetilde{S}(t^{\prime },t)=\int \mathrm{e}^{2\pi \mathrm{j}(ft^{\prime
}-f^{\prime }t)}S(f^{\prime },f)\,\mathrm{d}f^{\prime }\,\mathrm{d}f.
\end{equation*}%
Monochrome patterns are defined by a similar kernel $S(\mathbf{q}^{\prime },%
\mathbf{q})$, with $\iota (\mathbf{q})=S(\mathbf{q,q)}$ and $\widetilde{%
\iota }(\mathbf{p})=\widetilde{S}(\mathbf{p,p})$. Having in mind the
possibility of such mixing, we will describe signals in an abridged manner
by nonnegative definite operators of intensity density, $S$, with kernels $%
S(q,q^{\prime })$ that have a nonzero trace 
\begin{equation}
\mathrm{Tr}S=\int_{\Omega }S(q,q)\,\mathrm{d}q\equiv \left\langle
S,I\right\rangle ,  \label{1 1 17}
\end{equation}%
which determines the total intensity, $\iota (S)=\left\langle
S,I\right\rangle $, of the signal in $\Omega $. To each amplitude $\psi (q)$
we assign a one-dimensional operator $S=|\psi )(\psi |$ with a kernel 
\begin{equation}
S(q^{\prime },q)=\psi (q^{\prime })\psi ^{\ast }(q),  \label{1 1 18}
\end{equation}%
which defines the amplitude $\psi (q)$ to within a nonessential phase factor 
$\mathrm{e}^{\mathrm{j}\theta }$. The diagonal values $\iota (q)=S(q,q)$
describe the coordinate distribution of the intensity of such a signal,
while the momentum distribution is described by the diagonal values $%
\widetilde{\iota }(p)=\widetilde{S}(p,p)$ of the involution Fourier
transform 
\begin{equation}
\widetilde{S}(p^{\prime },p)=\int \mathrm{e}^{2\pi \mathrm{j}(p^{\prime
}q^{\intercal }-q^{\prime }p^{\intercal })}S(q^{\prime },q)\,\mathrm{d}%
q^{\prime }\,\mathrm{d}q.  \label{1 1 19}
\end{equation}%
Each such kernel can be obtained as a result of mixing 
\begin{equation}
S(q^{\prime },q)=\int \psi _{\alpha }(q^{\prime })\psi _{\alpha }^{\ast
}(q)\nu (\mathrm{d}\alpha )  \label{1 1 20}
\end{equation}%
of the one-dimensional kernels corresponding to the normalized amplitudes $%
\{\psi _{\alpha }\}$, $\left\Vert \psi _{\alpha }\right\Vert =1$,
parameterized by a space $A$ with a nonzero positive measure $\nu $ whose
mass determines the total intensity $\left\langle S,I\right\rangle =\nu (A)$%
. For example, the kernel $S(\mathbf{q}^{\prime },\mathbf{q})$ corresponding
to a monochrome pattern generated by amplitude $\varphi (f,\mathbf{q})$ can
be written in the form (\ref{1 1 20}) for%
\begin{equation*}
\psi _{f}(\mathbf{q})=\varphi (f,\mathbf{q})/\iota ^{1/2}(f)
\end{equation*}
on the set $A$ of frequencies $\Phi $ equipped with a nonzero measure $\nu (%
\mathrm{d}f)=\iota (f)\,\mathrm{d}f$.

In the case of an arbitrary Hilbert space $\mathcal{H}$, mixed signals are
described by density operators $S$ obtained as a result of weak integration 
\begin{equation}
S=\int |\psi _{\alpha })(\psi _{a}|\nu (d\alpha )  \label{1 1 21}
\end{equation}%
of one-dimensional density operators $S_{\psi }=|\psi )(\psi |$, 
\begin{equation}
|\psi )(\psi |:\chi \in \mathcal{H}\mapsto \psi (\psi \mid \chi )=(\psi \mid
\chi )\psi ,  \label{1 1 22}
\end{equation}%
corresponding to the normalized values $\psi _{\alpha }\in \mathcal{H}%
,\alpha \in A$, of the vector function $\alpha \mapsto \psi _{\alpha }$. The
operators (\ref{1 1 21}) are kernel-positive and have a finite trace $%
\mathrm{Tr\,}S=\nu (A)$, with each operator $S:\mathcal{H}\mapsto \mathcal{H}
$ being represented in the form (\ref{1 1 21}) via, say, the spectral
decomposition 
\begin{equation}
S=\sum_{i}|\psi _{i})(\psi _{i}|\nu _{i}.  \label{1 1 23}
\end{equation}%
Here $\{\psi _{i}\}$ is the maximal orthogonal set of normalized
eigenvectors $\psi _{i}\in \mathcal{H}$ corresponding to zero eigenvalues $%
\nu _{i}:S\psi _{i}=\nu _{i}\psi _{i}$, which determine the trace $\mathrm{Tr%
}\,S=\sum_{i}\nu _{i}$.

\subsection{Gaussian Signals}

As an example let us consider the important class of mixed canonical signals
defined by the integration 
\begin{equation}
S=\int |\alpha )(\alpha |\nu (d\xi \,d\eta ),\quad \alpha \in \mathbb{C}%
^{d+1},  \label{1 1 24}
\end{equation}%
of canonical projectors corresponding to the amplitudes $\psi _{\xi \eta
}=|\alpha )$, which in the coordinate representation have the following
general Gaussian form (c.f. (\ref{1 1 13})) 
\begin{equation}
\psi _{\xi \eta }(q)=C\exp \{2\pi \mathrm{j}(q-\tfrac{1}{2}\xi )\eta
^{\intercal }-\tfrac{1}{2}(q-\xi )\omega (q-\xi )^{\intercal }\}.
\label{1 1 25}
\end{equation}%
Here $\omega =\omega ^{\intercal }$ is a symmetric complex-valued $(d+1)$-by-%
$(d+1)$ matrix with a positive definite real part 
\begin{equation*}
\omega +\omega ^{\ast }=2\pi (\upsilon ^{\dagger }\upsilon )^{-1},
\end{equation*}%
$\xi =\sqrt{2/\pi }\func{Re}\alpha \upsilon $ and $\eta =\sqrt{2/\pi }\func{%
Re}\mathrm{j}\alpha ^{\dagger }\widetilde{\upsilon }$, with $\widetilde{%
\upsilon }=\upsilon ^{\dagger }\omega /(2\pi )$, are $(d+1)$-dimensional
rows, $\xi ^{\intercal }$ and $\eta ^{\intercal }$ are the corresponding
columns,%
\begin{equation*}
|C|^{2}=\det (\upsilon ^{\dagger }\upsilon )^{-1}=\left\vert \upsilon
\right\vert ^{-2}
\end{equation*}%
is the normalization constant, and $\upsilon ^{\ast \intercal }=\upsilon
^{\dagger }=\upsilon ^{\intercal \ast }$ is the Hermitian conjugate of
matrix $\upsilon $.

Let $\zeta =(\xi ,\eta )$ be a $2(d+1)$-dimensional row and $\nu (d\zeta
)=\nu (d\xi \,d\eta )$ a Gaussian measure on $A=\mathbb{R}^{2(d+1)}$
normalized to certain number $J<\infty $ (the Gaussian intensity) and
described (for $J$ positive) by the following moments:%
\begin{equation}
\overline{\xi }=J^{-1}\int \xi \nu (d\zeta )=\lambda ,\;\;\;\,\overline{\eta 
}=J^{-1}\int \eta \nu (d\zeta )=\varkappa ,  \label{1 1 26}
\end{equation}%
\begin{equation}
\left[ 
\begin{array}{ll}
\overline{\xi ^{\intercal }\xi } & \overline{\xi ^{\intercal }\eta } \\ 
\overline{\eta ^{\intercal }\xi } & \overline{\eta ^{\intercal \eta }}%
\end{array}%
\right] =J^{-1}\int \zeta ^{\intercal }\zeta \nu (d\zeta )=\left[ 
\begin{array}{ll}
\lambda ^{\intercal }\lambda & \lambda ^{\intercal }\varkappa \\ 
\varkappa ^{\intercal }\lambda & \varkappa ^{\intercal }\varkappa%
\end{array}%
\right] +\sigma _{\zeta \zeta },  \label{1 1 27}
\end{equation}%
where $\sigma _{\xi \xi }=\left[ 
\begin{array}{ll}
\sigma _{\xi \xi } & \sigma _{\xi \eta } \\ 
\sigma _{\eta \xi } & \sigma _{\eta \eta }%
\end{array}%
\right] $ is a nonnegative define $2(d+1)$-by-$2(d+1)$ matrix. The signals
that correspond to such a density operator (\ref{1 1 24}) are characterized
by the following first moments:%
\begin{equation}
\overline{q}=J^{-1}\int Q(\psi _{\zeta })\nu (d\zeta )=\lambda ,\,\overline{p%
}=J^{-1}\int P(\psi _{\zeta })\nu (d\zeta )=\varkappa ,  \label{1 1 28}
\end{equation}%
\begin{eqnarray}
\overline{q^{\intercal }q} &=&J^{-1}\int (Q^{\intercal }\psi _{\zeta }\mid
Q\psi _{\zeta })\nu (d\zeta )=\overline{\xi ^{\intercal }\xi }+(\omega
+\omega ^{\ast })^{-1},  \notag \\
\overline{q^{\intercal }p} &=&J^{-1}\int (Q^{\intercal }\psi _{\zeta }\mid
P\psi _{\zeta })\nu (d\zeta )=\overline{\xi ^{\intercal }\eta }+j\upsilon
^{\intercal }\widetilde{\upsilon }/(2\pi ),  \label{1 1 29} \\
\overline{p^{\intercal }q} &=&J^{-1}\int (P^{\intercal }\psi _{\zeta }\mid
Q\psi _{\zeta })\nu (d\zeta )=\overline{\eta ^{\intercal }\xi }-j\widetilde{%
\upsilon }^{\dagger }\upsilon ^{\ast }(2\pi ),  \notag \\
\overline{p^{\intercal }p} &=&J^{-1}\int (P^{\intercal }\psi _{\zeta }\mid
P\psi _{\zeta })\nu (d\zeta )=\overline{\eta ^{\intercal }\eta }+(\widetilde{%
\omega }+\widetilde{\omega }^{\ast })^{-1},  \notag
\end{eqnarray}%
where$\,\widetilde{\omega }/(2\pi )=2\pi /\omega ^{\ast }$, $Q$ and $P$ are
the rows of operators of position $Q_{k}$ and momentum $P_{k}$ defined in
the $q$-representation via multiplication by $q_{k}$ and differentiation
with respect to $q_{k}$, or $(2\pi \mathrm{j})^{-1}\partial /\partial q_{k}$%
, and we have allowed for the fact that%
\begin{equation*}
Q(\psi _{\zeta })=(\psi _{\xi }\mid Q\psi _{\zeta })=\xi ,\,P(\psi _{\zeta
})=(\psi _{\zeta }|P\psi _{\zeta })=\eta ,
\end{equation*}%
\begin{equation}
\left[ 
\begin{array}{ll}
(\widehat{q}^{\intercal }\psi _{\zeta }\mid \widehat{q}\psi _{\zeta }) & (%
\widehat{q}^{\intercal }\psi _{\zeta }\mid \widehat{p}\psi _{\zeta }) \\ 
(\widehat{p}^{\intercal }\psi _{\zeta }\mid \widehat{q}^{\intercal }\psi
_{\zeta }) & (\widehat{p}^{\intercal }\psi _{\zeta }\mid \widehat{p}\psi
_{\zeta })%
\end{array}%
\right] =(\upsilon ^{\ast },\mathrm{j}\widetilde{\upsilon })+(\upsilon
^{\ast },\widetilde{\upsilon })/2\pi )  \label{1 1 30}
\end{equation}%
for the \textquotedblleft shifted\textquotedblright\ operators $\widehat{q}%
=Q-\overline{q}I$ and $\widehat{p}=P-\overline{p}I$.

It can easily be demonstrated that for a nonsingular Gaussian measure
described by density $n(\zeta )=\nu (d\zeta )/\mathrm{d}\zeta $ of the form 
\begin{equation}
n(\zeta )=C\exp \left\{ -\tfrac{1}{2}(\zeta -\theta )\sigma _{\zeta \zeta
}^{-1}(\zeta -\theta )^{\intercal }\right\}  \label{1 1 31}
\end{equation}%
(with $\theta =(\varkappa ,\lambda )$ and $C=J/\sqrt{\det 2\pi \sigma
_{\zeta \zeta }}$) corresponding to a nonsingular correlation matrix $\sigma
_{\zeta \zeta }$ we can select representation (\ref{1 1 24}) of the density
operator $S$ by appropriate choice of matrix $\omega $ in such a manner that
the representation will be described by density (\ref{1 1 31}) with the
matrix $\sigma _{\zeta \zeta }$ of the form 
\begin{equation}
\sigma _{\zeta \zeta }=\frac{1}{\pi }\func{Re}\left[ (\upsilon ^{\ast },%
\mathrm{j}\widetilde{\upsilon })^{\dagger }s(\upsilon ^{\ast },\mathrm{j}%
\widetilde{\upsilon })\right]  \label{1 1 32}
\end{equation}%
where $s$ is a complex-valued positive definite $(d+1)$-by-$(d+1)$ matrix.
At this point it is expedient to introduce a complex-valued normal
representation characterized by the transition made from $2(d+1)$ real
variables $\zeta =(\xi ,\eta )$ to a $(d+1)$-dimensional complex variables%
\begin{equation*}
\alpha =\frac{1}{\gamma }(\xi \omega +2\pi \mathrm{j}\eta ),
\end{equation*}%
where $\gamma =(\omega +\omega ^{\ast })^{1/2}$ in terms of which the
density (\ref{1 1 31}) combined with (\ref{1 1 32}) can be written in the
following form: 
\begin{equation}
n(\alpha ,\alpha ^{\ast })=C\exp \{-(\alpha -\theta )^{\ast }s^{-1}(\alpha
-\theta )^{\intercal }\},  \label{1 1 33}
\end{equation}%
where $\theta =(\varkappa \omega +2\pi \mathrm{j}\lambda )\gamma ^{-1}$, and 
$C=J/\det s$ if density (\ref{1 1 33}) is normalized to $J$ with respect to $%
\mathrm{d}\alpha \,\mathrm{d}\alpha ^{\ast }=\mathrm{d}\xi \,\mathrm{d}\eta $%
. Note that, as in the case with (\ref{1 1 13}), amplitudes (\ref{1 1 25})
must be written in the form 
\begin{equation}
|\alpha )(q)=(\gamma /\sqrt{2\pi })^{1//2}\exp \left\{ (q\gamma -\func{Re}%
\alpha )\alpha ^{\dagger }-\tfrac{1}{2}q\omega \,q^{\intercal }\right\} ,
\label{1 1 34}
\end{equation}%
with the scalar product defined in(\ref{1 1 14}). However, in contrast to (%
\ref{1 1 13}), these amplitudes do not realize at $\func{Im}\omega \neq 0$
the lower bound in the uncertainty relation (\ref{1 1 11}) while they do
realize a more exact lower bound defined by the matrix inequality 
\begin{equation}
\det \left[ 
\begin{array}{ll}
\sigma _{qq} & \sigma _{qp} \\ 
\sigma _{pq} & \sigma _{pp}%
\end{array}%
\right] \geq 0,\text{ or }\sigma _{pp}\geq \sigma _{pq}\sigma
_{qq}^{-2}\sigma _{qp},  \label{1 1 35}
\end{equation}%
provided that matrix $\sigma _{qq}$ is nonsingular. Here 
\begin{eqnarray}
\sigma _{qq} &=&\sigma _{\xi \xi }+\gamma ^{-1},\,\;\;\;\;\;\;\;\sigma
_{pp}=\sigma _{\eta \eta }+\omega ^{\ast }\gamma ^{-2}\omega /(2\pi )^{2}, 
\notag \\
\sigma _{pq} &=&\sigma _{\eta \xi }+\omega ^{\ast }\gamma ^{-1}/(2\pi 
\mathrm{j}),\,\sigma _{qp}=\sigma _{\xi \eta }-\omega \gamma ^{-2}/(2\pi 
\mathrm{j})  \notag
\end{eqnarray}%
are the elements of the correlation matrices, 
\begin{eqnarray}
\sigma _{qq} &=&J^{-1}\int (\widehat{q}^{\intercal }\psi _{\alpha }\mid 
\widehat{q}\psi _{\alpha })\nu \left( d\alpha \right) ,  \notag \\
\sigma _{qp} &=&J^{-1}\int (q^{\intercal }\psi _{\alpha }\mid \widehat{p}%
\psi _{\alpha })\nu \left( d\alpha \right) ,  \notag \\
\sigma _{pq} &=&J^{-1}\int (\widehat{p}^{\intercal }\psi _{\alpha }\mid 
\widehat{q}\psi _{\alpha })\nu \left( d\alpha \right) ,  \label{1 1 37} \\
\sigma _{pp} &=&J^{-1}\int (\widehat{p}^{\intercal }\psi _{\alpha }\mid 
\widehat{p}\psi _{\alpha })\nu \left( d\alpha \right) ,  \notag
\end{eqnarray}%
which are defined for nonmixed canonical signals in (\ref{1 1 30}) and
satisfy, obviously, the following relation: 
\begin{equation}
\sigma _{q}(\sigma _{p}^{2}-\rho ^{\ast }\sigma _{p}^{2}\rho ^{\intercal
})\sigma _{q}=1/(4\pi )^{2},  \label{1 1 38}
\end{equation}%
with $\sigma _{p}^{2}=\sigma _{pp},\sigma _{q}^{2}=\sigma _{qq}$, and $\rho
=\omega ^{-1}\func{Im}\omega $; at $\rho =0$ this relation realizes the
bound of (\ref{1 1 11}). Otherwise, it realizes the bound to (\ref{1 1 35}).$%
\,$

\subsection{Canonical Representations}

In problems dealing with the recognition of complicated sound and visual
patterns, the most important information is usually contained in the
momentum representation as well as in the coordinate representation. For
example, in analyzing speech not only the frequency distribution of its
intensity is important but so is its temporal distribution, in the same way
as in colour pattern recognition it has proved important to know the wave
structure in addition to the spatial structure. although there can be no
joint coordinate-momentum representation that would enable calculating such
distributions simultaneously (due to noncommutativity of position and
momentum operators), the simultaneous estimate, say by the human ear, of the
frequency and temporal structures of sound points to the possibility of
building a mathematical model of such perception, which may prove extremely
important for automatic speech recognition.

The simplest models of such joint coordinate-momentum representations are
those whose densities are defined as the intensities 
\begin{equation}
k(z)=|(\psi _{z}\mid \varphi )|^{2},\,\;\;z=(x,y)\in \mathbb{R}^{2(d+1)},
\label{1 1 39}
\end{equation}%
of projections of amplitude $\varphi $ on the canonical amplitudes (\ref{1 1
24}) at $\zeta =z$, which are parametrized at a fixed $\omega $ by the
estimate vectors $\xi =x$ and $\eta =y$ of the generalized coordinates $%
x=(x_{0},\ldots ,x_{d})$ and momenta $y=(y_{0},\ldots ,y_{d})$. We can
directly verify that the density in $x$ has the form 
\begin{equation}
m(x)=\int k(x,y)dy=|\upsilon |^{-1/2}\int \mathrm{e}^{-\pi |(x-q)\upsilon
^{-1}|^{2}}|\varphi (q)|^{2}dq,  \label{1 1 40}
\end{equation}%
which in the limit of $|\upsilon |=\sqrt{\det \upsilon ^{\dagger }\upsilon }%
\rightarrow \infty $ coincides with the coordinate distribution $\iota
(q)=|\varphi (q)|^{2}$. Similarly, in the $p$-representation we find the
density in $y$: 
\begin{eqnarray}
\widetilde{m}(y) &=&\int k(x,y)dx  \notag \\
&=&|\widetilde{\upsilon }|^{-1/2}\int \mathrm{e}^{-\pi |(y-p)\widetilde{%
\upsilon }^{-1}|^{2}}|\widetilde{\varphi }(p)|^{2}dp.  \label{1 1 41}
\end{eqnarray}%
which in the limit of $|\widetilde{\upsilon }|\rightarrow \infty $ coincides
with the momentum distribution $\iota (p)=|\widetilde{\varphi }(p)|^{2}$.
Here for every amplitude $\varphi $ we have 
\begin{eqnarray}
\left\Vert \varphi \right\Vert ^{2} &=&\int m(x)dx=\iint k(x,y)\,dx\,dy 
\notag \\
&=&\int \widetilde{m}(y)\,dy=\left\Vert \widetilde{\varphi }\right\Vert ^{2},
\label{1 1 42}
\end{eqnarray}%
which means that the set of canonical amplitudes $\{\psi _{z}\mid z\in 
\mathbb{R}^{2d}\}$ is complete for every $\omega $, with $\func{Re}\omega $
positive. Thus, the $2(d+1)$-parametric set $\{\psi _{z}\}$ forms a
nonorthogonal base that defines for each $\omega $ a canonical
representation in which the diagonal elements of the kernel $(\psi _{z}\mid
S\psi _{2})$ of a signal described by density operator $S$ yield the density
of the distribution of the signal's intensity $\iota (S)=\left\langle
S,I\right\rangle $. For mixed canonical signals (\ref{1 1 14}) has the form 
\begin{equation}
k(x,y)=\int \exp \{-\left\vert c-\alpha \right\vert ^{2}\}n(\xi ,\eta )\,%
\mathrm{d}\xi ,\,\mathrm{d}\eta ,  \label{1 1 43}
\end{equation}%
where $c=(x\omega +2\pi \mathrm{j}y)\upsilon ^{\dagger }/2\pi $,
specifically, for Gaussian signals (\ref{1 1 25}) we arrive at the following
Gaussian density: 
\begin{equation}
k(z)=J\exp \left\{ -\tfrac{1}{2}(z-\theta )\sigma _{zz}^{-1}(z-\theta
)^{\intercal }\right\} /\sqrt{\det 2\pi \sigma _{zz}},  \label{1 1 44}
\end{equation}%
where 
\begin{equation}
\sigma _{zz}=\left[ 
\begin{array}{ll}
\sigma _{xx} & \sigma _{xy} \\ 
\sigma _{yx} & \sigma _{yy}%
\end{array}%
\right] =\sigma _{\zeta \zeta }+2(\upsilon ^{\dagger },\mathrm{j}\widetilde{%
\upsilon })+(\upsilon ^{\ast },\mathrm{j}\widetilde{\upsilon }^{\intercal
})/\pi  \label{1 1 45}
\end{equation}%
is a $(d+1)$-by-$(d+1)$ correlation matrix (as usually $\mathrm{j}=\sqrt{-1}$%
).

A remarkable property of canonical representations is the possibility of
calculating the intensity distribution in any representation knowing only
one canonical distribution by analytically continuing the density $%
k(c,c^{\ast })=k(x,y)$ to a kernel 
\begin{equation}
k(c^{\prime },c^{\ast })=(c|S|c^{\prime })  \label{1 1 46}
\end{equation}%
that is holomorphic in $c^{\prime }$ and $c^{\ast }\in \mathbb{C}^{d+1}$,
where $|c)=\psi _{z}$ and similarly, $|c^{\prime })=\psi _{z^{\prime }}$ at%
\begin{equation*}
c^{\prime }=(x^{\prime }\omega +2\pi \mathrm{j}y^{\prime })\upsilon
^{\dagger }/\sqrt{2\pi }
\end{equation*}%
for $z^{\prime }=(x^{\prime },y^{\prime })$. For one thing, for operator (%
\ref{1 1 24}) with a density of the form (\ref{1 1 33}) we obtain 
\begin{equation}
k(c^{\prime },c^{\ast })=J(c\mid c^{\prime })\mathrm{e}^{-(c^{\ast }-\theta
^{\ast })(s+1)^{-1}(c^{\prime }-\theta ^{\prime })^{\intercal }}/\left\vert
S_{\cdot }\right\vert .  \label{1 1 47}
\end{equation}%
The transition from kernels of the (\ref{1 1 46}) type to, say, the
coordinate representation of operator $S$ is carried out by the following
formula: 
\begin{equation}
S=\int |c)(c^{\prime }\mid k(c^{\prime },c^{\ast })(c\mid c^{\prime })\,%
\mathrm{d}c\,\mathrm{d}c^{\ast }\,\mathrm{d}c^{\prime }\,\mathrm{d}c^{\ast
\prime },  \label{1 1 48}
\end{equation}%
where $|c)(c^{\prime }|:\varphi \mapsto (c^{\prime }\left\vert \varphi
)\right\vert c)$ are one-dimensional operators defined by the canonical
amplitudes (\ref{1 1 24}), respectively, at $\alpha =c$ and $c^{\prime }\in 
\mathbb{C}^{d+1},\,\mathrm{d}c\,\mathrm{d}c^{\ast }=\mathrm{d}x\,\mathrm{d}%
y,\,\mathrm{d}c^{\prime }\,\mathrm{d}c^{\ast \prime }=\mathrm{d}x^{\prime }\,%
\mathrm{d}y^{\prime }$, 
\begin{equation}
(c\mid c^{\prime })=\exp \left\{ -\tfrac{1}{2}|c|^{2}+c^{\ast }c^{\prime }-%
\tfrac{1}{2}|c^{\prime }|^{2}\right\} .  \label{1 1 49}
\end{equation}%
Note that the canonical kernels (\ref{1 1 46}), as operators $S$, act in the
space of entire functions 
\begin{equation}
h(c)=\mathrm{e}^{|c|^{2}/2}(\varphi \mid c),  \label{1 1 50}
\end{equation}%
which define the representation of amplitudes $\chi \in \mathcal{H}$ in the
Bargmann space of all entire functions on $\mathbb{C}^{d+1}$, for which 
\begin{equation}
\int \left\vert h(c)\right\vert ^{2}\exp \left\{ -\left\vert c\right\vert
^{2}/2\right\} \,\mathrm{d}c\,\mathrm{d}c^{\ast }<\infty .  \label{1 1 51}
\end{equation}%
Specifically, one-dimensional operators of density $S=|\varphi )(\varphi |$
have kernels 
\begin{equation}
k(c^{\prime },c^{\ast })=(c\mid \varphi )(\varphi \mid c^{\prime }).
\label{1 1 52}
\end{equation}

\section{Mathematical Models of Wave Pattern Analyzers}

\label{Models Sound and Visual}In this section we will consecutively
introduce and describe mathematical models of an ideal filter, a
quasifilter, a disjoint selector, and a quasiselector that make it possible
to move to arbitrary representations necessary for solution of the problems
of best wave-pattern recognition based on measurements of pattern
intensities in a single representation. We will also discuss a dilation
theory, based on the work done by Halmos \cite{46} and Neumark \cite{47},
for designing ideal filters and selectors and their realization via indirect
measurements, an idea that originated in quantum theory \cite{3}.

\subsection{Ideal Filters}

The simplest measurement of a signal is the determination of the intensity
of the oscillations in the signal in a given mode described by a vector $%
\psi $ normalized to unity, $\left\| \psi \right\| =1$, belonging to the
Hilbert space $\mathcal{H}$ of amplitudes $\varphi $ admissible at the
``in'' terminals of the receiver and having a finite intensity $I(\varphi
)=(\varphi \mid \varphi )<\infty $. The oscillation amplitude in mode $\psi $
is determined by the projection $(\psi \mid \varphi )$ of the received
amplitude $\varphi $ on direction $\psi $, while the intensity is calculated
according to the formula 
\begin{equation}
E_{\psi }(\varphi )=(\varphi \mid \psi )(\psi \mid \varphi )=|(\varphi \mid
\psi )^{2},  \label{1 2 1}
\end{equation}
similar to the transition amplitude of a quantum mechanical system from
state $\varphi $ to state $\psi $. The intensity given by formula (\ref{1 2
1}) is a positive quantity, just as probability is; however, it can assume
values greater than unity (but not greater than the total intensity $%
I(\varphi ))$. The appropriate measuring device acts as an ideal filter if
it receives a signal $\varphi $ completely, provided that $\psi $ and $%
\varphi $ are collinear, and does not receive $\varphi $ if $\varphi $ and $%
\psi $ are orthogonal. A mixed signal described by a density operator $S$
excites in mode $\psi $ oscillations of intensity 
\begin{equation}
\varepsilon _{\psi }(S)=\left\langle S,E_{\psi }\right\rangle =(\psi \mid
S\psi ).  \label{1 2 2}
\end{equation}
More general analyzers carry out the measurement of the intensity 
\begin{equation}
E(\varphi )=(\varphi \mid E\varphi )=\left\| E\varphi \right\| ^{2}
\label{1 2 3}
\end{equation}
of the projection $E\varphi $ of the received signal on an arbitrary
subspace of $\mathcal{H}$ described by an orthoprojector $E=E^{\ast }E$. For
example, an audio frequency filter with a pass band $\Delta $ is determined
by an orthoprojector $E=I(\Delta )$ on the subspace of amplitudes $\psi (f)$
with support $\Delta =\Phi $ acting as the operator of multiplication by the
indicator $1(f,\Delta )$ of set $\Delta $. In a similar manner one can
define spatial optical filters that cut out a visual field in a certain
region of aperture $\Delta $.

The reader will recall that an orthoprojector is any linear operator $E:%
\mathcal{H}\rightarrow \mathcal{H}$ satisfying the condition $E^{\ast
}=E=E^{2}$, and to each Hilbert space $\mathcal{E}\subseteq \mathcal{H}$
there corresponds a unique orthoprojector for which $\mathcal{E}=E\mathcal{H}
$. Bearing in mind this one-to-one relation, we will describe such analyzers
by orthoprojectors and call them ideal filters, that is, as noted earlier,
filters that pass a signal without distortion while measuring its intensity
if $\varphi \in E\mathcal{H}$ and that do not pass a signal if it is
orthogonal to space $E\mathcal{H}$. The set of all such filters is partially
ordered with a smallest element and a greatest element for which we take the
null operator $0$ and the identity operator $I$; specifically, filter $A$ is
stronger than filter $B$, or $A\geq B$, if orthoprojector $B$ is greater
than $A$, or $AB=A$. Maximal filters, with the exception of the zero filter,
are described by one-dimensional orthoprojectors $E_{\chi }=|\chi )(\chi |$
acting according to the formula 
\begin{equation}
|\chi )(\chi |\varphi =\chi (\chi |\varphi )=(\chi |\varphi )\chi
\label{1 2 4}
\end{equation}%
and defining the normalized vector $\chi $ to within a phase factor $\mathrm{%
e}^{\mathrm{j}\theta }$. For every filter $A$ there is a unique
complementary filter $A^{\perp }$ such that $A+A^{\perp }=I$, with $A^{\perp
\perp }=A$, and $A^{\perp }\geq B^{\perp }$ if $B\geq A$. The orthogonal
complement $A\rightarrow A^{\perp }$ possesses all the properties of logical
negation: if $A$ \textquotedblleft passes\textquotedblright\ $\varphi
:A\varphi =\varphi $, then $A^{\perp }$ does not: $A^{\perp }\varphi =0$,
and vice versa, with $A\wedge A^{\perp }=0$ and $A\vee A^{\perp }=I$ with
respect to the operations of conjunction $\wedge $ and disjunction $\vee $,
for which we take the upper and lower bounds 
\begin{equation}
A\wedge B=\sup \{A,B\},\,\;\;\;A\wedge B=\inf \{A,b\},  \label{1 2 5}
\end{equation}%
and with the duality formula being valid, or $(A\vee B)^{\perp }=A^{\perp
}\vee B^{\perp }$.

Generally, filters $A$ and $B$ are said to be disjoint if $A\perp B$, that
is, $A\varphi =\varphi $ implies $B\varphi =0$ ($B\varphi =\varphi
\Rightarrow A\varphi =0$), or, which is equivalent, if $A^{\perp }\geq B$ ($%
B^{\perp }\geq A$); they are said to be incompatible if $A\wedge B=0$, that
is, if there is not a single signal that can pass completely through filter $%
A$ and filter $B$ in the sense that $A\varphi =\varphi $ and $B\varphi
=\varphi $.

It can easily be shown that disjoint filters are incompatible, but not the
other way round. For this reason the logic of filters is nondistributive,
similar to the logic of quantum theory, which is also nondistributive. It
satisfies a weaker condition of orthomodularity 
\begin{equation}
(A\vee B^{\perp })\wedge C=A\vee (B^{\perp }\wedge C)\text{ if }A\leq B\leq
C.  \label{1 2 6}
\end{equation}
It is in this respect that the logic of filters differs from Boolean logic,
where incompatibility and disjointness mean the same. The intensity of a
mixed signal $S$ measured by an ideal filter $E$ can be calculated via the
formula 
\begin{equation}
\varepsilon (S)=\left\langle S,E\right\rangle =\mathrm{Tr}\,(ES).
\label{1 2 7}
\end{equation}

\subsection{Disjoint Selectors}

Complicated analyzers measure the intensities $E_{i}(\varphi )=|(\chi
_{i}\mid \varphi )|^{2}$ of the received field $\varphi $ simultaneously in
several standard modes $\chi _{i}\in \mathcal{H},\,i=1,\ldots ,m$, which, if
the normalization conditions $\left\Vert \chi _{i}\right\Vert =1$ for all $i$%
's, are necessarily orthogonal in view of the condition%
\begin{equation*}
\sum_{i=1}^{m}E_{i}(\varphi )\leq I(\varphi ).
\end{equation*}
Otherwise, the total received intensity $\sum_{i=1}^{m}E_{i}(\varphi )$
could be greater than the total intensity $I(\varphi )=\left\Vert \varphi
\right\Vert ^{2}$ of the received signal $\varphi $. Such analyzers act as
disjoint selectors, or ideal selective filters that split the received
signal $\varphi $ into orthogonal components $\varphi _{i}=(\chi _{i}\mid
\varphi )\chi _{i},\,i=1,\ldots ,m$. Signal $\varphi $ is received
completely by a selective filter if $E\varphi =\varphi $, where $%
E=\sum_{i=1}^{m}E_{i}$ is the appropriate nonselective filter defined by the
one-dimensional orthoprojectors $E_{i}$ on the subspaces generated by the
standard modes $\chi _{i}$.

More general selective filters are specified by arbitrary sets (or families) 
$\{E_{i}\mid i=1,\ldots ,m\}$ of projectors $E_{i}:\mathcal{H}\rightarrow 
\mathcal{H}$ that satisfy the condition of pairwise orthogonality $%
E_{i}E_{k}=0$ for $i\neq k$. For example, a disjoint selector that measures
the intensity of a signal in each region $\Delta _{i}$ of a Borel partition%
\begin{equation*}
\Omega =\sum_{i}\Delta _{i}:\Delta _{i}\subseteq \Omega
\end{equation*}
is described by an orthogonal set of $E_{i}=I(\Delta _{i})$ of indicators $%
I(\Delta _{i})=\{1(q,\Delta _{i})\}$. Note that such a set $\{E_{i}\}$ may
have an infinite number of members if space $\mathcal{H}$ is not
finite-dimensional; in this case the received intensity is determined for
each $\varphi $ by an absolutely convergent series $\sum_{i=1}^{\infty
}E_{i}(\varphi )\leq I(\varphi )$, where 
\begin{equation}
E_{i}(\varphi )=(\varphi \mid E_{i}\varphi )=\left\Vert E_{i}\varphi
\right\Vert ^{2}.  \label{1 2 8}
\end{equation}%
A selective measurement is said to be complete if the inequality $E(\varphi
)\leq I(\varphi )$ is transformed into an equality for every $\varphi \in 
\mathcal{H}$, that is, if $\sum_{i=1}^{\infty }E_{i}=I$, in a strong
operator topology ($I$ is the identity operator in $\mathcal{H}$) and is
said to be maximal if all the $E_{i}$ are one-dimensional.

Complete filters are usually related to self-adjoint operators with a
nondegenerate discrete spectrum $\{x_{i}\}$ through the spectral
decomposition (or expansion) 
\begin{equation}
A=\sum_{i=1}^{\infty }x_{i}E_{i},  \label{1 2 9}
\end{equation}
with each set $\{E_{i}\}$ being assigned a numbering self-adjoint operator $%
N=\sum_{i}iE_{i}$.

In addition to discrete filters there is another important class of filters,
known as continuous filters, which are related to normal operators with a
continuous spectrum $X\subseteq \mathbb{C}^{1}$. In accordance with von
Neumann's theorem, to each such operator there is uniquely assigned a
projector-valued measure $E$ on $X$ that specifies the orthogonal expansion
(or decomposition) of unity $I=\int E(dx)$, so that 
\begin{equation}
A=\int xE(dx),\;\mathcal{D}(A)=\left\{ \chi \in \mathcal{H}:\int
|x|^{2}E(\chi ,dx)<\infty \right\} .  \label{1 2 10}
\end{equation}%
Here a family $\{A_{j}\mid j=1,\ldots ,n\}$ of pairwise commutative normal
operators $A_{j}:\mathcal{H}\rightarrow \mathcal{H}$ has corresponding to it
a selective filter described by a projector-valued measure $%
E=\bigotimes_{j=1}^{n}E_{j}$ on $X\subseteq \mathbb{C}^{m}$ that defines a
spectral representation $A=\int xE(dx)$ for the vector operator $A=(A_{j})$.
It is with these vector selective filters that the measurement of the
intensity distribution (\ref{1 1 1}) in the coordinate region is carried
out. Such a distribution is described by the orthogonal decomposition of
unity $I=\int I(dx)$ for the coordinate vector operator $Q=(Q_{k},k=0,1,%
\ldots ,d)$; the coordinates in the coordinate (or position) representation
are given by the respective operator of multiplication by $q=(q_{k})$, so
that 
\begin{equation}
I(\Delta )\varphi (q)=1(q,\Delta )\varphi (q),  \label{1 2 11}
\end{equation}%
where $1(\Delta )$ is the indicator of the Borel subset $\Delta \subseteq 
\mathbb{R}^{d+1}:1(\Delta ,q)=1$ for $q\in \Delta $ and $1(q,\Delta )=0$ for 
$q\not\in \Delta $. The result of such a measurement is the continuous
measure $I(\varphi ,dx)$ with a density 
\begin{equation}
\iota (\varphi ,x)=I(\varphi ,dx)/\mathrm{d}x=|\varphi (x)|^{2}.
\label{1 2 12}
\end{equation}%
Note that the self-adjoint position operator 
\begin{equation}
Q=\int qI(dq),\,\mathcal{D}(Q)=\left\{ \chi \in \mathcal{H}:\int q^{2}|\chi
(q)|^{2}\mathrm{d}q<\infty \right\}  \label{1 2 13}
\end{equation}%
has a domain of definition $\mathcal{D}(Q)$ coinciding with $\mathcal{H}%
=L^{2}(Q)\,$\ only in the case of a bounded region $\Omega $. Generally
speaking, operator $Q$ is only a densely definite operator, such as the
frequency operator $F=\int_{0}^{\infty }$ if $I(df)$ in the case of a
semi-infinite band $\Phi =[0,\infty \lbrack $ of the spectrum.

In general, let $X$ be an arbitrary set and $\mathcal{B}(X)$ the Borel
algebra of its subsets. Every measure $E:\Delta \in \mathcal{B}\mapsto
E(\Delta )$ with values in the orthoprojectors of the Hilbert space $%
\mathcal{H}$ is said to be a disjoint selector, it is called a complete
selector if $E(X)=I$. Disjoint selectors measure the intensity distribution
in the received signal $\varphi $ on $X$ according to the formula 
\begin{equation}
E(\varphi ,\Delta )=(\varphi \mid E(\Delta )\varphi )=\left\| E(\Delta
)\varphi \right\| ^{2},  \label{1 2 14}
\end{equation}
and define for each $\varphi $ a positive measure on $X$ of finite mass $%
E(\varphi ,X)\leq I(\varphi )$, coinciding with $I(\varphi )=\left\| \varphi
\right\| ^{2}$ in the case of a complete selector.

We will say that selector $E^{\prime }$ on $X^{\prime }$ majorizes selector $%
E$ on $X$ (denoted $E^{\prime }\gtrsim E$) if there exists a measurable
mapping $f:X^{\prime }\rightarrow X$ with respect to which 
\begin{equation}
E(\Delta )=E^{\prime }(f^{-1}(\Delta ))\text{ for every }\Delta \in \mathcal{%
B}(X)  \label{1 2 15}
\end{equation}%
where $f^{-1}(\Delta )=\{x^{\prime }\in X^{\prime }\mid f(x^{\prime })\in
\Delta \}$ is the inverse image of set $\Delta \subseteq X$, and the
selectors $E^{\prime }$ and $E$ are equivalent, $E^{\prime }\simeq E$, if $%
E^{\prime }\lesssim E$, too.

\subsection{Successive Filters and Quasifilters}

The common practice in processing sound and visual patterns is to use
analyzers that act not in the initial Hilbert space $\mathcal{H}$ generated
by the amplitudes $\varphi $ on the \textquotedblleft in\textquotedblright\
terminals but in an extension of this space. For example, temporal
measurements of sound signals $\varphi (f)$ with a restricted frequency band 
$\Phi $ are reduced to determining the intensity of these signals in this or
that temporal interval $\Delta \in \mathbb{R}_{+}^{1}$ via the
orthoprojector $\widetilde{I}(\Delta )$ of multiplication of the signal in
the temporal representation by the indicator function $1(t,\Delta )$: 
\begin{equation}
\widetilde{I}(\varphi ,\Delta )=\int 1(t,\Delta )|\widetilde{\varphi }%
(t)|^{2}\mathrm{d}t=\left\Vert \widetilde{I}(\Delta )\varphi \right\Vert ^{2}
\label{1 2 16}
\end{equation}%
where $\widetilde{I}(\Delta )$ acts in the space of signals of unlimited
bandwidth, $L^{2}(\mathbb{R})$. In a similar manner space $\mathcal{H}$ is
extended to $L^{2}(\mathbb{R}^{d})$ in the wave processing of optical fields
observed on a limited aperture $S\subset \mathbb{R}^{d}$, which results in
determining the intensities of the fields in this or another momentum
interval $\Delta \in \mathbb{R}^{d}$.

In general, such an extension is described by an isometric embedding $F:%
\mathcal{H}\rightarrow \mathcal{H}^{\prime }$ of Hilbert space $\mathcal{H}$
into another space $\mathcal{H}^{\prime }$, for which for examples
considered here we can take the Hilbert space $\mathcal{H}^{\prime }=L^{2}(%
\mathbb{R}^{d+1})$ into which the space $\mathcal{H}=L^{2}(\Omega )$ is
isometrically embedded via the Fourier transform $F:\varphi \mapsto 
\widetilde{\varphi }^{2}$. An ideal filter described in $\mathcal{H}^{\prime
}$ by the orthoprojector $E$ measures the intensity of amplitude $\varphi
\in \mathcal{H}$ defined by the Hermitian form 
\begin{equation}
D(\varphi )=\left\Vert EF\varphi \right\Vert ^{2}=(F\varphi \mid EF\varphi
)=(\varphi \mid D\varphi ),  \label{1 2 17}
\end{equation}%
where $D=F^{\ast }EF$ is a positive contraction operator in $\mathcal{H}$.
Formula (\ref{1 2 17}) shows that this intensity can be considered the
result of successive action of two ideal filters, $F$ and $E$, with $%
\mathcal{H}$ being identified with a subspace $F\mathcal{H}\subset \mathcal{H%
}^{\prime }$, where filter $F$ is described by the orthoprojector $F$ that
cuts subspace $\mathcal{H}$ out of $\mathcal{H}^{\prime }$. For example,
temporal measurement of narrow-band signals is the result of noncommutative
action of a frequency filter $F=I(\Delta f)$ and a temporal filter $E=%
\widetilde{I}(\Delta t)$, the result is effectively described by the
Hermitian form (\ref{1 2 17}) defined by the operator%
\begin{equation*}
D=I(\Delta f)\widetilde{I}(\Delta t)I(\Delta f).
\end{equation*}
It is, therefore, advisable to generalize the concept of a filter by
describing it in space $\mathcal{H}$ by any operator $D:\mathcal{H}%
\rightarrow \mathcal{H}$ that satisfies the condition 
\begin{equation}
I\geq D^{\ast }=D\geq 0,  \label{1 2 18}
\end{equation}%
and calling it a quasifilter if $D\neq D^{2}$.

The basis for this extension is the Halmos theorem \cite{46}, according to
which every quasifilter described by operator (\ref{1 2 18}) can be
considered as a reduction (projection) $D=F^{\ast }EF$ on $\mathcal{H}$ of
an ideal filter $E$ acting in an extension $\mathcal{H}^{\prime }$. For the
Hilbert space $\mathcal{H}^{\prime }$ we can always take the doubling $%
\mathcal{H}^{\prime }=\mathcal{H}\oplus \mathcal{H}=\mathbb{C}^{2}\otimes 
\mathcal{H}$ of space $\mathcal{H}$ with embedding $F:\varphi \mapsto
(\varphi ,0)$, selecting the operators 
\begin{equation}
E_{11}=D,\,E_{12}=\sqrt{D(I-D)}=E_{21},\,E_{22}=I-D  \label{1 2 19}
\end{equation}%
for the blocks of orthoprojector $E$. Note that allowance for consecutive
action of several noncommutative ideal filters $E_{1},\ldots ,E_{n}$ in the
initial Hilbert space also leads to the notion of a quasifilter. These ideal
filters then measure the intensity 
\begin{eqnarray}
&&D(\varphi )=\left\Vert E_{n}\ldots E_{1}\varphi \right\Vert ^{2}=(\varphi
\mid D\varphi ),  \label{1 2 20} \\
&&D=E_{1}\ldots E_{n-1}E_{n}E_{n-1}\ldots E_{1},  \label{1 2 21}
\end{eqnarray}%
and the result can be considered the effect of linear nonideal filters not
necessarily described by Hermitian contraction operators $A:\mathcal{H}%
\rightarrow \mathcal{H},\,\left\Vert A\right\Vert \leq 1$, with the nonideal
filters damping and distorting the amplitudes and with $D=A^{\ast }A$ in the
formula for the appropriate intensity: 
\begin{equation}
D(\varphi )=\left\Vert A\varphi \right\Vert ^{2}=(\varphi \mid D\varphi ).
\label{1 2 22}
\end{equation}

\subsection{Quasiselectors and Indirect Measurements}

In a similar manner we can introduce generalized selectors, which are
defined on a Borel space $X$ by a positive operator-valued measure $M:\Delta
\in \mathcal{B}(X)\rightarrow M(\Delta )$ specifying in the Hilbert space $%
\mathcal{H}$ a weak decomposition $D=\int M(dx)$ of an operator $D$
satisfying condition (\ref{1 2 18}) in the following sense: 
\begin{equation}
D(\varphi )=\int M(\varphi ,dx)\text{ for every }\varphi .  \label{1 2 23}
\end{equation}
Here, as usual, $D(\varphi )=(\varphi \mid D\varphi )$ is a Hermitian form
defined by the operator of total effect $D$, while 
\begin{equation}
M(\varphi ,dx)=(\varphi \mid M(dx)\varphi )  \label{1 2 24}
\end{equation}
is the distribution of intensity of $X$ corresponding to amplitude $\varphi $
and measured by such a selector. Note that the expansion of operator $D$
defined by measure $M(dx)$ may not necessarily be orthogonal even if the
operator is a projector, that is, if the total filter is ideal, such
nondisjoint selective filters will be called quasiselective filters, or
simply quasiselectors. A quasiselector is said to be complete if $M(X)=I$
and maximal if $M\lesssim M^{\prime }\Rightarrow M^{\prime }\simeq M$ in the
same sense as in (\ref{1 2 15}).

An example of a complete maximal quasiselector for sound signals and optical
fields observed in a restricted region $\Omega $ of coordinates $\mathbb{R}%
^{d+1}\ni q=(f,\mathbf{q})$ is the analyzer of the momentum distribution (%
\ref{1 1 4}), which in the $q$-representation is defined by formula (\ref{1
2 24}) via the operator-valued measure 
\begin{equation}
M(dp)=F^{\ast }I(dp)F\equiv \widetilde{I}(dp),  \label{1 2 25}
\end{equation}%
where $F$ is the Fourier transform (\ref{1 1 5}), and $I(D)=\{1(p,D)\}$ is
the projector-valued measure on $\mathbb{R}^{d+1}$ described in the space $%
\mathcal{H}^{\prime }=L^{2}(\mathbb{R}^{d+1})$ of the proper representation
of the generalized-momentum operator $p=(t,\mathbf{p})$ by the indicator
measure $1(p,dx)$.

Note that the momentum operator defined in $\mathcal{H}=L^{2}(\Omega )$ by
the nonorthogonal integral expansion 
\begin{equation*}
P=\int p\widetilde{I}(dp),\,\mathcal{D}(P)=\left\{ \chi \in \mathcal{H}:\int
p^{2}|\widetilde{\chi }(p)|^{2}\mathrm{d}p<\infty \right\}
\end{equation*}%
is always unbounded with a spectrum $\mathbb{R}^{d+1}$ and, for $\Omega \neq 
\mathbb{R}^{d+1}$, non-selfadjoint, notwithstanding the fact that the form
of the total momentum $P(\varphi )=\int p\widetilde{I}(\varphi ,dp)$ is
always Hermitian. Nevertheless, this operator always uniquely defines a
nonorthogonal expansion $I=\int \widetilde{I}(dp)$ via the condition 
\begin{equation*}
\int p^{2}(\chi \mid \widetilde{I}(dp)\chi )=(P\chi \mid P\chi
)\,\;\;\forall \chi \in \mathcal{D}(P),
\end{equation*}%
and is a restriction to functions $\varphi (q)=0$ for $q\notin \Omega $ of
the operator $(2\pi \mathrm{j})^{-1}\partial /\partial q$ that is
self-adjoint in $L^{2}(\mathbb{R}^{d+1})$ with a domain of definition%
\begin{equation*}
\mathcal{D}(P)=\{\varphi \in L^{2}(\mathbb{R}^{d+1}):\left\Vert \partial
^{2}\varphi /\partial q^{2}\right\Vert ^{2}<\infty \}.
\end{equation*}
For instance, time operator $T=\int t\widetilde{I}(dt)$ in the space $%
\mathcal{H}=L^{2}(\Phi )$ with a semi-infinite band $\Phi =[0,\infty \lbrack 
$ is a symmetric but not a self-adjoint operator $(2\pi \mathrm{j}%
)^{-1}\partial /\partial f$ with a domain of definition 
\begin{equation*}
\mathcal{D}(T)=\left\{ \chi \in L^{2}(0,\infty ):\chi
(0)=0,\,\int_{0}^{\infty }|\partial \chi (f)/\partial f|^{2}\mathrm{d}%
f<\infty \right\} .
\end{equation*}%
Similarly, the validity of the representation 
\begin{equation}
M(dx)=F^{\ast }E(dx)F,\,F^{\ast }F=I,  \label{1 2 26}
\end{equation}%
for an arbitrary quasiselector $M$ in the form of the projection of the
disjoint selector $E$ described in the extended space $\mathcal{H}^{\prime }$
by an orthogonal projector-valued measure $E(dx)$ is ensured by the Neumark
theorem \cite{47}. For mixed signals the intensity, as a function of the
density operator $S$, is described by a measure $\mu (S,dx)$ defined by the
following linear form: 
\begin{equation}
\mu (S,dx)=\left\langle S,M(dx)\right\rangle .  \label{1 2 27}
\end{equation}

Quasiselective filters also emerge as a result of reducing the description
of indirect measurement of the received signal via the disjoint selection $%
E_{0}(dx)$ of the initially uncorrelated reference signal interacting with
the received signal; this reference signal generates a Hilbert space $%
\mathcal{H}_{0}$. Specifically, if $S_{0}$ is the density operator of the
normalized reference signal $(\mathrm{Tr}\,S_{0}=1)$ and $U$ is a unitary
operator describing in the tensor product $\mathcal{H}\otimes \mathcal{H}%
_{0} $ the result of the interaction $S^{\prime }=U(S\otimes S_{0})U^{\ast }$
with the received signal $S$, then the intensity distribution corresponding
to such indirect measurement may be effectively calculated via formula (\ref%
{1 2 27}) as a result of the quasimeasurement 
\begin{equation*}
\left\langle S,M(dx)\right\rangle =\left\langle S^{\prime },I\otimes
S_{0}(dx)\right\rangle =\left\langle S\otimes S_{0},E^{\prime
}(dx)\right\rangle
\end{equation*}%
described by the operator-valued measure 
\begin{equation}
M(dx)=\mathrm{Tr}\;\left[ (I\otimes S_{0})E^{\prime }(dx)|\mathcal{H}\right]
,  \label{1 2 28}
\end{equation}%
where $E^{\prime }(dx)=U^{\ast }(I\otimes E_{0}(dx))U$, and $\mathrm{Tr}%
\,\{\cdot |\mathcal{H}\}$ is the partial trace in $\mathcal{H}\otimes 
\mathcal{H}_{0}$ defined for factorable density operators $S\otimes S_{0}$
via the formula 
\begin{equation*}
\mathrm{Tr}\,\{S\otimes S_{0}|\mathcal{H}\}=S\mathrm{Tr}\,S_{0}.
\end{equation*}%
The indirect calculation of the intensity distribution over the frequency
(colour) $f\in \Phi $ of static monochrome patterns is an example of the
above-mentioned type of measurement. It can be carried out as a result of
the wave processing of the patterns in which the intensity distribution over
the momenta $p$ in such patterns is calculated.

Using the Neumark theorem as a basis, let us give an explicit description of
a construction that makes it possible to reduce any quasimeasurement to an
indirect measurement. To this end we take for $\mathcal{H}_{0}$ the space $%
\mathcal{H}^{\prime }$ of the Neumark construction and for the reference
signal a normalized amplitude $\psi ^{\prime }=F\psi ,\,\left\Vert \psi
\right\Vert =1$, and introduce the linear operator $U$ in $\mathcal{H}%
\otimes \mathcal{H}^{\prime }$ (which is defined by the Neumark isometry $F:%
\mathcal{H}\rightarrow \mathcal{H}^{\prime },\,F^{\ast }F=I$) in the
following manner: 
\begin{equation}
U:\varphi \otimes \varphi ^{\prime }\mapsto F^{\ast }\varphi ^{\prime
}\otimes F\varphi +\varphi \otimes (1-FF^{\ast })\varphi ^{\prime }
\label{1 2 29}
\end{equation}%
with the generating elements being $\varphi \otimes \varphi ^{\prime
},\,\varphi \in \mathcal{H},\varphi ^{\prime }\in \mathcal{H}^{\prime }$. It
can be directly verified that $U=U^{\ast }$ and $U^{2}=I$ and, hence, $%
U^{\ast }U=I=UU^{\ast }$. Taking for $E_{0}(dx)$ the Neumark expansion $%
E(dx) $ in $\mathcal{H}^{\prime }$, assuming that $S_{0}=|\psi )(\psi |$ and
allowing for (\ref{1 2 26}), we arrive at an indirect measurement whose
reduction (\ref{1 2 28}) yields the initial measure $M(dx)$: 
\begin{equation*}
\mathrm{Tr}\,\{(I\otimes |\psi ^{\prime })(\psi ^{\prime }|)E^{\prime
}(dx)\}=(\psi \mid \psi )F^{\ast }E(dx)F=M(dx).
\end{equation*}

\subsection{Canonical Operators and Measurements}

Bearing in mind the invariance of the domains of definitions of operators $Q$
and $P$ with respect to the self-adjoint operators of multiplication by $q$
and differentiation with respect to $q$, or $(2\pi \mathrm{j}%
)^{-1}\,\partial /\partial q$, which on $\mathcal{H}=L^{2}(\Omega )$
coincide with $Q$ and $P$, respectively, in what follows we will take for $Q$
and $P$ in $L^{2}(\mathbb{R}^{d+1})$ their extensions, while always assuming
that the region where these operators act is $\Omega \subset \mathbb{R}%
^{d+1} $. Such operators are known as canonical and satisfy the commutation
relations (\ref{1 1 9}) in the common domain $\mathcal{D}(P)\cap \mathcal{D}%
(Q)$.

Let us now discuss simultaneous measurement of the coordinate (or position)
and momentum distributions. In view of the noncommutativity of $Q$ and $P$,
there can be no joint orthogonal decomposition of unity for these operators;
there can even be no joint nonorthogonal decomposition of $M(dx\,dy)$ for
which the following would be true 
\begin{equation}
I(dq)=\int M(dq\,dy),\,\widetilde{I}(dp)=\int M(dx\,dp).  \label{1 2 30}
\end{equation}
Otherwise, in view of the Neumark theorem, there would be commutative
self-adjoint operators in $\mathcal{H}^{\prime }\supset \mathcal{H}$
coinciding on $\mathcal{H}$ with the noncommutative operators $Q$ and $P$,
which is impossible.

Another interesting question is the relation to these operators of the
measurements of the canonical distributions (\ref{1 1 39}). Such canonical
measurements are described, obviously, by continuous with respect to $%
\mathrm{d}z=\mathrm{d}x\,\mathrm{d}y$ nonorthogonal measures $K(dz)=k(z)%
\mathrm{d}z$ with projector-valued densities 
\begin{equation}
k(z)=|\psi _{z})(\psi _{z}|=|c)(c|\equiv k(c,c^{\ast }),  \label{1 2 31}
\end{equation}%
which are defined by canonical amplitudes (\ref{1 1 25}) at $\zeta =z$ and a
certain $\omega $ or, in complex variables, by (\ref{1 1 34}) at $\alpha =c$%
. The respective quasiselective filters, which are parameterized by
symmetric $\omega $ matrices with a nonsingular real part $\omega +\omega
^{\ast }$ and now will be called canonical filters, are, obviously, maximal
and, because of (\ref{1 1 42}), complete: 
\begin{equation}
\int |\psi _{z})(\psi _{z}|\mathrm{d}z=I=\iint |c)(c)\,\mathrm{d}c\,\mathrm{d%
}c^{\ast },  \label{1 2 32}
\end{equation}%
where $\mathrm{d}c\,\mathrm{d}c^{\ast }=\mathrm{d}x\,\mathrm{d}y$.

By fixing $\omega $ and directly integrating we find that the
quasimeasurement of an intensity distribution in $x\in \mathbb{R}^{d+1}$ is
described by a continuous measure%
\begin{equation*}
M(dx)=\int K(dx\,dy)=m(x)\mathrm{d}x
\end{equation*}%
diagonal in the $q$-representation, 
\begin{equation}
m(x)=\int k(x,y)\mathrm{d}y=|\upsilon |^{-1}\int \mathrm{e}^{-\pi
|(x-q)\upsilon ^{-1}|^{2}}I(dq).  \label{1 2 33}
\end{equation}%
with a Gaussian density and $\upsilon ^{\ast }\upsilon =2\pi (\omega +\omega
^{\ast })^{-1}$, while the quasimeasurement of an intensity distribution in $%
y\in \mathbb{R}^{d+1}$ is described by an operator measure $\widetilde{M}%
(dy)=\int K(dx\,dy)=\widetilde{m}(y)\,dy$, where 
\begin{equation}
\widetilde{m}(y)=\int k(x,y)dx=|\widetilde{\upsilon }|^{-1}\int \mathrm{e}%
^{-\pi |(y-p)\widetilde{\upsilon }^{-1}|^{2}}\widetilde{I}(dp),
\label{1 2 34}
\end{equation}%
with a Gaussian density and%
\begin{equation*}
\widetilde{\upsilon }^{\dagger }\widetilde{\upsilon }=2\pi (\widetilde{%
\omega }+\widetilde{\omega }^{\ast })^{-1},
\end{equation*}%
where $\widetilde{\omega }/2\pi =2\pi /\omega ^{\ast }$. The operator
measures $M$ and $\widetilde{M}$ on $\mathbb{R}^{d+1}$, which define
nonorthogonal expansions of operators $Q$ and $P$, that is, 
\begin{equation}
Q=\int xm(x)\mathrm{d}x\text{ and }P=\int y\widetilde{m}(y)\,\mathrm{d}y,
\label{1 2 35}
\end{equation}%
describe, in contrast to the spectral measures $I$ and $\widetilde{I}$,
inaccurate measurements of position and momentum distributions, which are
obtained by smoothing out (\ref{1 1 40}) and (\ref{1 1 41}) with Gaussian
weighting functions $m$ and $\widetilde{m}$. Nevertheless, the canonical
operator measure $K$ that generates the two spectral measures possesses
certain spectral properties with respect to the complex-valued combinations
of the two respective operators: 
\begin{equation}
A=\frac{1}{\sqrt{2\pi }}(Q\omega +2\pi \mathrm{j}P)\upsilon ^{\dagger }\text{%
.}  \label{1 2 36}
\end{equation}%
Namely, applying $A$ directly to the canonical amplitudes (\ref{1 1 34}), we
can easily verify that it is well-defined on these amplitudes: 
\begin{equation}
A|\alpha )=\alpha |\alpha ),\,\alpha \in \mathbb{C}^{d+1},  \label{1 2 37}
\end{equation}%
which, therefore, form a proper base for $A$ in $\mathcal{H}=L^{2}(\mathbb{R}%
^{d+1})$.

Hermitian conjugate operators $C=A^{\ast }$ are diagonal in the Bargmann
representation, 
\begin{equation*}
(\widehat{c}h)(c)=\mathrm{e}^{|c|^{2}/2}(C\chi |c)=\mathrm{e}%
^{|c|^{2}/2}c(\chi |c)=ch(c),
\end{equation*}%
with a domain of definition $\mathcal{D}(\widehat{c})=\{\chi \in \mathcal{H}%
:\left\Vert \widehat{c}h\right\Vert <\infty \}$, where 
\begin{equation}
\left\Vert \widehat{c}h\right\Vert ^{2}=\int |c|^{2}h(c)|^{2}\mathrm{e}%
^{-|c|^{2}}\,\mathrm{d}c\,\mathrm{d}c^{\ast },  \label{1 2 38}
\end{equation}%
on which domain there is also defined the operator $A$ by differentiation $(%
\widehat{a}h)(c)=\partial h(c)/\partial c$. Thus, in the initial
representation we obtain the nonorthogonal "spectral" decompositions 
\begin{eqnarray*}
A &=&\int cK(dz),\;\;A^{\ast }=\int c^{\ast }K(dz)\, \\
\mathcal{D}(A^{\ast }) &=&\left\{ \chi \in \mathcal{H}:\int |c|^{2}|(\chi
\mid c)^{2}\,\mathrm{d}c\,\mathrm{d}c^{\ast }<\infty \right\} .
\end{eqnarray*}

Now let us describe a simple realization of a canonical measurement by an
indirect measurement defined in the tensor product $\mathcal{H}\otimes 
\mathcal{H}_{0}$, where $\mathcal{H}_{0}$ is a copy of $\mathcal{H}$. To
this end we take the commutative self-adjoint operators 
\begin{equation}
X=Q\otimes I_{0}+I\otimes Q_{0},\,Y=P\otimes I_{0}-I\otimes P_{0},
\label{1 2 39}
\end{equation}%
where $Q_{0}=q_{0},\,P_{0}=(2\pi \mathrm{j})^{-1}\partial /\partial q_{0}$,
and $I_{0}$ is the identity operator in $\mathcal{H}_{0}=L^{2}(\mathbb{R}%
^{d+1})$. Suppose that $E(dz)$ is the orthogonal spectral measure of the set 
$Z=(X,Y)$ and that 
\begin{equation}
\psi _{0}(q_{0})=|\frac{\omega +\omega ^{\ast }}{2\pi }|^{1/4}\exp \left\{ -%
\tfrac{1}{2}q_{0}\omega q_{0}^{\intercal }\right\} =|0)_{0}  \label{1 2 40}
\end{equation}%
is the basic canonical amplitude in $\mathcal{H}_{0}$. We take an arbitrary
amplitude $\chi \in \mathcal{H}$ and the corresponding tensor product%
\begin{equation*}
(\chi \otimes \psi _{0}^{\ast })(q,q_{0})=\chi (q)\psi _{0}^{\ast }(q_{0})
\end{equation*}%
in $\mathcal{H}\oplus \mathcal{H}_{0}$ and define the characteristic
function of the corresponding distribution thus: 
\begin{eqnarray}
&&\Upsilon (u,u^{\ast })=\int \mathrm{e}^{\mathrm{j}(u^{\ast }c^{\intercal
}+c^{\ast }u^{\intercal })}(\chi \otimes \psi _{0}^{\ast }\mid E(dz)\chi
\otimes \psi _{0}^{\ast }),  \label{1 2 41} \\
&&z=(x,y)\in \mathbb{R}^{2(d+1)}  \notag
\end{eqnarray}%
where $u,u^{\ast }\in \mathbb{C}^{2(d+1)}$ and as usual,%
\begin{equation*}
c=\frac{1}{\sqrt{2\pi }}(x\omega +2\pi \mathrm{j}y)\upsilon ^{\dagger }.
\end{equation*}%
We write this function in terms of normal operators 
\begin{equation}
B=\int cE(dz)=\frac{1}{\sqrt{2\pi }}(X\omega +2\pi \mathrm{j}Y)\upsilon
^{\dagger }=A\otimes I_{0}+I\otimes C_{0},  \label{1 2 42}
\end{equation}%
with $A$ the operators (\ref{1 2 36}) and%
\begin{equation*}
C_{0}=\frac{1}{\sqrt{2\pi }}(Q_{0}\omega -2\pi \mathrm{j}P_{0})\upsilon
^{\dagger },
\end{equation*}%
in the form 
\begin{equation*}
\Upsilon (u,u^{\ast })=(\mathrm{e}^{\mathrm{j}B^{\ast }u^{\intercal }}\chi
\otimes \psi _{0}^{\ast }|\mathrm{e}^{\mathrm{j}B^{\ast }u^{\intercal }}\chi
\otimes \psi _{0}^{\ast })=(\mathrm{e}^{\mathrm{j}A^{\ast }u^{\intercal
}}\chi |\mathrm{e}^{\mathrm{j}A^{\ast }u^{\intercal }}\chi ),
\end{equation*}%
where we have allowed for the property $C_{0}\psi _{0}^{\ast }=0$ for the
basic amplitude (\ref{1 2 40}). Employing now the completeness property (\ref%
{1 2 32}) of canonical amplitudes, we obtain 
\begin{eqnarray*}
\Upsilon (u,u^{\ast }) &=&\int (\mathrm{e}^{\mathrm{j}A^{\ast }u^{\prime
}}\chi |c)(c|\mathrm{e}^{\mathrm{j}A^{\ast }u^{\prime }}\chi )\,\mathrm{d}c\,%
\mathrm{d}c^{\ast } \\
&=&\int \mathrm{e}^{\mathrm{j}u^{\ast }c^{\prime }+\mathrm{j}c^{\ast
}u^{\prime }}|(\chi |c)|^{2}\,\mathrm{d}c\,\mathrm{d}c^{\ast }=\widetilde{k}%
(u,u^{\ast }).
\end{eqnarray*}%
Thus, the characteristic function (\ref{1 2 41}) of the indirect measurement
of intensity of amplitude $\chi \in \mathcal{H}$ coincides for the ground
state $\psi _{0}^{\ast }$, when calculated in the $z$-representation of
operators (\ref{1 2 39}), with the characteristic function $\widetilde{k}$
of the canonical distribution $k(c,c^{\ast })=|(\chi \mid c)|^{2}$ for this
amplitude.

\chapter{Optimal Wave Detection and Discrimination}

In this chapter we develop the wave theory of hypothesis testing for solving
problems of optimal recognition of sound and visual patterns. We formulate
the necessary and sufficient conditions for the optimality of
two-alternative and multialternative detection of wave patterns according to
the maximum criterion for the measured intensity of acoustic signals and
optical fields. We consider problems involving the discrimination of a wave
pattern against an acoustic or optical background, problems involving the
discrimination of pure nonorthogonal signals and fields, and problems
involving the recognition of mixed patterns described by noncommutative
density operators. Complete solution of the last type of problem is then
obtained for the case of mixing two pure patterns. The discussed results of
solution of the corresponding extremal problems follow from the methods of
linear programming in Banach partially ordered operator spaces \cite{48}.
The results generalize the corresponding results of the quantum detection
and estimation theory, which have been obtained for the two-alternative case
by Helstrom \cite{11} and for the multialternative case by Belavkin \cite{4}%
, \cite{5}. The necessary and sufficient optimality conditions for the
quantum theory of hypothesis testing have been discussed by Kennedy \cite{9}%
, Yuen and Lax \cite{15}, Kholevo \cite{24}, Belavkin \cite{4}, and Belavkin
and Vancjan \cite{27}.

\section{Optimal Detection of Sound and Visual Patterns}

\label{sound and visual}In this section we will discuss the problem of
detecting wave patterns that are in a partially coherent superposition with
an acoustic or optical background. The problem is complicated by the
presence of interference. We start by considering the superposition
principle for generalized mixed amplitudes. We then formulate the necessary
and sufficient conditions for the optimality of detection and give solutions
to a number of problems considered in the quantum case in the review \cite%
{30}.

\subsection{The Superposition Principle}

The problem of detecting a sound or visual pattern described by a wave
amplitude $\varphi (q)$ taken from the Hilbert space $\mathcal{H}%
=L^{2}(\Omega )$ can be solved in a trivial manner by measuring the total
intensity $I(\varphi )=\left\Vert \varphi \right\Vert ^{2}$ only in the
absence of an acoustic or optical background consisting of other signals and
fields in the frequency-spatial region $\Omega $ considered. If in the
region of measurement there is another signal or field described by
amplitude $\varphi _{0}\in \mathcal{H}$ the question of whether a wave
pattern $\varphi $ is present cannot generally be unambiguously solved by
simply measuring the total intensity of the resulting amplitude $\psi $,
which may be higher or lower than the background intensity. Such a
phenomenon is called interference and is the result of the wave
superposition principle $\psi +\varphi +\varphi _{0}$, according to which
the complex-valued amplitudes of the coherent signals $\varphi $ and $%
\varphi _{0}$ rather than the intensities of these signals, are added. The
intensity of the resulting signal has the form 
\begin{equation}
\left\Vert \psi \right\Vert ^{2}=\left\Vert \varphi \right\Vert ^{2}+2\func{%
Re}(\varphi \mid \varphi _{0})+\left\Vert \varphi _{0}\right\Vert ^{2}.
\label{2 1 1}
\end{equation}

To describe the result of the superposition of a mixed pattern and a
partially coherent background caused, say, by thermal fluctuations that have
an infinite total intensity of the acoustic or optical field, we can employ
the correlation theory by considering generalized random amplitudes within
the second-order statistical theory.

Partially coherent signals and fields determined in a similar manner in the
framework of the classical or the quantum theory are commonly described by
bounded operators $F$ from $\mathcal{H}$ to another Hilbert space $\mathcal{K%
}$; for ordinary nonrandom amplitudes $\psi \in \mathcal{H}$ these operators
are usually represented by the functional $F_{\psi \chi }=(\psi \mid \chi )$%
, denoted by $F_{\psi }=(\psi |$ and acting from $\mathcal{H}$ to $\mathcal{K%
}=\mathbb{C}$. The mean intensity of random oscillations excited in mode $%
\chi \in \mathcal{H},\left\Vert \chi \right\Vert =1$, is determined by a
Hermitian form in $F$: 
\begin{equation}
E_{\chi }(F)=(F\chi \mid F\chi )=(\chi \mid F^{\ast }F\chi ),  \label{2 1 2}
\end{equation}%
and is calculated for common mixed signals via formula (\ref{1 2 2}) with
the aid of a (generally infinite trace) density operator $P=F^{\ast }F$ of
the intensity $\iota (P)\in \lbrack 0,\infty ]$, where $F^{\ast }$ is the
Hermitian conjugate operator $\mathcal{K}\rightarrow \mathcal{H}$ acting for 
$F=(\psi |$ as an operator of multiplication $c\mapsto \psi c$ from $K=%
\mathbb{C}$ to $\mathcal{H}$. The intensity (\ref{2 1 2}) is a measurable
quantity bounded by the norm of the positive operator $P$ and equal, via the
duality theorem, to 
\begin{equation}
\varepsilon _{\chi }(P)=(\chi \mid P\chi )\leq \left\Vert P\right\Vert =\inf
\{\varepsilon \mid \varepsilon I\geq P\},  \label{2 1 3}
\end{equation}%
which for the case of white noise $P=\varepsilon I$, described by the
isometric operator $T=F/\sqrt{\varepsilon }$, $T^{\ast }T=I$, determines the
local intensity $\varepsilon =\varepsilon _{\chi }(P)$, the same for all
modes $\chi \in \mathcal{H}$. Note that every partially coherent signal $F$
can be considered as the result of action of a contraction filter $%
D=P/\left\Vert P\right\Vert $ on white noise of local intensity $\varepsilon
=\left\Vert P\right\Vert $ if we employ the polar decomposition $F=TP^{1/2}$%
, which determines uniquely the isometry operator $T$ on the range of values 
$F^{\ast }\mathcal{K}$ of operator $F^{\ast }$.

For generalized signals with infinite trace density operators $P$ it proves
expedient, however, to consider only such quasimeasurements for which the
operators $D$ of the total effect lead to finite intensities: 
\begin{equation}
D(F)\equiv \mathrm{Tr}\,(FDF^{\ast })=\mathrm{Tr}\,(PD).  \label{2 1 4}
\end{equation}%
In addition to one-dimensional projector $D=E_{\chi }=|\chi )(\chi |$, for
which the intensity (\ref{2 1 2}) is determined by the bounded form (\ref{2
1 3}), we can always consider finite-dimensional operators $D=\sum_{i}|\chi
_{i})(\chi _{i}|$ as well as any trace class operator $0\leq D<I$, since $%
\mathrm{Tr}\,(SD)\leq \varepsilon \mathrm{Tr}\,D$ if $S<\varepsilon I$.

Extending the superposition principle to generalized amplitudes $F,F_{0}:%
\mathcal{H}\rightarrow \mathcal{K}$, we find that the result $G=F+F_{0}$ of
addition of the generalized signal $F$ and the background $F_{0}$ is
described by a density operator $R=G^{\ast }G$, that is the sum of operators 
$P=F^{\ast }F$ and $P_{0}=F_{0}^{\ast }F_{0}$ only if $\func{Re}F^{\ast
}F_{0}=0$. The latter condition, which defines the incoherence relation
between $F$ and $F_{0}$, cannot be met for nonrandom amplitudes $F=(\varphi
| $ and $F_{0}=(\varphi _{0}|$ since $F^{\ast }F_{0}=|\varphi )(\varphi
_{0}|\neq 0$ even in the event of orthogonality $(\varphi \mid \varphi
_{0})=0$ if $\varphi \neq 0$ or $\varphi _{0}\neq 0$, although the total
intensity (\ref{2 1 1}) is equal to the sum $\left\Vert \varphi \right\Vert
^{2}+\left\Vert \varphi _{0}\right\Vert ^{2}$.

Generally, the resulting density operator $R$ can be represented in the form 
\begin{equation}
R=P+P^{1/2}CP_{0}^{1/2}+P_{0}^{1/2}C^{\ast }P^{1/2}+P_{0},  \label{2 1 5}
\end{equation}%
where $C=T^{\ast }T_{0}$ is the operator of mutual coherence of signal $%
F=TF^{1/2}$ and noise $F_{0}=T_{0}P^{1/2}$ determined by the partial
isometries $T:F^{\ast }\mathcal{K}\rightarrow \mathcal{H}$ and $%
T_{0}:F_{0}^{\ast }\mathcal{K}\rightarrow \mathcal{H}$.

Note that $C$ is a contracting operator:%
\begin{equation*}
\left\Vert C\right\Vert \leq \left\Vert T^{\ast }\right\Vert \left\Vert
T_{0}\right\Vert =1,
\end{equation*}%
and a partially isometric operator if $F_{0}\mathcal{H}\subseteq F\mathcal{H}
$. The latter condition determines the coherence relation between the
generalized amplitude $F_{0}$ and amplitude $F$, which is always met for
nonrandom amplitudes $F=(\varphi |$ and $F_{0}=(\varphi _{0}|$ for which $%
F_{0}\mathcal{H}=\mathbb{C}=F\mathcal{H}$. Representing the partially
coherent amplitude $F_{0}$ in the form of a sum of the component $%
H_{0}=TCP_{0}^{1/2}$ coherent with $F$ and the component $W=F_{0}-H_{0}$
that is incoherent and doing the same with the resulting amplitude $G$, or%
\begin{equation*}
G=H_{1}+W,
\end{equation*}%
where $H_{1}=F+H_{0}$, we can isolate from operators $P_{0}$ and $R$ a
common density operator of the incoherent background $N=W^{\ast }W$ by
writing the two operators, with allowance made for the fact that $W^{\ast
}H_{i}=0$, in the form $P_{0}=S_{0}+N$ and $R=S_{1}+N$, where $%
S_{i}=H_{i}^{\ast }H_{i}$ are the operators $S_{0}=P_{0}^{1/2}C^{\ast
}CP_{0}^{1/2}$ and 
\begin{equation}
S_{1}=P+P^{1/2}US_{0}^{1/2}+S_{0}^{1/2}U^{\ast }D^{1/2}+S_{0},  \label{2 1 6}
\end{equation}%
with $U$ the partially isometric operator of polar expansion, and $%
CP_{0}^{1/2}=US_{0}^{1/2}$. In contrast to $P_{0}$ and $R$, for a trace
class operator $P$ the operators $S_{i}$ are usually also trace class
operators of rank $r(S_{i})\leq r(P)$ and one-dimensional operators if $%
r(P)=1$.

Infinite trace operators $P$ may also be replaced with trace class operators
if we consider finite total intensities (\ref{2 1 4}) with respect to a
fixed $D$, assuming that $S=D^{1/2}PD^{1/2}$. The effective operator $D$ is
then replaced with the orthoprojector $E$ on the subspace $\mathcal{E}=D%
\mathcal{H}$ that determines the total intensity $\varepsilon (S)=\mathrm{%
Tr\,}\,S$ by taking the trace $\varepsilon (S)=\mathrm{Tr}\,(ES)$ on $%
\mathcal{E}$.

\subsection{Classical Detection}

\label{class det}The simplest detection problem, that of isolating a pattern
described by a kernel operator $P>0$ from an incoherent mixture $R=P+N$ of
this pattern with the background $N$, is solved by measuring the intensity
of one of the possible signals, $R_{0}=N$ or $R_{1}=R$, by comparing this
signal with the background level $\iota (N)=\mathrm{Tr}\,N$. To this end it
has proved sufficient to limit oneself to measuring the total degree of
contrast $\iota (C)=\mathrm{Tr}\,C$ of the received signal by calculating
the trace $\left\langle C,E\right\rangle =\mathrm{Tr}\,(CE)$ of the
appropriate operator $C_{i}=R_{i}-N$, $i=0,1$, on any subspace $\mathcal{E}=E%
\mathcal{H}$, $CE=C$, with the trace assuming finite values $\left\langle
C_{0},E\right\rangle =0$ in the absence of a pattern, $i=0$, and $%
\left\langle C_{1},E\right\rangle =\mathrm{Tr}\,P$, $i=1$, in the presence
of a pattern even for an infinitely high level of the background $\iota
(N)=\infty $.

In the case of a partially coherent superposition $R$ of pattern $P$ and
background $P_{0}$, the difference $C=R-P_{0}$ may be a nonpositive trace
class operator with a zero or even negative trace, with the result that the
detection criterion, which is based on the condition that the total degree
of contrast $\iota (C)$ is positive, may lead to incorrect results. Even if $%
\iota (C)$ is positive, which is the case when the superposition $\psi
=\varphi _{0}+\varphi $ of orthogonal amplitudes, $(\varphi \mid \varphi
_{0})=0$, is coherent, that is,%
\begin{equation*}
\iota (C)=\left\Vert \psi \right\Vert ^{2}-\left\Vert \varphi
_{0}\right\Vert ^{2}=\left\Vert \varphi \right\Vert ^{2},
\end{equation*}
we can considerably increase the degree of contrast of amplitudes $\varphi
_{0}$ and $\psi $ if we sum, say, the coordinate distribution of the degree
of contrast, 
\begin{equation}
c(x)=|\psi (x)|^{2}-|\varphi _{0}(x)|^{2}=|\varphi (x)|^{2}+2\func{Re}%
\varphi ^{\ast }(x)\varphi _{0}(x),  \label{2 1 7}
\end{equation}%
not over the entire region $\Omega $ but only that part of the region where $%
c(x)$ is positive. As a result we arrive at the following classical problem
of optimal detection of a pattern in a coordinate (frequency-spatial) region 
$\Omega $: we must find a measurable subregion $\Delta ^{\mathrm{o}%
}\subseteq \Omega $ in which the upper bound 
\begin{equation}
\varkappa _{I}^{\mathrm{o}}(C)=\sup_{\Delta \subseteq \Omega }\left\langle
C,I(\Delta )\right\rangle =\int_{\Delta ^{\ast }}c(x)\,\mathrm{d}x
\label{2 1 8}
\end{equation}%
of the integral of the contrast function $c(x)=C(x,x)$ is attained.

This function is determined by the diagonal values of the kernel $%
C(x^{\prime },x)$, which is the difference between the generalized matrix
elements $R(x^{\prime },x)$ and $N(x^{\prime },x)$ of operators $R$ and $N$
in the coordinate representation.

It is sufficient to consider the supremum (\ref{2 1 8}) in the class of
measurable subsets $\Delta \subseteq \Omega $ of the coordinate region $%
\Omega =\{x\in \mathbb{R}^{d+1}|c(x)\neq 0\}$, the support of the integrable
function $c(x)$, in which the supremum is attained only on the set 
\begin{equation}
\Delta ^{\mathrm{o}}=\{x\in \Omega \mid c(x)>0\}\equiv \Omega _{+}.
\label{2 1 9}
\end{equation}%
Its value, $\varkappa _{I}^{\mathrm{o}}(C)=\int_{\Omega ^{+}}c(x)\,\mathrm{d}%
x$, coincides, obviously, with the integral over $\Omega $ of the positive
part 
\begin{equation}
c_{+}(x)=\max \{0,c(x)\}=\tfrac{1}{2}(c(x)+|c(x)|),  \label{2 1 10}
\end{equation}%
where the functions $c$ determine the solution to the duality problem 
\begin{equation}
\left\langle c\right\rangle _{+}=\inf_{b\geq 0}\left\{ \int_{\Omega }b(x)\,%
\mathrm{d}x\mid b\geq c\right\} =\int_{\Omega }c_{+}(x)\,\mathrm{d}x.
\label{2 1 11}
\end{equation}%
The lower bound (\ref{2 1 11}) over all positive integrable functions $%
b(x)\in L_{+}^{1}(\Omega )$, majorizing almost everywhere the function $c$,
is attained at $b^{\mathrm{o}}=0\vee c=c_{+}$ and determines on the space of
integrable functions $c$ a positive gauge $\left\langle c\right\rangle _{+}$%
, which is zero only when $c\leq 0$. The set (\ref{2 1 9}) specifies the
optimal band of the frequency-spatial filter in which the best quality of
detection, (\ref{2 1 8}), is achieved.

Reasoning along similar lines, we can solve the problem of optimal detection
in the momentum (or temporal-wave) space $X=\mathbb{R}^{d+1}$, 
\begin{equation}
\varkappa _{I}^{\mathrm{o}}(C)=\sup_{\Delta \subseteq X}\left\langle C,%
\widetilde{I}(\Delta )\right\rangle =\int_{\Delta ^{\mathrm{o}}}\widetilde{c}%
(x)\,\mathrm{d}x,  \label{2 1 12}
\end{equation}%
where $\widetilde{c}(x)=\widetilde{C}(x,x)$ are the diagonal elements of the
difference $\widetilde{R}(x^{\prime },x)-\widetilde{N}(x^{\prime },x)$ of
the operators $R$ and $N$ in the momentum representation; in coherent
superposition these diagonal elements are 
\begin{equation}
\widetilde{c}(x)=|\widetilde{\psi }(x)|^{2}-|\widetilde{\varphi }%
_{0}(x)|^{2}=|\varphi (x)|^{2}+2\func{Re}\widetilde{\varphi }^{\ast }(x)%
\widetilde{\varphi }_{0}(x).  \label{2 1 13}
\end{equation}%
The quality of such detection, $\varkappa _{I}^{\mathrm{o}}(C)=\left\langle 
\widetilde{C}\right\rangle _{+}$, based on a momentum quasimeasurement may
differ considerably from (\ref{2 1 11}). For example, the canonical
amplitudes (\ref{1 1 25}) $\varphi _{0}=\psi _{00}$ and $\psi =\psi _{0\eta
} $, which are similarly localized in the coordinate representation, differ
by their momenta, $\eta \neq 0$, and can be thought of as two hypotheses,
corresponding to the absence and presence of a complex-valued amplitude $%
\varphi =\psi _{0\eta }-\psi _{00}$ in the coherent superposition $\psi
=\varphi +\varphi _{0}$, that cannot be distinguished by the measurement of%
\begin{equation*}
|\varphi _{0}(x)|^{2}=|\psi (x)|^{2}
\end{equation*}%
($c_{+}=0$ since $c(x)=0$ for all $x\in \Omega $). At the same time, such
wave packets are easily distinguished in the momentum representation: 
\begin{equation*}
\left\langle \widetilde{C}\right\rangle _{+}=|\widetilde{\upsilon }%
|^{-1}\int (\mathrm{e}^{-\pi |(x-\eta )\upsilon ^{-1}|^{2}}-\mathrm{e}^{-\pi
|x\widetilde{\upsilon }^{-1}|^{2}})\,\mathrm{d}x\simeq 1
\end{equation*}%
if $|\eta \upsilon ^{-1}|\gg 1$ since in this case $\widetilde{c}%
_{+}(x)\simeq |\widetilde{\psi }_{0\eta }(x)|^{2}$.

In general, for every quasiselective measurement of intensity on a Borel
space $X$ with a positive operator measure $M(\Delta )\leq I,\Delta
\subseteq X$, optimal detection is determined by the solution to the problem 
\begin{equation}
\varkappa _{M}^{\mathrm{o}}\left( C\right) =\sup_{\Delta \subseteq
X}\left\langle C,M(\Delta )\right\rangle =\varkappa (\Delta ^{\mathrm{o}})
\label{2 1 14}
\end{equation}%
of finding the upper bound of the degree-of-contrast measure $\varkappa
(\Delta )=\left\langle C,M(\Delta )\right\rangle $. The supremum (\ref{2 1
14}) is attained on the $|\varkappa |$-measurable set $\Delta ^{\mathrm{o}}$%
, the support of the positive part%
\begin{equation*}
\varkappa _{+}=0\vee \varkappa =(\varkappa +|\varkappa |)/2
\end{equation*}%
of measure $\varkappa $: 
\begin{equation}
\Delta ^{\mathrm{o}}=\cap \{\overline{\Delta }:\varkappa _{+}(\Delta
)=0\}\equiv \Delta _{+},  \label{2 1 15}
\end{equation}%
which realizes the lower bound in the positive measures $\lambda \geq
\varkappa $ of finite variation: 
\begin{equation}
\left\langle \varkappa \right\rangle _{+}=\inf_{\lambda \geq 0}\{\lambda
(X)\mid \lambda \geq \varkappa \}=\varkappa _{+}(X),  \label{2 1 16}
\end{equation}%
which determines the gauge $\left\langle \varkappa \right\rangle
_{+}=0\Leftrightarrow \varkappa \leq 0$ of measure $\varkappa $.

\subsection{Optimal Detection}

As the example in Section \ref{class det} shows, the quality of detection,
which is determined for a given intensity distribution on $X$ by the
degree-of-contrast measure $\mu (C,\Delta )=\left\langle C,M(\Delta
)\right\rangle $, must be optimized not only with respect to measurement
regions $\Delta \subseteq X$ but also with respect to the methods of
measurement of this quantity. These methods are determined by the ways in
which the positive operator-valued measure $M(\Delta )\leq E$ is specified,
where $E$ is any orthoprojector in $\mathcal{H}$ satisfying the condition $%
CE=C$. Here it is sufficient to find at least one resolving operator $%
D=M(\Delta )$ that realizes the upper bound of the maximal degree of
contrast (\ref{2 1 14}): 
\begin{equation}
\varkappa ^{\mathrm{o}}(C)=\sup_{D\geq 0}\{\left\langle D,D\right\rangle
\mid D\leq E\}.  \label{2 1 17}
\end{equation}%
Employing the methods of linear programming in partially ordered Banach
spaces \cite{48}, we can formulate the necessary and sufficient conditions
for the optimality of the detection operator $D$ employing criterion (\ref{2
1 17}), which is determined by the trace class degree-of-contrast operator $%
C=R-P_{0}$.

\begin{theorem}
\label{T 2 1 1}The upper bound \textup{(\ref{2 1 17})} is attained on
operator $0\leq D^{\mathrm{o}}\leq E$ if and only if 
\begin{equation}
B^{\mathrm{o}}(E-D^{\mathrm{o}})=0,\quad (B^{\mathrm{o}}-C)\,D^{\mathrm{o}%
}=0,  \label{2 1 18}
\end{equation}%
where $B^{\mathrm{o}}\geq 0,\,C$. The operator $B^{\mathrm{o}}$ here is the
solution to the duality 
\begin{equation}
\left\langle C\right\rangle _{+}=\inf_{B\geq 0}\left\{ \left\langle
B,E\right\rangle \mid B\geq C\right\}  \label{2 1 19}
\end{equation}%
for which the conditions \textup{(\ref{2 1 18})} for admissible $D^{\mathrm{o%
}}$ are also necessary and sufficient, with $\varkappa ^{\mathrm{o}%
}(C)=\left\langle C\right\rangle _{+}$.
\end{theorem}

\begin{proof}
The sufficiency of the optimality conditions (\ref{2 1 18}) for solving
problems (\ref{2 1 17}) and (\ref{2 1 19}) can be verified directly by
employing the property of the monotonicity of the trace,%
\begin{equation*}
B\geq C\Rightarrow \mathrm{Tr}\,(BD)\mathrm{Tr}\,(CD),
\end{equation*}%
for every positive operator $D$. Allowing for the fact that $B^{\mathrm{o}%
}E=B^{\mathrm{o}}D^{\mathrm{o}}=CD^{\mathrm{o}}$ for every $0\leq D\leq E$,
we obtain 
\begin{equation*}
\left\langle C,D\right\rangle =\mathrm{Tr}\,(CD)\leq \mathrm{Tr}\,(B^{%
\mathrm{o}}D)\leq \mathrm{Tr}\,(B^{\mathrm{o}}E)=\left\langle C,D^{\mathrm{o}%
}\right\rangle .
\end{equation*}%
Similarly, for every $B\geq 0$ and every $C$ we obtain 
\begin{equation*}
\left\langle B,E\right\rangle =\mathrm{Tr}\,(BE)\geq \mathrm{Tr}\,(BD^{%
\mathrm{o}})\geq \mathrm{Tr}\,(CD)^{\mathrm{o}}=\left\langle B^{\mathrm{o}%
},E\right\rangle .
\end{equation*}%
The necessity of the optimality conditions (\ref{2 1 18}) follows from the
fact that the inequality 
\begin{equation}
\left\langle C,D\right\rangle =\mathrm{Tr}\,(CD)\leq \mathrm{Tr\,}(BD)\leq 
\mathrm{Tr\,}(BE)=\left\langle B,E\right\rangle ,  \label{2 1 20}
\end{equation}%
which is valid for all operators $D$ and $B$ admissible in problems (\ref{2
1 17}) and (\ref{2 1 19}), must transform into the equality%
\begin{equation*}
\left\langle C,D^{\mathrm{o}}\right\rangle =\left\langle B^{\mathrm{o}%
},E\right\rangle
\end{equation*}
on the extremal operators $D^{\mathrm{o}}$ and $B^{\mathrm{o}}$, in
accordance with Lagrange's principle of duality: 
\begin{eqnarray*}
\sup_{D\geq 0}\left\{ \left\langle C,D\right\rangle \mid D\leq E\right\}
&=&\sup_{D\geq 0}\inf_{R\geq 0}\left\{ \left\langle C,D\right\rangle
+\left\langle B,E-D\right\rangle \right\} \\
&=&\inf_{E\geq 0}\sup_{D\geq 0}\left\{ \left\langle C-B,D\right\rangle
+\left\langle B,E\right\rangle \right\} \\
&=&\inf_{B\geq 0}\left\{ \left\langle B,E\right\rangle \mid B\geq C\right\} .
\end{eqnarray*}%
Whereby, allowing for the fact that the trace of the product of positive
operators is zero if and only if the product itself is zero, we arrive at
conditions (\ref{2 1 18}) via the following relation: 
\begin{equation*}
\mathrm{Tr}\,[B^{\mathrm{o}}(E-D^{\mathrm{o}})]+\mathrm{Tr}\,[(B^{\mathrm{o}%
}-c)D^{\mathrm{o}}]=\mathrm{Tr}\,(B^{\mathrm{o}}E)-\mathrm{Tr}\;(CD^{\mathrm{%
o}})=0.
\end{equation*}%
The proof of the theorem is complete.
\end{proof}

Note that the solutions to problems (\ref{2 1 17}) and (\ref{2 1 19}) exist
for every Hermitian trace class operator $C$ and every bounded positive
orthoprojector $E$; the solution to problem (\ref{2 1 17}) is unique only if 
$E$ is the minimal of the orthoprojectors for which $CE=C$, while the
solution to problem (\ref{2 1 19}) is unique only if $E$ is the maximal $E=I$
of the orthoprojector $E$. Indeed, employing the spectral representation of
operator $C$, we write this operator in the form of the orthogonal sum 
\begin{equation}
C=\sum \varkappa _{n}|\chi _{n})(\chi _{n}|=C_{+}+C_{-}  \label{2 1 21}
\end{equation}%
of the positive and negative operators 
\begin{equation}
C_{+}=\sum_{\varkappa _{n}>0}\varkappa _{n}|\chi _{n})(\chi
_{n}|,C_{-}=\sum_{\varkappa _{n}<0}\varkappa _{n}|\chi _{n})(\chi _{n}|,
\label{2 1 22}
\end{equation}%
where we have allowed for the fact that a Hermitian trace class operator has
a discrete spectrum of finite multiplicity, $\varkappa _{n}\in \mathbb{R}$,
which can be found by solving the eigenvalue problem $C\chi =\varkappa \chi $%
. The orthoprojector $E$ satisfying condition $CE=C$ can be written in the
form of the orthogonal sum 
\begin{equation}
E=E_{+}+E_{0}+E_{-},  \label{2 1 23}
\end{equation}%
where $E_{0}=E-E_{+}-E_{-}$ with 
\begin{equation*}
E_{+}=\sum_{\varkappa _{n}>0}|\chi _{n})(\chi _{n}|,E_{-}=\sum_{\varkappa
_{n}<0}|\chi _{n})(\chi _{n}|.
\end{equation*}%
The operators $D^{\mathrm{o}}=E_{+}$, $B^{\mathrm{o}}=C_{+}$ are, obviously,
admissible: $0\leq E_{+}\leq E$, $C_{+}\geq 0$, $C$ and optimal: 
\begin{eqnarray}
C_{+}(E-E_{+}) &=&C_{+}(E_{0}+E)=0,  \notag \\
(C_{+}-C)E_{+} &=&-C_{-}E_{+}=0.  \label{2 1 24}
\end{eqnarray}%
Every other solution $D^{\mathrm{o}}$ to problem (\ref{2 1 17}) satisfies
conditions (\ref{2 1 18}) for $B^{\mathrm{o}}=C_{+}$; 
\begin{eqnarray*}
C_{+}(E-D^{\mathrm{o}}) &=&C_{+}-C_{+}D^{\mathrm{o}}=0, \\
(C_{+}-C)D^{\mathrm{o}} &=&-C_{-}D^{\mathrm{o}}=0,
\end{eqnarray*}%
in view of which $E_{+}=E_{+}D^{\mathrm{o}}$ and $E_{-}D^{\mathrm{o}}=0$,
that is, 
\begin{equation}
E_{+}\leq D^{\mathrm{o}}\leq E-E_{-}=E_{+}+E_{0}.  \label{2 1 25}
\end{equation}

Similarly, every solution $B^{\mathrm{o}}$ to problem (\ref{2 1 19})
satisfies conditions (\ref{2 1 18}) for $D^{\mathrm{o}}=E_{+}$: 
\begin{equation*}
B^{\mathrm{o}}(E-E_{+})=0,\quad (B^{\mathrm{o}}-C)E_{+}=B^{\mathrm{o}%
}E_{+}-C_{+}=0,
\end{equation*}%
which imply that $B$ is commutative with $E_{+}$ and, hence can be
represented in the form of the orthogonal sum $B^{\mathrm{o}}=B_{+}+B_{0}$,
with 
\begin{equation*}
B_{+}=B^{\mathrm{o}}E_{+}=C_{+},\quad B_{0}(E-E_{+})=0,
\end{equation*}%
that is, 
\begin{equation}
B^{\mathrm{o}}=C_{+}+B_{0},\quad B_{0}\geq 0,\quad B_{0}E=0.  \label{2 1 26}
\end{equation}

Thus, the general solution to the problem of optimal detection is determined
by the quasifilter (\ref{2 1 25}) of the form $D^{\mathrm{o}}=E_{+}+D_{0}$,
where $D_{0}$ is an arbitrary operator, $0\leq D_{0}\leq E_{0}$, and an
ideal filter $D^{\mathrm{o}}=E_{+}$ if $E=E_{+}+E_{-}$. The general solution
to the duality problem (\ref{2 1 19}) is determined by the operator of the
form (\ref{2 1 26}), with $B_{0}=0$ at $E=I$. The maximal possible degree of
contrast realized by the optimal detector $D^{\mathrm{o}}$ is given by the
expression 
\begin{equation}
\varkappa _{+}(R-P_{0})=\mathrm{Tr}\;(R-P_{0})_{+}=\sum_{\varkappa
_{n}>0}\varkappa _{n}.
\end{equation}

\subsection{Coherent and Quasioptimal Detection}

Let us consider the particular problem of optimal detection of a wave
pattern described by a common amplitude $\varphi \in \mathcal{H}$ in a
partially coherent mixture with a generalized random amplitude $H_{0}:%
\mathcal{H}\rightarrow \mathcal{K}$. The resulting amplitude%
\begin{equation*}
G=|\xi )(\varphi |+H_{0},
\end{equation*}%
with $\xi \in \mathcal{K}$ a normalized vector $\left\Vert \xi \right\Vert
=1 $, defines a density operator $R=G^{\ast }G$ of the form 
\begin{equation}
R=|\varphi )(\varphi |+|\varphi )(\varphi _{0}|+|\varphi _{0})(\varphi
|+P_{0},  \label{2 1 38}
\end{equation}%
with $\varphi _{0}=F_{0}^{\ast }\xi \in \mathcal{H}$ and $P_{0}=F_{0}^{\ast
}F_{0}$ the background-density operator. Thus, we are required to solve the
extremal problem (\ref{2 1 17}) for the two-dimensional degree-of-contrast
operator $C=R-P_{0}$ of the form 
\begin{eqnarray*}
C &=&|\varphi )(\varphi |+|\varphi )(\varphi _{0}|+|\varphi _{0})(\varphi |
\\
&=&|\psi )(\psi |-|\varphi _{0})(\varphi _{0}|,
\end{eqnarray*}%
which corresponds to the coherent superposition $\psi =\varphi +\varphi _{0}$
of the common amplitudes $\varphi $ and $\varphi _{0}$. We will consider
this problem in the minimal subspace $\mathcal{E}\subset \mathcal{H}$
generated by the amplitudes $\psi _{0}=\varphi _{0}$ and $\psi _{1}=\varphi
_{0}+\varphi $. For its solution we find the eigenvectors and eigenvalues of
operator $C$ by constructing the secular equation $C\chi =\varkappa \chi $
for the coefficients of the expansion%
\begin{equation*}
\chi =\alpha _{0}\psi _{0}+\alpha _{1}\psi _{1}
\end{equation*}%
in the base $\{\psi _{0},\psi _{1}\}$ of space $\mathcal{E}$: 
\begin{equation}
\psi _{1}(\psi _{1}\mid \alpha _{0}\psi _{0}+\alpha _{1}\psi _{1})-\psi
_{0}(\psi _{0}\mid \alpha _{0}\psi _{0}+\alpha _{1}\psi _{1})=\varkappa
(\alpha _{0}\psi _{0}+\alpha _{1}\psi _{1}).  \label{2 1 29}
\end{equation}%
Introducing the notation $\nu _{i}=\left\Vert \psi _{i}\right\Vert
^{2},\;i=0,1$,$\;\beta =(\psi _{0}\mid \psi _{1})$, and equating the
coefficients of $\psi _{i},\;i=0,1$, in (\ref{2 1 19}), we arrive at a
system of two homogeneous equations, 
\begin{equation}
(\nu _{0}+\varkappa )\alpha _{0}+\beta \alpha _{1}=0,\quad \bar{\beta}\alpha
_{0}+(\nu _{1}-\varkappa )\alpha _{1}=0.  \label{2 1 30}
\end{equation}%
This system has nonzero solutions only if the system determinant is zero, or 
\begin{equation}
(\nu _{0}+\varkappa )(\nu _{1}-\varkappa )-|\beta |^{2}=0.  \label{2 1 31}
\end{equation}%
Solving this quadratic equation for $\varkappa $, we obtain the eigenvalues: 
\begin{equation}
\varkappa _{\pm }=\frac{\nu _{1}-\nu _{2}}{2}\pm \sqrt{\left( \frac{\nu
_{1}+\nu _{0}}{2}\right) ^{2}-|\beta |^{2},}  \label{2 1 32}
\end{equation}%
which are real in view of the Schwarz inequality 
\begin{equation*}
|\beta |^{2}=|(\psi _{0}\mid \psi _{1})|^{2}\leq |\psi _{0}|^{2}|\psi
_{1}|^{2}=\nu _{0}\nu _{1},
\end{equation*}%
and, obviously, have opposite signs: $\pm \varkappa _{\pm }\geq 0$. At $%
\beta =0$ the amplitudes $\psi _{1}$ and $\psi _{0}$ by measuring the degree
of contrast of oscillations in the resulting mode $\chi _{+}=\psi _{1}/\sqrt{%
\nu _{1}}$, which is equal to the intensity of oscillations at $\varkappa
=\nu _{1}$ in this mode if the received signal is $\psi _{1}$ and to zero if
the received signal is $\psi _{0}$. In the opposite case $|\beta |^{2}=\nu
_{0}\nu _{1}$ of the colinearity of $\psi _{1}$ and $\psi _{0}$, the values $%
\varkappa _{\pm }$ are equal respectively, to the positive and negative
parts of the difference $\nu _{1}-\nu _{0}$: 
\begin{equation*}
\varkappa _{\pm }=\tfrac{1}{2}(\nu _{1}-\nu _{0}\pm |\nu _{1}-\nu
_{0}|)=(\nu _{1}-\nu _{0})_{\pm }.
\end{equation*}%
The corresponding optimal detection is reduced to the measurement of the
positive degree of contrast $\varkappa _{+}=\nu _{1}-\nu _{0}$ in the mode $%
\chi =\psi _{1}/\sqrt{\nu _{1}}=\psi _{0}/\sqrt{\nu _{0}}$ if $\nu _{1}>\nu
_{0}$, in the opposite case, $\nu _{0}\geq \nu _{1}$, the degree of contrast 
$\varkappa _{+}$ is zero and no measurement is carried out, or $\chi _{+}=0$%
. The optimal detection of a wave pattern $\varphi $ of intensity $\mu
=\left\Vert \varphi \right\Vert ^{2}\neq 0$ in the coherent superposition $%
\psi =\varphi +\varphi _{0}$ is therefore reduced to the measurement of the
maximal degree of contrast 
\begin{equation}
\varkappa _{+}=\sqrt{\mu \nu _{0}}\left( \func{Re}\gamma +\sqrt{\gamma }+%
\sqrt{(\func{Re}\gamma +\sqrt{\gamma })^{2}+1-|\gamma |^{2}}\right) ,
\label{2 1 33}
\end{equation}%
where $\gamma =(\varphi _{0}\mid \varphi )/\sqrt{\mu \nu _{0}}$ is the
coefficient of colinearity of amplitudes $\varphi $ and $\varphi _{0}$, and $%
\lambda $ is the signal-to-noise ratio. The corresponding ideal filter $%
E_{+}=|\chi _{+})(\chi _{+}|$ is defined at $\varkappa _{+}\neq 0$ by the
eigenvector $\chi _{+}=\alpha _{+}\varphi +\alpha _{0}\varphi _{0}$ with
coefficient 
\begin{equation}
\alpha _{+}=\sqrt{\nu _{0}/\mu \alpha _{0}}\left( \mathrm{j}\func{Im}\gamma +%
\sqrt{\gamma }+\sqrt{(\func{Re}\gamma +\sqrt{\lambda })^{2}+1-|\gamma |^{2}}%
\right) ,  \label{2 1 34}
\end{equation}%
$\alpha _{0}>0$, found from the normalization condition $\left\Vert \chi
_{+}\right\Vert =1$. The case where $\varkappa _{-}=0$, and therefore $\chi
_{+}=0$, is possible in the minimal subspace $\mathcal{E}$ only if $\varphi $
and $\varphi _{0}$ are colinear, when $|\gamma |=1$, and 
\begin{equation}
\varkappa _{+}=\sqrt{\mu \nu _{0}}(\cos \theta +\sqrt{\lambda })_{+},
\label{2 1 35}
\end{equation}%
where $\cos \theta =\func{Re}\gamma $. The optimal filter in this case is
matched with the signal mode $\chi _{+}=\varphi /\sqrt{\mu }$ if $\cos
\theta >-\sqrt{\lambda }$ and $\chi =0$ in the opposite case if $\cos \theta
\leq -\sqrt{\lambda }$ which is possible only if $\lambda \leq 1$.

The same filter $\chi _{0}=\varphi /\sqrt{\mu }$ matched with $\psi $ is
used to describe the asymptotically optimal detection at large
signal-to-noise ratios%
\begin{equation*}
\lambda =1/\varepsilon \gg (\func{Re}\gamma )^{2}.
\end{equation*}%
The degree of contrast is then 
\begin{equation}
\varkappa _{0}=\mu (1+\sqrt{\varepsilon }\func{Re}\gamma ),  \label{2 1 36}
\end{equation}%
which coincides with (\ref{2 1 33}) to within $\varepsilon $. In the next
order we obtain a filter matched with the resulting mode $\chi _{1}=\psi /%
\sqrt{\nu _{1}}$ and realizing the degree of contrast 
\begin{equation}
\varkappa _{1}=\mu \left( 1+\frac{\sqrt{\varepsilon }}{2}\func{Re}\gamma +%
\frac{\varepsilon }{4}-\frac{\varepsilon |\sqrt{\varepsilon }/2+\gamma |^{2}%
}{4[1+(\sqrt{\varepsilon }/2)\func{Re}\gamma +\varepsilon /4]}\right) .
\label{2 1 37}
\end{equation}

For an orthogonal background, $\varphi \perp \varphi _{0}$, we have $\gamma
=0$, and the normalized eigenvector $\chi _{+}$ corresponding to the
eigenvalue 
\begin{equation}
\varkappa _{+}=\sqrt{\mu \nu _{0}}\left( \sqrt{1+\lambda }+\sqrt{\lambda }%
\right)
\end{equation}%
can be written in the form 
\begin{equation*}
\chi _{+}=\frac{\left( \varphi _{0}/\sqrt{\nu _{0}}+\left( \sqrt{1+\lambda }+%
\sqrt{\lambda }\right) \varphi /\sqrt{\mu }\right) }{2(1+\lambda )+\sqrt{%
\lambda (1+\lambda )}}.
\end{equation*}%
For $\func{Im}\lambda \neq 1$ and a low signal-to-noise ratio $\lambda \ll
1-(\func{Im}\gamma )^{2}$, the maximal degree of contrast is realized at 
\begin{equation}
\chi _{+}=\frac{1}{\sqrt{2}}(\frac{1}{\sqrt{\mu }}\mathrm{e}^{\mathrm{j}%
\theta }\varphi +\frac{1}{\sqrt{\nu _{0}}}\varphi _{0}),  \label{2 1 39}
\end{equation}%
with $\sin \theta =\func{Im}\gamma $, and is determined asyptotically by the
expression 
\begin{equation}
\varkappa _{+}\cong \sqrt{\mu \nu _{0}}\left( \func{Re}\lambda +\cos \theta
+\varepsilon \lambda (\func{Re}\gamma +1)\right) ,  \label{2 1 40}
\end{equation}%
with $\cos \theta =\sqrt{1-(\func{Im}\gamma )^{2}}$. For one, at $\gamma =0$
we get $\varkappa _{+}\cong \sqrt{\mu \nu _{0}}(1+\sqrt{\lambda }).$

In the general case of the partially coherent superposition $G=F+F_{0}$, the
solution of the eigenvalue problem for the degree-of-contrast operator $%
C=G^{\ast }G-F_{0}^{\ast }F_{0}$ constitutes a complicated mathematical
problem. If we isolate the coherent component from the generalized amplitude 
$F_{0}$, we can represent the latter in the form 
\begin{equation*}
F_{0}=\tfrac{1}{2}\sqrt{\varepsilon }FA^{\ast }+W,
\end{equation*}%
with $W$ the incoherent component, $F^{\ast }W=0$, and $A$ is an operator in 
$\mathcal{H}$, which we assume to be bounded: $\left\Vert A\right\Vert \leq
1 $. Next we select the positive constant $\varepsilon $ in an appropriate
manner. Operator $C$ then assumes the form 
\begin{equation}
C=P+\tfrac{1}{2}\sqrt{\varepsilon }(PA^{\ast }+AP)  \label{2 1 41}
\end{equation}%
and is a trace class operator if $P=F^{\ast }F$ is an operator with a finite
trace.

For high signal-to-noise ratios $\lambda =1/\varepsilon \gg 1$, the
eigenvectors $\chi _{n}$ and the corresponding eigenvalues $\varkappa _{n}$
of operator $C$ can be found via perturbation theory methods. In the first
order in $\sqrt{\varepsilon }$ the eigenvectors coincide with the
eigenvectors $\varphi _{n}$ of the signal density operator $P$, that is, $%
P\varphi _{0n}=\mu _{n}\varphi _{0n}$, and realize the following degrees of
contrast: 
\begin{equation}
\varkappa _{0n}=(\varphi _{0n}\mid C\varphi _{0n})=\mu _{n}(1+\sqrt{%
\varepsilon }\func{Re}\gamma _{n}),
\end{equation}%
with $\gamma _{n}=(A\varphi _{n}\mid \varphi _{n})$. The corresponding
quasioptimal detection is reduced, therefore, to measuring the total degree
of contrast 
\begin{equation}
\varkappa _{0}=\mathrm{Tr}(CE_{0})=\sum_{n\in N_{+}}\mu _{n}(1+\sqrt{%
\varepsilon }\func{Re}\gamma _{n})  \label{2 1 43}
\end{equation}%
matched with the signal orthogonal modes $\varphi _{n}$ of the ideal filter 
\begin{equation}
E_{0}=\sum_{n\in N_{+}}|\varphi _{n})(\varphi _{n}|,\;\;\;\;N_{+}=\{n:\func{%
Re}\gamma _{n}>-1/\sqrt{\varepsilon }\}.  \label{2 1 44}
\end{equation}%
When the intensities of the signal and the noise are comparable, $%
\varepsilon \approx 1$, the quality of such detection may be considerably
lower than that of optimal detection. In particular, for the above example
of orthogonal $\varphi $ and $\varphi _{0}$ we have $\gamma =0$ and $%
\varkappa _{0}=\mu $, while the quality of optimal detection (\ref{2 1 38})
equal to%
\begin{equation*}
\varkappa _{+}=\mu (1+\sqrt{1+\varepsilon })>\mu
\end{equation*}%
is more than twice as great as $\varkappa _{0}$ if the signal intensity is
less than half of the intensity of the noise, and we have 
\begin{equation}
\frac{\varkappa _{+}}{\varkappa _{0}}=\frac{1}{2}(1+\sqrt{1+\varepsilon }%
)\rightarrow \infty \text{ as }\varepsilon =\frac{\nu _{0}}{4\mu }%
\rightarrow \infty .  \label{2 1 45}
\end{equation}

\section{Multialternative Detection of Wave Patterns}

\label{Multi}In this section we will consider the problem of detecting one
of several simple or mixed wave patterns that is in a partially coherent
super-position with the background. We will introduce the necessary and
sufficient conditions for the optimality of such detection, using the
criterion of the maximum degree of contrast, and give these conditions a
concrete meaning for the problem of separating such patterns from an
incoherent background. We will also give the complete solution to the
problem of identifying nonorthogonal waves of the same intensity. This
solution formally coincides with that of the problem of optimal
discrimination between pure quantum states obtained in \cite{4}, \cite{5}.
Finally, we will discuss the quasioptimal method of multialternative
detection based on the perturbation theory for degree-of-contrast operators.

\subsection{Statement of the Pattern Identification Problem}

The problem of $m$-alternative detection of sound and visual patterns
described in a given spatial-frequency region $\Omega $ by the wave
amplitudes $\varphi _{i}(q)$, $i=1,\ldots ,m$ belonging to the Hilbert space 
$\mathcal{H}=L^{2}(\Omega )$ can be solved in a trivial manner in the
absence of an audio or optical background, $\varphi _{0}=0$, only on the
assumption that these amplitudes are pairwise orthogonal, $(\varphi _{i}\mid
\varphi _{k})=0$ for $i\neq k$. It is sufficient to measure the intensity
distribution $\varepsilon _{i}=|(\psi \mid \chi )|^{2}$ in the received
signal $\psi \in \{\varphi _{i}\}_{i=1}^{m}$ over the orthogonal modes $\chi
_{k}=\varphi _{k}/\left\Vert \varphi _{k}\right\Vert $, $i=1,\ldots ,m$, to
determine correctly the pattern $\varphi =\varphi _{i}$ with a nonzero
intensity $\mu _{i}=\left\Vert \varphi _{i}\right\Vert ^{2}$ by specifying
the number of the excited mode $i:\varepsilon _{i}=\mu _{i}\neq 0$. The
other modes $\chi _{k},\,k\neq i$, remain unexcited in the process, and the
case $\varepsilon _{i}=0$ for all $i=1,\ldots ,m$ means that these patterns
are absent from the measurement region $\Omega $.

The simplest problem of multialternative detection in noise, the problem of
isolating one of a set of orthogonal amplitudes $\{\varphi _{i}\}$ from an
incoherent mixture%
\begin{equation*}
R_{i}=|\varphi _{i})(\varphi _{i}|+N
\end{equation*}%
with an optical or acoustic background not necessarily described by a trace
class density operator $N$, has the same solution if we compare the
intensities%
\begin{equation*}
\varepsilon _{i}=(\chi _{i}\mid R\chi _{i})=\mu _{i}+\nu _{i}
\end{equation*}
of the received signal $R\in \{R_{i}\}_{i=0}^{m}$ not with zero but with the
background level $\nu _{i}=(\chi _{i}\mid N\chi _{i})$ in the orthogonal
modes $\chi _{i}=\varphi _{i}/\left\Vert \varphi _{i}\right\Vert $, $%
i=1,\ldots ,m$.

In the case of nonorthogonal amplitudes $\{\varphi _{i}\}_{i=1}^{m}$ there
is no way of measuring the intensity distribution in the received signal
directly over the modes $\varphi _{i}/\left\Vert \varphi _{i}\right\Vert $.
We must therefore find a set $\{\chi _{i}\}_{i=1}^{m}\subset \mathcal{H}$
that satisfies the condition%
\begin{equation*}
\sum_{i=1}^{m}|\chi _{i})(\chi _{i}|\leq I
\end{equation*}
and for which a pattern $\varphi _{i}$ can be confidently reconstructed from
the distribution $\varkappa _{i}=|(\psi \mid \chi _{i})|^{2}$ corresponding
to the received signal $\psi \in \{\varphi _{i}\}_{i=1}^{m}$.

The same problem emerges in the multialternative detection of patterns $%
\{\varphi _{i}\}_{i=1}^{m}$ in the coherent superposition $\psi _{i}=\varphi
_{i}+\varphi _{0}$ with a nonzero background amplitude $\varphi _{0}$ even
when all the amplitudes $\{\varphi _{i}\}_{i=1}^{m}$ are mutually orthogonal
and orthogonal to $\varphi _{0}$. Although in the latter case we can still
measure the intensities in the orthogonal signal modes $\chi _{i}=\varphi
_{i}/\left\| \varphi _{i}\right\| $ and this yields a total degree of
contrast $\varkappa _{0}=\sum_{i=1}^{m}\mu _{i}$, we can achieve a higher
quality of detection if we minimize the expression 
\begin{equation}
\varkappa =\sum_{i=1}^{m}(\chi _{i}\mid C_{i}\chi _{i})=\sum_{i=1}^{m}|(\chi
_{i}\mid \psi _{i})|^{2}-|(\chi _{i}\mid \varphi _{0})|^{2},  \label{2 2 1}
\end{equation}
where 
\begin{eqnarray}
C_{i} &=&|\varphi _{i})(\varphi _{i}|+|\varphi _{0})(\varphi _{i}|+|\varphi
_{i})(\varphi _{0}|  \notag \\
&=&|\psi _{i})(\psi _{i}|-|\varphi _{0})(\varphi _{0}|,  \label{2 2 2}
\end{eqnarray}
as we did in the case with $m=1$ in the example at the end of Section \ref%
{sound and visual}.

Note that in the coordinate representation the wave patterns $\{\varphi
_{i}\}_{i=1}^{m}$ may be indistinguishable even if they are orthogonal, as
is the case, say, for harmonic amplitudes%
\begin{equation*}
\varphi _{i}(f)=\exp \{2\pi \mathrm{j}fi/\Phi \}
\end{equation*}
which are orthogonal in the frequency interval $[0,\Phi ]$ and have the same
homogeneous distributions, $|\varphi _{i}(f)|^{2}=1$. The maximal quality of 
$m$-alternative detection achievable through measurements of the coordinate
distribution of the contrast degree 
\begin{equation}
c_{i}(x)=|\varphi _{i}(x)|^{2}+2\func{Re}\varphi _{i}^{\ast }(x)\varphi
_{0}=|\psi _{i}(x)|^{2}-|\varphi _{0}(x)|^{2}  \label{2 2 3}
\end{equation}%
is determined by the solution to the extremal problem 
\begin{equation}
\varkappa _{I}^{\mathrm{o}}(C)=\sup_{\sum_{i=1}^{m}\Delta _{i}\subseteq
\Omega }\;\sum_{i=1}^{m}\left\langle C_{i},I(\Delta _{i})\right\rangle
=\sum_{i=1}^{m}\int_{\Delta _{i}^{\mathrm{o}}}c_{i}(x)\;\mathrm{d}x,
\label{2 2 4}
\end{equation}%
where the supremum is taken over the measurable nonintersecting subsets $%
\Delta _{i}\subset \Omega $ of a coordinate region $\Omega $ that can be
bound by the union of the supports of the integrable functions $c_{i}(x)$, $%
i=1,\ldots ,m$. This limit is attained, obviously, on the partitions%
\begin{equation*}
\Omega _{+}=\sum_{i=1}^{m}\Delta _{i}^{\mathrm{o}}
\end{equation*}
of the measurable set $\Omega _{+}$ in every point of which at least one of
the functions $c_{i}(x)$ is positive and coincides on $\Delta _{c}^{\mathrm{o%
}}$ with the upper envelope%
\begin{equation*}
c_{\vee }(x)=\max_{i=1,\ldots ,m}c_{i}(x).
\end{equation*}
Thus, the total degree of contrast (\ref{2 2 4}) coincides with the integral
over $\Omega $ of the positive part%
\begin{equation*}
c_{+}(x)=\max (0,c_{\vee }(x))
\end{equation*}
of $c_{\vee }$, which determines the solution of the duality problem 
\begin{equation}
\left\langle c\right\rangle _{+}=\inf_{b\geq 0}\left\{ \int_{\Omega }b(x)\;%
\mathrm{d}x|b\geq c_{i},\;i=1,\ldots ,m\right\} =\int_{\Omega }c_{+}(x)\;%
\mathrm{d}x.  \label{2 2 5}
\end{equation}%
The lower bound (\ref{2 2 5}) over all positive integrable functions $%
b(x)\geq 0$, which almost everywhere majorize every function $c_{i}(x)$,
defines the positive gauge of the vector function $c=\{c_{i}\}_{i=1}^{m}$, $%
\left\langle c\right\rangle _{+}=0\Leftrightarrow c_{i}(x)\leq 0$. The best $%
m$-alternative detection in the coordinate region $\Omega $ is reduced,
therefore, to a search among the $\Delta _{i}$ for the regions $\Delta _{i}^{%
\mathrm{o}}$ on which the measured degree of contrast $c(x)$ is positive and
reaches $c_{+}(x)$; in the opposite case of $c(x)<c_{+}(x)$ the pattern may
not be detected for all $x\in \Omega _{+}$ and it can be assumed to be
undetected if $c(x)\leq 0$ for all $x\in \Omega $.

\subsection{The Optimality Conditions}

\label{Optimality Cond}Let us consider the problem of maximizing the quality
of $m$-alternative detection of mixed patterns $F_{i}:\mathcal{H}\rightarrow 
\mathcal{K}$ with trace class density operators $P_{i}=F_{i}^{\ast }F_{i}$, $%
i=1,\ldots ,m$, in a partially coherent superposition $G_{i}=F_{i}+F_{0}$
for which the mutual densities%
\begin{equation*}
F_{i}^{\ast }F_{0}=\sqrt{\varepsilon }P_{i}A_{i}^{\ast }/2
\end{equation*}%
are determined by the contraction operators $A_{i}:\mathcal{H}\rightarrow 
\mathcal{H}$ ($\varepsilon >0$ is a parameter). The respective extremal
problem is formulated for trace class operators $\mathbf{C}=\left(
C_{i}\right) _{i=1}^{m}$ of the degree of contrast, 
\begin{equation}
C_{i}=R_{i}-P_{0}=P_{i}+\frac{1}{2}\sqrt{\varepsilon }(P_{i}A_{i}^{\ast
}+A_{i}P_{i}),\;i=1,\ldots ,m.  \label{2 2 6}
\end{equation}%
$R_{i}=G_{i}^{\ast }G_{i}$, in the class of quasiselective measurements
described by any resolving operators $\{D_{i}\}_{i=1}^{m}$: 
\begin{equation}
\varkappa ^{\mathrm{o}}(\mathbf{C})=\sup_{D_{i}\geq 0}\left\{
\sum_{i=1}^{m}\left\langle C_{i},\;D_{i}\right\rangle \Big|%
\sum_{i=1}^{m}\;D_{i}\leq E\right\} ,  \label{2 2 7}
\end{equation}%
where $E$ is an operator for which $C_{i}E=C_{i}$ for all $i=1,\ldots ,m$.

\begin{theorem}
\label{T 2 2 1}The upper bound \textup{(\ref{2 2 7})} is attained on the
admissible operators $D_{i},\;i=1,\ldots ,m$, if and only if there is a
trace class operator $B^{\mathrm{o}}\geq 0,\,C_{i},\;i=1,\ldots ,m$, such
that 
\begin{equation}
B^{\mathrm{o}}(E-D^{\mathrm{o}})=0,\;(B^{\mathrm{o}}-C_{i})D_{i}^{\mathrm{o}%
}=0,\;i=1,\ldots ,m,  \label{2 2 8}
\end{equation}%
with $D^{\mathrm{o}}=\sum_{i=1}^{m}D_{i}^{\mathrm{o}}$. The operator $B^{%
\mathrm{o}}$ then is the solution to the duality problem 
\begin{equation}
\left\langle \mathbf{C}\right\rangle _{+}=\inf_{B\geq 0}\left\{ \left\langle
B,E\right\rangle |B\geq C,\;i=1,\ldots ,m\right\} ,  \label{2 2 9}
\end{equation}
with $\varkappa ^{\mathrm{o}}(\mathbf{C})=\left\langle \mathbf{C}%
\right\rangle _{+}$, for which conditions \textup{(\ref{2 2 8})} are also
necessary and sufficient when $D_{i}^{\mathrm{o}}\geq 0$, $%
\sum_{i=1}^{m}D_{i}^{\mathrm{o}}\leq E$.
\end{theorem}

\begin{proof}
The proof, which is similar to the proof of a particular case of this
theorem, Theorem \ref{T 2 1 1}, will be found as a corollary of a more
general theorem, Theorem \ref{T 2 3 1}.

It can also be easily proved that the solution to problem (\ref{2 2 7})
exists for all trace class operators $C_{i}$ and determines on subspace $%
\mathcal{E}=E\mathcal{H}$ for which $C_{i}E=C_{i}$ for all $i=1,\ldots ,m$,
a unique solution $B^{\mathrm{o}}=B^{\mathrm{o}}E$ to problem (\ref{2 2 9}).

Indeed, the lower bound (\ref{2 2 9}) determines for the vector operators $%
\mathbf{C}=\left( C_{i}\right) _{i=1}^{m}$ be the gauge $\left\langle 
\mathbf{C}\right\rangle _{+}$, a positive homogeneous sublinear functional
on the space of families $\left( C_{i}\right) _{i=1}^{m}$ of the trace class
operators $C_{i,}\;i=1,\ldots ,m$, that possesses the property $\left\langle 
\mathbf{C}\right\rangle _{+}=0\Leftrightarrow C_{i}\leq 0$ for all $i$'s.
Bearing in mind that every linear functional $\mathbf{C\rightarrow
\left\langle \mathbf{C,D}\right\rangle }$ satisfying the condition $%
\left\langle \mathbf{C,D}\right\rangle \leq \left\langle \mathbf{C}%
\right\rangle _{+}$ is positive and has the form 
\begin{equation*}
\left\langle \mathbf{C,D}\right\rangle =\sum_{i=1}^{m}\mathrm{Tr}%
\;(C_{i},D_{i}),
\end{equation*}%
where $\sum_{i=1}^{m}D_{i}=E$, we find that the set that is conjugate to $%
\left\{ \mathbf{C|\left\langle C\right\rangle }_{+}\leq 1\right\} $ consists
of the resolving families $\mathbf{D}=\{D_{i}\}_{i=1}^{m}$ of bound
operators that are admissible in problem (\ref{2 2 7}). The existence of a
solution to problem (\ref{2 2 7}) follows, therefore, from the Hahn-Banach
theorem, according to which for every vector $\mathbf{C}^{\mathrm{o}}$ of a
calibrated space there exists a supporting functional $\mathbf{D}^{\mathrm{o}%
}$ defined by the conditions $\left\langle \mathbf{C,D}^{\mathrm{o}%
}\right\rangle \leq \left\langle \mathbf{C}\right\rangle _{+}$ and $%
\left\langle \mathbf{C}^{\mathrm{o}},\mathbf{D}^{\mathrm{o}}\right\rangle
=\left\langle \mathbf{C}^{\mathrm{o}}\right\rangle _{+}$. For every solution 
$\mathbf{D}^{\mathrm{o}}$ to problem (\ref{2 2 7}) the solution of the
conjugate problem (\ref{2 2 9}) on the subspace $\mathcal{E}=E\mathcal{H}$
is determined uniquely by the formula 
\begin{equation}
B^{\mathrm{o}}E=B^{\mathrm{o}}D^{\mathrm{o}}=\sum_{i=1}^{m}C_{i}D_{i}^{%
\mathrm{o}}  \label{2 2 10}
\end{equation}%
which is obtained by adding (\ref{2 2 8}) over $i=1,\ldots ,m$. Note that
the above proof of the existence of a solution to problem (\ref{2 2 7}) and
of the uniqueness of the solution to problem (\ref{2 2 9}) remains valid for
the case of an infinite number of patterns $m=\infty $ if we require that $%
\left\langle C_{i},E\right\rangle =\mathrm{Tr}\,C_{i}\rightarrow 0$ as $%
i\rightarrow \infty $.

Conditions (\ref{2 2 8}) can easily be met for $m>1$ by analogy with the
case of $m=1$ only for commutative $C_{i}$, when these operators have a
joint spectral representation 
\begin{equation}
C_{i}=\sum_{i=1}^{m}\varkappa _{in}|\chi _{n})(\chi _{n}|,\;(\chi _{n}\mid
\chi _{m})=\delta _{mn}.  \label{2 2 11}
\end{equation}%
The orthoprojector $E$ can be resolved into an orthogonal sum $%
E=E_{0}+\sum_{i=1}^{m}E_{i}$, where 
\begin{equation*}
E_{i}=\sum_{n\in \mathbb{N}_{i}}|\chi _{n})(\chi _{n}|,\;\;\mathbb{N}%
_{i}\subseteq \{n\in \mathbb{N}|\varkappa _{in}\geq \left\{ 0,\;\varkappa
_{kn}:\;k\neq i\right\} \}
\end{equation*}%
(points $n$ at which $\varkappa _{in}=\max_{j=1,\ldots ,m}\varkappa
_{jn}=\varkappa _{kn}$ refer to any one of the nonintersecting sets $\mathbb{%
N}_{i},\;\mathbb{N}_{k}$). The operators 
\begin{equation}
D_{i}^{\mathrm{o}}=E_{i},\;B^{\mathrm{o}}=\sum_{i=1}^{m}\sum_{n\in \mathbb{N}%
_{i}}\varkappa _{in}|\chi _{n})(\chi _{n}|=\sum_{i=1}^{m}C_{i}E_{i}
\label{2 2 12}
\end{equation}%
are, therefore, admissible and optimal: 
\begin{equation*}
B^{\mathrm{o}}(E-D^{\mathrm{o}})=B^{\mathrm{o}}E_{0}=0,\;(B^{\mathrm{o}%
}-C_{i})E_{i}=C_{i}E_{i}-C_{i}E_{i}=0.
\end{equation*}%
Thus, optimal $m$-alternative detection in the commutative case is reduced
to measuring the discrete distribution of the degree of contrast $\varkappa $
in the proper representation of operators $C_{i}$. The total maximal degree
of contrast in this case is determined from the formula 
\begin{equation}
\varkappa ^{\mathrm{o}}(\mathbf{C)=}\sum_{i=1}^{m}\sum_{n\in \mathbb{N}%
_{i}}\varkappa _{in}=\sum_{n\in \mathbb{N}}\max_{i}\{\varkappa _{in}\vee 0\}.
\label{2 2 13}
\end{equation}
\end{proof}

\subsection{Solution of Optimal Identification}

Let us consider the important case of positive operators $%
C_{i}=H_{i}^{+}H_{i}=S_{i}$ which occur, say, in the case of an incoherent
superposition of wave patterns $F_{i}$ with a background $F_{0}$, when the
degrees of contrast (\ref{2 2 6}) are the density operators $%
P_{i}=F_{i}^{+}F_{i}$. The corresponding extremal problem (\ref{2 2 7}) of
pattern recognition, which is known as the optimal identification problem,
is not trivial for noncommutative $S_{i},\;i=1,\ldots ,m$, for $m>1$ even if
these patterns are pure, that is, are described by nonorthogonal amplitudes $%
\psi _{i},\;i=1,\ldots ,m$. For the case of $m=2$, however, the optimal
identification problem can easily be solved by reducing it to the problem of
optimal detection with one degree-of-contrast operator $C=S_{1}-S_{2}$.
Indeed, allowing for the fact that the admissible operators $B=L$ in the
duality problem (\ref{2 2 9}) are determined by the conditions $L\geq
B>0,\;i=1,2$, we can proceed from (\ref{2 2 9}) to (\ref{2 1 19}) by
carrying out the substitution%
\begin{equation*}
\inf \left\langle L,E\right\rangle =\left\langle S_{2},E\right\rangle +\inf
\left\langle B,E\right\rangle
\end{equation*}%
where $B=L-S_{2}$ is the admissible operator of problem (\ref{2 1 19}):%
\begin{equation*}
B\geq \left\{ S_{1}-S_{2},S_{2}-S_{2}\right\} =\{C,0\}.
\end{equation*}%
Thus, the solution $D^{\mathrm{o}}$ to problem (\ref{2 1 17}) makes it
possible to represent the solution to problem (\ref{2 2 7}) in the form 
\begin{equation*}
\varkappa ^{\mathrm{o}}(\mathbf{S})=\left\langle C,D^{\mathrm{o}%
}\right\rangle +\left\langle S_{2},E\right\rangle =\left\langle S_{1},D^{%
\mathrm{o}}\right\rangle +\left\langle S_{2},E-D^{\mathrm{o}}\right\rangle ,
\end{equation*}%
which yields the optimal decision operators $D_{1}^{\mathrm{o}}=D^{\mathrm{o}%
}$ and $D_{2}^{\mathrm{o}}=E-D^{\mathrm{o}}$.

To investigate the problem of identifying wave patterns in the
multialternative case with $m>2$, we restrict the space $\mathcal{H}$ by the
minimal space $\mathcal{E}^{\mathrm{o}}\subseteq \mathcal{H}$ containing all
the ranges of values $\mathcal{H}_{i}=S_{i}\mathcal{H}$. Since the $%
S_{i}\geq 0$, every operator $B$ admissible to problem (\ref{2 2 9}) is
determined by the conditions $B\geq S_{i}$, $i=1,\ldots ,m$, in view of
which it is nonsingular on the subspace $\mathcal{E}^{\mathrm{o}}$ in the
sense that $BD=0\Rightarrow D=0$ for every operator $D$ in $\mathcal{E}^{%
\mathrm{o}}$. Otherwise, operator $D^{+}(B-S_{i})D$ could be negative for at
least one $i\in 1,\ldots ,m$. This last fact means that the first condition
in (\ref{2 2 8}) is met only if $E=D^{\mathrm{o}}$, that is, the optimal
decision operators $D_{i}^{\mathrm{o}}$ determine the decomposition of unity 
$E^{\mathrm{o}}=\sum_{i=1}^{m}D_{i}^{\mathrm{o}}$, the orthoprojector on
subspace $\mathcal{E}^{\mathrm{o}}$, and operator $B^{\mathrm{o}}$ can be
found uniquely by summation of the remaining optimality conditions in (\ref%
{2 2 8}).

When the subspaces $\mathcal{K}_{1}=H_{i}\mathcal{H}$ have a low
dimensionality, say, ordinary amplitudes $H_{i}=(\psi _{i}|$ for which $%
\mathcal{K}_{i}=\mathbb{C}$, it has proved expedient to represent the
solution to problem (\ref{2 2 7}) via the following.

\begin{theorem}
\label{T 2 2 2}The optimal decision operators $D^{\mathrm{o}}$ determined by
conditions \textup{(\ref{2 2 8})} for $C_{i}=H_{i}^{\ast }H_{i},\;i=1,\ldots
,m$, have the following form in space $\mathcal{E}^{\mathrm{o}}$%
\begin{equation}
D_{i}^{\mathrm{o}}=(L^{\mathrm{o}})^{-1}H_{i}^{\ast }\mu _{i}^{\mathrm{o}%
}H_{i}(L^{\mathrm{o}})^{-1},\;i=1,\ldots ,m,  \label{2 2 14}
\end{equation}%
where $L^{\mathrm{o}}=\left( \sum_{i=1}^{m}H_{i}^{\ast }\mu _{i}H_{i}\right)
^{1/2}=B^{\mathrm{o}}$ is the solution to problem \textup{(\ref{2 2 9})},
and the $\mu _{i}$ are trace class positive operators in $\mathcal{K}_{i}$
defined by the conditions 
\begin{equation}
(1_{i}-H_{i}(L^{\mathrm{o}})^{-1}H_{i}^{\ast })\mu _{i}^{\mathrm{o}%
}=0,\;1_{i}\geq H_{i}(L^{\mathrm{o}})^{-1}H_{i}^{\ast }  \label{2 2 15}
\end{equation}%
\textup{(}$1_{i}$ are the identity operators in $\mathcal{K}_{i}$\textup{)}.
If these conditions are met, maximal intensity of graded signals $\varkappa
^{\mathrm{o}}=\sum_{i=1}^{m}\mathrm{Tr}\;\mu _{i}^{\mathrm{o}}$ is achieved.
\end{theorem}

\begin{proof}
Multiplying the remaining equations in (\ref{2 2 8}) from the right by $(L^{%
\mathrm{o}})^{1/2}$ and from the left by $(L^{\mathrm{o}})^{-1/2}$, where $%
L^{\mathrm{o}}=B^{\mathrm{o}}$, we can rewrite the optimality conditions in
the form 
\begin{equation}
(E^{\mathrm{o}}-F_{i}^{\ast }F_{i})M_{i}^{\mathrm{o}}=0,\;E^{\mathrm{o}}\geq
F_{i}^{\ast }F_{i},\;i=1,\ldots ,m,  \label{2 2 16}
\end{equation}%
where $F_{i}=H_{i}(L^{\mathrm{o}})^{1/2}$ and%
\begin{equation*}
M_{i}^{\mathrm{o}}=(L^{\mathrm{o}})^{-1/2}D_{i}^{\mathrm{o}}(L^{\mathrm{o}%
})^{1/2}.
\end{equation*}%
Thus, 
\begin{equation}
M_{i}^{\mathrm{o}}=F_{i}^{\ast }F_{i}M_{i}^{\mathrm{o}}=M_{i}^{\mathrm{o}%
}F_{i}^{\ast }F_{i}=F_{i}^{\ast }\mu _{i}^{\mathrm{o}}F_{i},  \label{2 2 17}
\end{equation}%
where $\mu _{i}^{\mathrm{o}}=F_{i}M_{i}^{\mathrm{o}}F_{i}^{\ast }$, which
leads to (\ref{2 2 14}) if we carry out the inverse transformation. If we
substitute (\ref{2 2 17}) into (\ref{2 2 16}) and multiply the result from
the right by $F_{i}^{\ast }$ and from the left by $(F_{i}^{\ast })^{-1}$, we
arrive at (\ref{2 2 15}) if we allow for the reversibility of the operators $%
F^{\ast }:\mathcal{K}_{i}\rightarrow \mathcal{E}$. The inequalities (\ref{2
2 15}) are simply the inequality (\ref{2 2 16}) in the form $%
F_{i}F_{i}^{\ast }\leq 1_{i}$. The operator $L^{\mathrm{o}}$ is determined
by summation $\sum_{i=1}^{m}D_{i}^{\mathrm{o}}=E^{\mathrm{o}}$ of the
optimal decision operators (\ref{2 2 14}), which yields $L^{\mathrm{o}%
}=\sum_{i=1}^{m}H_{i}^{\mathrm{o}}\mu _{i}^{\mathrm{o}}H_{i}$, and this
determines uniquely the positive operator $B^{\mathrm{o}}=L^{\mathrm{o}}$.

The proved theorem reduces the solution of the optimal identification
problem to finding the operators $\mu _{i}^{\mathrm{o}}$ that satisfy
conditions (\ref{2 2 15}), which in the case of finitely mixed patterns $%
H_{i}$ constitute finite-dimensional algebraic equations and inequalities.
For one, for pure patterns $H_{i}=(\psi _{i}|$, conditions (\ref{2 2 15})
have the scalar form 
\begin{equation}
\mu _{i}^{\mathrm{o}}=(\psi _{i}\mid (L^{\mathrm{o}})^{-1}\psi _{i})\mu
_{i}^{\mathrm{o}},\;1\geq (\psi _{i}\mid (L^{\mathrm{o}})^{-1}\psi
_{i}),\;i=1,\ldots ,m.  \label{2 2 18}
\end{equation}%
where $L^{\mathrm{o}}=(|\psi _{i})\mu _{i}^{\mathrm{o}}(\psi _{i}|)^{1/2}$.
The numerical positive solutions of the system of algebraic equations (\ref%
{2 2 18}) determine the one-dimensional decision operators 
\begin{equation}
D_{i}^{\mathrm{o}}=|\chi _{i})(\chi _{i}|,\;\chi _{i}=(L^{\mathrm{o}%
})^{-1}\psi _{i}\sqrt{\mu _{i}^{\mathrm{o}}}  \label{2 2 19}
\end{equation}%
(which are equal to zero for those $i$'s for which $(\psi _{i}\mid (L^{%
\mathrm{o}})^{-1}\psi _{i}<1)$ and the quality of the optimal solution, $%
\varkappa ^{\mathrm{o}}=\sum_{i=1}^{m}\mu _{i}^{\mathrm{o}}$).

Solution of the pattern identification problem makes it possible to
establish the quasioptimal multialternative detection scheme using the
maximum criterion of the total degree of contrast (\ref{2 2 6}) as the first
approximation in $\sqrt{\varepsilon }$ for decision operators of the form 
\begin{equation}
D_{i}=(F_{0i}+\sqrt{\varepsilon }F_{1i})^{\ast }(F_{0i}+\sqrt{\varepsilon }%
F_{1i})=F_{\varepsilon i}^{\ast }F_{\varepsilon i}.  \label{2 2 20}
\end{equation}%
Assuming that $F_{0i}=\sqrt{\mu _{i}^{\mathrm{o}}}H_{i}(L^{\mathrm{o}})^{-1}$
and $D_{0i}=F_{0i}^{\ast }F_{0i}$, in the first order in the signal-to-noise
ratio $\varepsilon \ll 1$ we obtain the following formula for the degree of
contrast of quasioptimal detection $\varkappa _{0}=\sum_{i=1}^{m}S_{i}D_{0i}$%
\begin{equation}
\varkappa _{0}=\sum_{i=1}^{m}\mathrm{Tr}_{\mathcal{K}_{i}}\mu _{i}^{\mathrm{o%
}}(1+\sqrt{\varepsilon }(\gamma _{i}+\gamma _{i}^{\ast })/2),\;
\label{2 2 21}
\end{equation}%
where $\gamma _{i}=H_{i}A_{i}^{\ast }(L^{\mathrm{o}})^{-1}H_{i}^{\ast }$, or 
$\gamma _{i}=(A_{i}\psi _{i}|(L^{\mathrm{o}})^{-1}\psi _{i})$ when $\mathcal{%
K}_{i}=\mathbb{C}$.
\end{proof}

\subsection{The Signal Representation}

It has proved expedient to represent solution (\ref{2 2 14}) to the problem
of optimal identification of wave patterns in the so-called signal space, $%
\mathcal{K}^{m}=\bigoplus_{i=1}^{m}\mathcal{K}_{i}$, which is the direct sum
of Hilbert spaces $\mathcal{K}_{i}=H_{i}\mathcal{H}$ and which, in the case
of ordinary amplitudes $H_{i}=(\psi _{i}|$, is equal to $\mathbb{C}^{m}$.
Such decomposition is carried out via the partially isometric operator $V:%
\mathcal{H}\rightarrow \mathcal{K}^{m}$ of the polar expansion $H=\sigma
^{1/2}V,\;\sigma =HH^{\ast }$, for the operator $H:\varphi \in \mathcal{H}%
\mapsto \lbrack H_{i}\varphi ]_{i=1}^{m}$ from $\mathcal{H}$ into $\mathcal{K%
}^{m}$, which is defined uniquely on the subspace $\mathcal{E}^{\mathrm{o}}$
by the conditions $V^{\ast }V=E^{\mathrm{o}}$ and $VV^{\ast }=\varepsilon ^{%
\mathrm{o}}$, where $\varepsilon ^{\mathrm{o}}$ is the support of the
correlation matrix $\sigma =\sigma \varepsilon ^{\mathrm{o}}$. Note that the 
$m$-by-$m$ matrix $\sigma =[\sigma _{ik}]$ consisting of operator components 
$\sigma _{ik}=H_{i}H_{k}^{\ast }$,$\;i,k=1,\ldots ,m\;$($\sigma _{ik}=(\psi
_{i}|\psi _{k})$ if $H_{i}=(\psi _{i}|$), is positive and, in the case of
the linear independence of the signals $H_{i}$, nonsingular with support $%
\varepsilon ^{\mathrm{o}}=\bigoplus_{i=1}^{m}1_{i}\equiv 1^{m}$. The
components $V_{i}=1_{i}V,\;i=1,\ldots ,m$, of the isometric operator $V:%
\mathcal{E}^{\mathrm{o}}\rightarrow \mathcal{K}^{m}$ determined by the
diagonal projectors $1_{i}$ from $\mathcal{K}^{m}$ onto $\mathcal{K}_{i}$
bring about, obviously, the decomposition of the unit element 
\begin{equation}
E^{\mathrm{o}}=V^{\ast }V=\sum_{i=1}^{m}V^{\ast
}1_{i}V=\sum_{i=1}^{m}V_{i}^{\ast }V_{i}  \label{2 2 22}
\end{equation}%
of space $\mathcal{E}^{\mathrm{o}}$ and are orthogonal if $\varepsilon ^{%
\mathrm{o}}=1^{m}$: 
\begin{equation*}
V_{i}V_{k}^{\ast }=\varepsilon _{ik}^{\mathrm{o}}=1_{i}\varepsilon ^{\mathrm{%
o}}1_{k}=\delta _{ik}1_{k}.
\end{equation*}%
Representing the operators $H_{i}$ in the form $H_{i}=h_{i}V$, with $%
h_{i}=1_{i}\sigma ^{1/2}$, we can write the necessary and sufficient
conditions for the optimality of the separating operators $D_{i}$ in the
following form: 
\begin{equation}
(\lambda ^{\mathrm{o}}-\sigma _{i})\delta _{i}^{\mathrm{o}}=0,\;\lambda ^{%
\mathrm{o}}\geq \sigma _{i}:=h_{i}^{\ast }h_{i},\;i=1,\ldots ,m,
\label{2 2 23}
\end{equation}%
which are simply conditions for the decomposition of the $m$-by-$m$
projection matrix $\varepsilon ^{\mathrm{o}}=\sum_{i=1}^{m}\delta _{1}^{%
\mathrm{o}},\;\delta _{i}^{\mathrm{o}}=VD_{i}^{\mathrm{o}}V^{\ast }$.
Theorem \ref{T 2 2 2} in this case assumes the form of

\begin{theorem}
\label{T 2 2 3}The optimal decomposition of the support $\varepsilon ^{%
\mathrm{o}}$ of the correlation matrix $\sigma $ defined by conditions 
\textup{(\ref{2 2 23})} has the form 
\begin{equation}
\delta _{i}^{\mathrm{o}}=\lambda ^{\mathrm{o}-1}\;h\mu _{i}^{\mathrm{o}%
}h\lambda ^{\mathrm{o}-1},\;i=1,\ldots ,m,  \label{2 2 24}
\end{equation}%
where $h=\sigma ^{1/2},\;\lambda ^{\mathrm{o}}=(h\mu ^{\mathrm{o}}h)^{1/2}$,
and $\mu ^{\mathrm{o}}=\bigoplus_{i=1}^{m}\mu _{i}$ is a diagonal matrix $%
\mu ^{\mathrm{o}}=[\mu _{i}^{\mathrm{o}}\delta _{ik}]$ consisting of the
positive operators $\mu _{i}^{\mathrm{o}}:\mathcal{K}_{i}\rightarrow 
\mathcal{K}_{i}$ and defined by the conditions 
\begin{equation}
\mu ^{\mathrm{o}}=\epsilon (h\lambda ^{\mathrm{o}-1}h)\mu ^{\mathrm{o}%
},\;1^{m}\geq \epsilon (h\lambda ^{\mathrm{o}-1}h),  \label{2 2 25}
\end{equation}%
or $\mu ^{\mathrm{o}}=\epsilon (\sqrt{\sigma \mu ^{\mathrm{o}}})$ if $\sigma 
$ is nonsingular $(\varepsilon ^{\mathrm{o}}=1)$, where $\epsilon :a\mapsto
\sum_{i=1}^{m}1_{i}a1_{i}$ is the partial diagonalization operation $%
[a_{ik}]\mapsto \lbrack a_{ik}\delta _{ik}]$ of the $m$-by-$m$
block-matrices $a=[a_{ik}]$ consisting of operators $a_{ik}:\mathcal{K}%
_{k}\rightarrow \mathcal{K}_{i}$. The quality of optimal identification is
determined by the trace in $\mathcal{K}^{m}$, or $\varkappa ^{\mathrm{o}}=%
\mathrm{Tr}$ $\mu ^{\mathrm{o}}$.
\end{theorem}

\begin{proof}
Representation (\ref{2 2 24}) can be obtained directly via the isomorphism $%
V $ of spaces $E^{\mathrm{o}}\mathcal{H}$ and $\varepsilon ^{\mathrm{o}}%
\mathcal{K}^{m}$. Here $\left( \lambda ^{\mathrm{o}}\right) ^{-1}=V(L^{%
\mathrm{o}})^{-1}V^{\ast }$, an $m$-by-$m$ matrix with elements $(\lambda
^{ki})^{-1}:\mathcal{K}_{1}\rightarrow \mathcal{K}_{k}$, is the inverse of
matrix $\lambda ^{\mathrm{o}}=VL^{\mathrm{o}}V^{\ast }$ with respect to $%
\varepsilon ^{\mathrm{o}}$: $\lambda ^{\mathrm{o}-1}\lambda ^{\mathrm{o}%
}=\varepsilon ^{\mathrm{o}}=\lambda ^{\mathrm{o}}\lambda ^{\mathrm{o}-1}$.
Matrix $\lambda ^{\mathrm{o}}=VL^{\mathrm{o}}V^{\ast }$ consisting of
operator elements $\lambda _{ik}^{\mathrm{o}}:\mathcal{K}_{k}\rightarrow 
\mathcal{K}_{i}$ is directly expressible in terms of the square root $h=%
\sqrt{\sigma }$ of the correlation matrix 
\begin{equation}
\lambda ^{\mathrm{o}}=\left( \sum_{i=1}^{m}h_{i}^{\ast }\mu _{i}^{\mathrm{o}%
}h_{i}\right) ^{1/2}=\sqrt{h\mu ^{\mathrm{o}}h},  \label{2 2 26}
\end{equation}%
while the conditions (\ref{2 2 15}) for determining the operators $\mu _{i}^{%
\mathrm{o}}$, which in the signal representation have form 
\begin{equation}
(1_{i}h_{i}\lambda ^{\mathrm{o}-1}h_{i}^{\ast })\mu _{i}^{\mathrm{o}%
}=0,\quad 1_{i}\geq h_{i}\lambda ^{\mathrm{o}-1}h_{i}^{\ast },
\label{2 2 27}
\end{equation}%
represent the element-by-element notation for the conditions (\ref{2 2 25})
imposed on the diagonal elements in $\mathcal{K}^{m}$. If $\sigma $ is
nonsingular (which means that $h$ is nonsingular, too), we can rewrite (\ref%
{2 2 25}) in the following simple form: 
\begin{equation}
\mu ^{\mathrm{o}}=\epsilon (h\lambda ^{\mathrm{o}-1}h\mu ^{\mathrm{o}%
})=\epsilon (h\lambda ^{\mathrm{o}}h^{-1})=\epsilon (\sqrt{\delta \mu ^{%
\mathrm{o}}}),  \label{2 2 28}
\end{equation}%
where we have allowed for the fact that $\epsilon (a)\mu ^{\mathrm{o}%
}=\epsilon (a\mu ^{\mathrm{o}})$ (because $\mu ^{\mathrm{o}}$ is diagonal)
and that $\sigma \mu ^{\mathrm{o}}=(h\lambda ^{\mathrm{o}}h^{-1})^{2}$, in
accordance with (\ref{2 2 25}) can be resolved explicitly. Let us assume
that the diagonal part $\epsilon (h)$ of matrix $h=\sigma ^{1/2}$ is
commutative with $h$. Then conditions (\ref{2 2 25}) are met at $\mu ^{%
\mathrm{o}}=\epsilon (\sqrt{\sigma })^{2}$, that is, at $\mu _{i}^{\mathrm{o}%
}=(h_{ii})^{2}$, $i=1,\ldots ,m$. Indeed, the diagonal matrix $\mu ^{\mathrm{%
o}}$ in this case is commutative with $h$ and 
\begin{equation*}
\lambda ^{\mathrm{o}}=\sqrt{h\mu ^{\mathrm{o}}h}=\sqrt{h^{2}\mu ^{\mathrm{o}}%
}=h\sqrt{\mu ^{\mathrm{o}}}=h\epsilon (h)=\sqrt{\sigma }\epsilon (\sqrt{%
\sigma }).
\end{equation*}%
Moreover, $h\lambda ^{\mathrm{o}-1}h=h\epsilon (h)^{-1}$, where $\epsilon
(h)^{-1}$ is the diagonal that is the inverse of $\epsilon (h)$, which
always exists because the diagonal elements $\sigma _{ii}=H_{i}H_{i}^{\ast }$
of the correlation matrix $\sigma $ are nonsingular and, hence, so are the
diagonal elements $h_{ii}$ of the matrix $h=\sqrt{\sigma }$ on the spaces $%
\mathcal{K}_{i}=H_{i}\mathcal{H}$. Thus, 
\begin{equation*}
\epsilon (h\lambda ^{\mathrm{o}-1}h)=\epsilon (h\epsilon (h)^{-1})=\epsilon
(h)\epsilon (h)^{-1}=1^{m},
\end{equation*}%
and conditions (\ref{2 2 25}) are satisfied. The optimal decision operators
then assume the form 
\begin{equation}
\delta _{i}^{\mathrm{o}}=1_{i},\;D_{i}^{\mathrm{o}}=V^{\ast }1,=V_{i}^{\ast
}V_{i},\;i=1,\ldots ,m,  \label{2 2 29}
\end{equation}%
where $V=h^{-1}H$, while the quality of optimal separation is determined by
the total intensity: 
\begin{equation}
\varkappa ^{\mathrm{o}}=\sum_{i=1}^{m}\mathrm{Tr}\;(h_{ii})^{2}=%
\sum_{i=1}^{m}\mathrm{Tr}\,(\sigma _{ii}^{1/2})^{2}.  \label{2 2 30}
\end{equation}%
The above-noted property of commutativity manifests itself, for one thing,
in the case where all diagonal operators $\sigma _{ii}$ coincide and are
multiples of the identity element $1_{i}=1$ of space $\mathcal{K}_{i}=%
\mathcal{K}$, which is the same for all $i=1,\ldots ,m$. In view of the
assumption that $\sigma _{ii}$ is a trace class operator and, hence, $\mu
_{i}^{\mathrm{o}}=(\sigma _{ii})^{2}$, this is possible only for a
finite-dimensional $\mathcal{K}$. In Section \ref{Cyclic Systems} we
consider concrete equidiagonal families of ordinary amplitudes $H_{i}=(\psi
_{i}|$, for which $\mathcal{K}=\mathbb{C}$.
\end{proof}

\subsection{Separation of Cyclic Systems}

\label{Cyclic Systems}Let $\left\{ \psi _{i}\right\} _{i=1}^{m}$ be a family
(or set) of nonorthogonal wave amplitudes $\psi _{i}\in \mathcal{H}$ that
describe sound or visual patterns with a correlation matrix $\sigma =[(\psi
_{i}|\psi _{k})]$ whose square root, $h=\sqrt{\sigma }$, has the same
diagonal elements $h_{ii}=a=h_{kk}$ for all $i,k=1,\ldots ,m$. The optimal
identification of wave patterns $\left\{ \psi _{i}\right\} $ is described by
the one-dimensional separating operators $\left\{ D_{k}^{\mathrm{o}}\right\}
_{k=1}^{m}$ of the form (\ref{2 2 12}), where $\chi _{k}^{\mathrm{o}%
}=C_{k}^{\ast },\;k=1,\ldots ,m$, is generally an overcomplete system of
polar decomposition, 
\begin{eqnarray*}
&&\psi _{k}=\sum_{i=1}^{m}V_{k}^{\ast }\sigma _{ki}^{1/2}=\sum_{i=1}^{m}\chi
_{k}^{\mathrm{o}}h_{ki}, \\
&&\sum_{k=1}^{m}|\chi _{k}^{\mathrm{o}})(\chi _{k}^{\mathrm{o}%
}|=\sum_{k=1}^{m}V_{k}^{\ast }V_{k}=V^{\ast }V=E^{\mathrm{o}},
\end{eqnarray*}%
in the space $\mathcal{E}^{\mathrm{o}}$ induced by the set $\left\{ \psi
_{k}\right\} $. Bearing in mind that $\mu =a^{2}=(\mathrm{Tr}\,h/m)^{2}$, we
can represent the maximal intensity $\varkappa ^{\mathrm{o}}=ma^{2}$ of
optimally separated amplitudes $\left\{ \chi _{i}^{\mathrm{o}}\right\} $ in
the following invariant form: 
\begin{equation*}
\varkappa ^{\mathrm{o}}=\frac{1}{m}(\mathrm{Tr}\,h)^{2}=\frac{1}{m}(\mathrm{%
Tr}\,\sigma ^{1/2})^{2}.
\end{equation*}

Let us consider the following example when the above-mentioned condition of
the equidiagonality of matrix $h=\sqrt{\sigma }$ is met. We will call the
system $\left\{ \psi _{i}\right\} $ of amplitudes of equal intensity $%
\left\Vert \psi _{i}\right\Vert ^{2}=\nu $ equiangular if $(\psi _{i}|\psi
_{k})=\nu \gamma $ for every $i\neq k$, that is, if the cosines of all
mutual angles are equal to $\gamma $. This is possible in the case when $%
\gamma \geq 1/(1-m)$, say, when $\psi _{i}=\varphi _{0}+\varphi _{i}$, where 
$\left\{ \varphi _{i}\right\} _{i=0}^{m}$ is an orthogonal system of
amplitudes with intensities $\left\Vert \varphi _{0}\right\Vert ^{2}=\nu
\gamma $ and $\left\Vert \varphi _{i}\right\Vert ^{2}=\nu (1-\gamma )$ at $%
i\neq 0$. Representing the respective correlation matrix $\sigma $ in the
form 
\begin{equation}
\sigma =\nu ((1-\gamma )1^{m}+\gamma x^{\intercal }x),\;x=(1,\ldots ,1)\in 
\mathbb{C}^{m},  \label{2 2 31}
\end{equation}%
and using the formula 
\begin{equation*}
f(1^{m}+\tau x^{\intercal }x)=f(1)1^{m}+\frac{1}{xx^{\intercal }}[f(1+\tau
xx^{\intercal })-f(1))x^{\intercal }x]
\end{equation*}%
to invert it and extract a square root, we can write out the optimal system $%
\left\{ \chi _{i}^{\mathrm{o}}\right\} $ for $\gamma \in ](1-m)^{-1},1[$
explicitly: 
\begin{equation*}
\chi _{k}^{\mathrm{o}}=\frac{1}{\sqrt{1-\gamma }}\left( \frac{1}{\sqrt{\mu }}%
\psi _{k}-(1-(1+\frac{m\gamma }{1-\gamma })^{-1/2})\frac{1}{m}\sum_{i=1}^{m}%
\frac{1}{\sqrt{\mu }}\psi _{i}\right) .
\end{equation*}%
The intensity of the signals separated by this orthogonal system is 
\begin{equation*}
\varkappa ^{\mathrm{o}}=\nu (m-(1-\frac{1}{m})(\sqrt{1-\gamma +m\gamma }-%
\sqrt{1-\gamma }))^{2},
\end{equation*}%
and admits the maximal value $\varkappa ^{\mathrm{o}}=m\nu $ in the event of
orthogonality $\gamma =0$ of the family $\left\{ \psi _{i}\right\} $ and the
value $\varkappa ^{\mathrm{o}}=\mu $ in the case of colinearity of $\left\{
\psi _{k}\right\} $.

For one, when the $\psi _{i}=|\alpha _{i})$ are canonical equiangular
amplitudes $\alpha _{i}$ defined by a $(d+1)$-by-$(d+1)$ matrix of the
scalar products of vectors $\alpha _{i}\in \mathbb{C}^{d+1}$ of the form $%
\alpha _{i}^{\ast }\alpha _{k}^{\intercal }=\lambda \delta $ for $i\neq k$, $%
|\alpha _{i}|^{2}=\lambda $ for all $i=1,\ldots ,m$, the quantity $\gamma
=\exp \left\{ \lambda (\delta -1)\right\} $ does not vanish and the maximal
intensity of optimal separation is always lower than $m\mu $ even if the
vectors $\left\{ \alpha _{i}\right\} $ are orthogonal $(\delta =0)$ and
tends to $m\gamma $ only as $\lambda \rightarrow \infty $. Note that the
maximal intensity of separation of canonical amplitudes is reached on
simplex vectors $\alpha _{i}\in \mathbb{C}^{d+1}$ defined by the condition $%
\delta =(1-m)^{-1}$; for one thing, at $m=2$ the intensity of separation of
a pair of canonical amplitudes, 
\begin{equation*}
\varkappa ^{\mathrm{o}}=\nu (1-\sqrt{1-\gamma ^{2}})=1-\sqrt{1-\mathrm{e}%
^{2\lambda (\delta -1)}},
\end{equation*}%
can be attained at $\delta =-1$ by employing orthogonal vectors $\alpha _{i}$%
, while at $\delta =0$ this can be done only by doubling $\lambda =|\alpha
_{i}|^{2}$.

Equiangular systems constitute a particular case of cyclic systems, which
are defined by the condition that the correlation matrix $\sigma _{ik}$
remain invariant under translations $s\in \mathbb{Z}:(i,k)\mapsto
(i+s,\;k+s) $, that is, at $\sigma _{ik}=\sigma (i-k)$. Such translation
invariant systems as containing only a finite number $m$ of distinct
amplitudes must satisfy also the cyclicity condition $\sigma (l)=\sigma
(l+s) $ for $l=i-k<0$. Since the matrix $h=\sqrt{\sigma }$, as any other
matrix function of $\sigma $, also depends solely on the difference in the
indices, or $h_{i,k}=h(i-k)$, the equidiagonality condition $h_{i,k}=a=h(0)$
is certain to be met and the solution to the problem of separating any
cyclic system can be written explicitly.

Let us take the case of cyclic canonical amplitudes $\psi _{k}=|\alpha _{k})$
defined by complex numbers $\alpha _{k}\in \mathbb{C}$ whose real and
imaginary parts can be interpreted as the mean frequency and duration of the
wave packet $|\alpha _{i})$. There can be only two cases of the cyclicity of
amplitudes $|\alpha _{i})$ corresponding to the equidistant distribution of
points $\alpha _{i}$ along a circle or a straight line with the center at $%
\alpha =0$.

(1) \textit{Optimal estimation of phase} Let $\alpha _{i}=\sqrt{\lambda }%
\mathrm{e}^{2\pi i\mathrm{j}k/m},\mathrm{j}=\sqrt{-1}$. In this case we have
a cyclic system 
\begin{equation*}
\sigma _{ik}=\exp \left\{ \lambda (\mathrm{e}^{-2\pi \mathrm{j}%
(i-k)/m}-1\right\} =\sigma (i-k).
\end{equation*}%
To extract the square root of matrix $\sigma $ one should diagonalize it by
a discrete Fourier transformation applying the unitary matrix 
\begin{equation*}
U_{in}=\exp \left\{ 2\pi i\mathrm{j}n/m\right\} /\sqrt{m},\;n=0,\ldots ,m-1.
\end{equation*}%
A continuous analog of this problem, to which one can pass if $m$ is sent to
infinity, is the estimation of phase $\theta $ of the vector $\alpha _{0}=%
\sqrt{\lambda }\exp \{2\pi \mathrm{j}\theta \}$ of the canonical amplitude $%
\psi _{\theta }=|\alpha _{\theta })$ on the interval $[0,1]$. Diagonalizing
matrix 
\begin{equation*}
\sigma _{x\theta }=(\alpha _{x}|\alpha _{\theta })=\exp \left\{ \lambda (%
\mathrm{e}^{-2\pi \mathrm{j}(x-\theta )}-1\right\}
\end{equation*}%
via a discrete-continuous Fourier transformation $\mu _{xn}=\exp \{2\pi 
\mathrm{j}xn\}$, $n\in \mathbb{Z}$, we obtain its eigenvalues 
\begin{equation*}
\lambda _{n}=\lambda ^{n}\mathrm{e}^{-\lambda }/n!,\;n=0,1,\ldots
;\;\;\;\;\;\lambda _{n}=0,\;n<0.
\end{equation*}%
The optimal system of decision vectors $\chi _{x}^{\mathrm{o}},\;x\in
\lbrack 0,1]$, has the form 
\begin{equation*}
\chi _{x}^{\mathrm{o}}=\sum_{n=0}^{\infty }\mathrm{e}^{2\pi \mathrm{j}xn}|n),%
\text{ where }|n)=\frac{1}{\sqrt{n!}}(A^{\ast })^{n}|0),
\end{equation*}%
with $A^{\ast }$ the creation operator in $\mathcal{H}=L^{2}(\mathbb{R})$.
It can easily be verified that the system $\chi _{x}^{\mathrm{o}}$ defines
the decomposition of unity, 
\begin{equation*}
I=\int_{0}^{1}|\chi _{x}^{\mathrm{o}})(\chi _{x}^{\mathrm{o}}|\;\mathrm{d}%
x=\sum_{n,m=0}^{\infty }|n)(m|\int_{0}^{1}\mathrm{e}^{2\pi \mathrm{j}%
x(n-m)}\;\mathrm{d}x=\sum_{n=0}^{\infty }|n)(n|,
\end{equation*}%
but is not orthogonal.

(2) \textit{Optimal estimation of amplitude}. Let us take $\alpha
_{i}=i\Delta \mathrm{e}^{\mathrm{j}\theta }$, where $i\in \mathbb{Z}%
,\;\Delta >0$, and $\mathrm{j}=\sqrt{-1}$. In this case the cyclicity
condition is satisfied: 
\begin{equation*}
\sigma _{ik}=\exp \{-\Delta ^{2}(i-k)^{2}/2\}=\sigma (i-k).
\end{equation*}%
The matrix $\sigma =[\sigma _{ik}]$ is diagonalized by the
discrete-continuous Fourier transformation $U_{i\lambda }=\exp \{2\pi 
\mathrm{j}i\lambda \}$, $\lambda \in \lbrack 0,1]$. For $\Delta \ll 1$ the
problem of optimal separation of the respective coherent amplitudes is
reduced to the problem of optimal estimation of the real parameter $x\in 
\mathbb{R}$ of the coherent amplitude $\psi _{x}=|x\mathrm{e}^{\mathrm{j}%
\theta })$. This estimation is realized by measuring the intensity in the
proper representation of the self-adjoint operator $\func{Re}A\mathrm{e}^{-%
\mathrm{j}\theta }$ in space $\mathcal{H}=L^{2}(\mathbb{R})$. At $\theta =0$
this is the frequency representation, while at $\theta =\pi /2$ it is the
temporal representation.

\section{Optimal Discrimination of Mixed Waves}

In this section we will take up the problem of testing wave hypotheses based
on measuring the appropriate intensity distributions. We will derive the
necessary and sufficient conditions for optimal testing of such hypotheses
by the minimum criterion of parasitic contrast at a fixed level of the
received signal by employing a method of linear programming in partially
ordered Banach operator spaces. In a specific case these conditions formally
coincide with conditions obtained earlier in \cite{15} on the optimality of
quantum measurements by the minimum criterion for the error probability. A
general geometric solution will be given for the case of a two-dimensional
space, which is sufficient for describing the recognition of the
polarization of a plane wave. This solution is similar to the solution of
the problem of measuring quantum mechanical spin \cite{4}.

\subsection{Wave Pattern Hypotheses}

The problem of recognizing sound and visual patterns based on measurements
of the intensity of the received audio or optical wave can be formulated
within the framework of the wave theory of hypothesis testing discussed
below.

Let $H_{i},\;i=1,\ldots ,m$, be bounded operators from the Hilbert space $%
\mathcal{H}$ to another Hilbert space $\mathcal{K}$ describing the possible
generalized random amplitudes at the \textquotedblleft in\textquotedblright\
terminals of the receiver with density operators $S_{i}=H_{i}^{\ast }H_{i}$
with a trace $\mathrm{Tr\;}S_{i}<\infty $. The reader will recall that at $%
\mathcal{K}=\mathbb{C}$ the operators $H_{i}$ correspond to ordinary
amplitudes $\psi _{i}\in \mathcal{H},\;i=0,\ldots ,m$, which define the
bounded functionals $H_{i}=(\psi _{i}|:\chi \in \mathcal{H}\mapsto (\psi
_{i}|\chi )$. Each operator $H_{i},\;i=0,\ldots ,m$, can be thought of as a
hypothesis, according to which at the \textquotedblleft
in\textquotedblright\ terminals of the receiver there is one of the possible
simple or mixed patterns $G_{i},\;i=1,\ldots ,m$, in a partially coherent
superposition $H_{i}=G_{i}+H_{0}$ with the absence of wave patterns $G_{i}$.
The problem of $m$-alternative detection of wave patterns $%
G_{i},\;i=1,\ldots ,m$, may, therefore, be considered as a problem of
testing $m+1$ hypotheses $H_{i},\;i=0,\ldots ,m$, and vice versa.

The optimal testing of the hypotheses $H_{i},\;i=0,\ldots ,m$ is determined
by the solution to the problem of finding a quasiselective measurement $%
D=\{D_{i}\}_{i=0}^{m}$ that maximizes the quality functional 
\begin{equation*}
\varkappa (R,D)=\sum_{i=0}^{m}\left\langle R_{i},D_{i}\right\rangle
,\;D_{i}\geq 0,\;\sum_{i=0}^{m}D_{i}=E,
\end{equation*}%
where $R=\left( R_{i}\right) _{i=0}^{m}$ are trace class operators with a
common support $E:R_{i}E=R_{i}$ for all $i=0,\ldots ,m$, operators that are
usually represented by linear combinations $R_{i}=%
\sum_{k=0}^{m}c_{i}^{k}S_{k}$ of density operators $S_{i}=H_{i}^{\ast }H_{i}$%
. For instance, in the problem of $m$-alternative optimal detection by
maximum of the total contrast (\ref{2 2 7}) criterion, the operators $R_{i}$
are in effect the degrees of contrast $R_{0}=0$, $R_{i}=S_{i}-S_{0}=C_{i}$, $%
i=1,\ldots ,m$, and the admissible operators $\{D_{i}\}_{i=0}^{m}$ are
determined by the decision operators $D_{i},\;i=1,\ldots ,m$ and $D_{0}=E-D$%
, where $D=\sum_{i=1}^{m}D_{i}$. For the problem of discriminating between
the hypotheses $H_{i}$ we can consider more general criteria defined, say,
by the operators 
\begin{equation}
R_{0}=\sum_{i=1}^{m}C_{i},\;R_{i}=(1+\lambda _{i})C_{i},\;\lambda _{i}\geq
0,\;i=1,\ldots ,m,  \label{2 3 1}
\end{equation}%
that appear in the problem of suppressing parasitic degrees of contrast $%
\left\langle C_{i},D_{k}\right\rangle $ for $i\neq k$: 
\begin{equation}
\tau ^{\mathrm{o}}(\mathbb{C)}=\inf_{D_{i}\geq 0}\left\{
\sum_{i=1}^{m}\left\langle C_{i},\sum_{k\neq i}D_{k}\right\rangle \Big|%
\left\langle C_{i},D_{i}\right\rangle \geq \varepsilon
_{i},\;\sum_{i=1}^{m}D_{i}\leq E\right\}  \label{2 3 2}
\end{equation}%
under the condition that the useful degrees of contrast $\left\langle
C_{i},D_{i}\right\rangle $, $i=1,\ldots ,m$, are not lower than given levels 
$\varepsilon _{i}$. Indeed, if we solve the extremal problem 
\begin{eqnarray}
\varkappa ^{\mathrm{o}}(R) &=&\left\{ \sum_{i=0}^{m}\left\langle
R_{i},D_{i}\right\rangle \Big|\sum_{i=1}^{m}D_{i}=E\right\}  \notag \\
&=&\sup_{D_{i}\geq 0}\Big\{\sum_{i=1}^{m}\Big(\left\langle
C_{i},D_{i}\right\rangle (1+\lambda _{i})  \notag \\
&&+\left\langle C_{i},E-D\right\rangle \Big)\Big|\sum_{i=1}^{m}D_{i}\leq E%
\Big\},  \label{2 3 3}
\end{eqnarray}%
for the operators $R_{i}=R_{i}\left( \lambda \right) $ defined in (\ref{2 3
1}), we can write the solution to the problem (\ref{2 3 2}) in the form 
\begin{eqnarray}
\tau ^{\mathrm{o}}(\mathbb{C}) &=&\sum_{i=1}^{m}\left\langle
C_{i},E\right\rangle +\sup_{\lambda _{i}\geq 0}\left\{ \sum_{i=1}^{m}\lambda
_{i}\varepsilon _{i}-\varkappa ^{\mathrm{o}}(R)\right\}  \notag \\
&=&\sup_{\lambda _{i}\geq 0}\inf_{D_{i}\geq 0}\Big\{\sum_{i=1}^{m}\Big(%
\left\langle C_{i},\sum_{k\neq i}D_{k}\right\rangle  \notag \\
&&+\lambda _{i}(\varepsilon _{i}-\left\langle C_{i},D_{i}\right\rangle )\Big)%
\Big |\sum_{i=1}^{m}D_{i}\leq E\Big\},  \label{2 3 4}
\end{eqnarray}%
provided that we employ Lagrange's method of multipliers $\lambda
_{i},\;i=1,\ldots ,m$.

Let us start with the classical variant 
\begin{eqnarray}
\varkappa _{M}^{\mathrm{o}}(R) &=&\sup_{\Delta _{i}>0}\left\{
\sum_{i=0}^{m}\left\langle R_{i},M(\Delta _{i})\right\rangle \Big|%
\sum_{i=0}^{m}\Delta _{i}=X\right\}  \notag \\
&=&\sum_{i=0}^{m}\mu _{i}(\Delta _{i}^{\mathrm{o}})  \label{2 3 5}
\end{eqnarray}%
of the problem (\ref{2 3 3}) of optimal testing of hypotheses $H$ in a fixed
measurement described by the decomposition $E=\int M(dx)$ of an
orthoprojector $E,\;R_{i}E=R_{i}$, on a Borel space $X$. This may be the
coordinate selective measurement $M(dx)=I(dx),\;X=\Omega $, or $x$ the
momentum quasiselective measurement $M(dx)=\widetilde{I}(dx)$, $X=\mathbb{R}%
^{d+1}$, or the canonical quasimeasurement $M(dx)=|x)(x|\mathrm{d}x,\;X=%
\mathbb{R}^{d+1}$, described in Section \ref{Models Sound and Visual}. The
upper bound (\ref{2 3 5}) in measurable partitions $X=\sum_{i=0}^{m}\Delta
_{i}$ reaches the gauge 
\begin{equation}
\left\langle \mu \right\rangle =\inf \left\{ \lambda (X)|\lambda \geq \mu
_{i},\;i=0,\ldots ,m\right\} =\mu _{\vee }(X)  \label{2 3 6}
\end{equation}%
of the family $\left( \mu _{i}\right) _{i=0}^{m}$ of measures $\mu
_{i}(\Delta )=\left\langle C_{i},M(\Delta )\right\rangle $, where the
infimum is taken over all the measures of finite variation $|\lambda
|(X)<\infty $ that majorize all $\mu _{i}$.

Indeed, $\varkappa _{M}^{\mathrm{o}}(C)\leq \varkappa _{\vee }(\mu )$, since
for every measurable partition $\Omega =\sum_{i=0}^{m}\Delta _{i}\subseteq X$%
, obviously, 
\begin{equation*}
\sum_{i=0}^{m}\mu _{i}(\Delta _{i})\leq \sum_{i=0}^{m}\lambda (\Delta
_{i})=\lambda (\Omega )\leq \lambda (X).
\end{equation*}%
The lower bound (\ref{2 3 6}) is attained at the upper bound $\mu _{\vee
}=\vee _{i=0}^{m}\mu _{i}$ of the family of measures $\{\mu _{i}\}_{i=0}^{m}$%
, defined as $\mu _{\vee }\geq \{\mu _{i}\}_{i=0}^{m},\;\lambda \geq \mu
_{i}\Rightarrow \lambda \geq \mu _{\vee }$, and is equal to the supremum (%
\ref{2 2 5}) reached on the partitions $\Omega =\sum_{i=0}^{m}\Delta _{i}^{%
\mathrm{o}}$ of the support $\Omega \subseteq X$ of measure $\mu _{\vee }$
into regions $\Delta _{i}^{\mathrm{o}}$, on which it coincides with the
respective measure $\mu _{i}:$%
\begin{equation}
\mu _{\vee }(\Delta _{i}^{\mathrm{o}})=\max_{k=0,\ldots ,m}\mu _{k}(\Delta
_{i}^{\mathrm{o}})=\mu _{i}(\Delta _{i}^{\mathrm{o}}).  \label{2 3 7}
\end{equation}%
In view of the last relationship, determining a hypothesis $H_{i}$ for a
given measurement $M$ is reduced to searching for the number of the nonempty
region $\Delta _{i}^{\mathrm{o}}$ on which the measured degree of contrast
reaches the envelope $\mu _{\vee }(\Delta _{i}^{\mathrm{o}})$ of the family $%
\{\mu _{i}\}$. However, this method does not enable us to find the wave
patterns for which $\mu (\Delta _{i}^{\mathrm{o}})<\mu _{\vee }(\Delta _{i}^{%
\mathrm{o}})$ for all $i=0,\ldots ,m$.

\subsection{Optimal Multialternative Testing}

To obtain a satisfactory solution to the problem of wave pattern recognition
one must look for the supremum (\ref{2 3 5}) not only over the measurement
regions $\Delta _{i}$ but also over the various methods of such a
measurement, which are described by the resolving operators $D_{i}=M(\Delta
_{i})$. Thus, there emerges a nonclassical extremal problem (\ref{2 3 3}),
which may be considered as part of the conditional problem (\ref{2 3 2}) of
testing the hypotheses $H_{i}$ in the degrees of contrast $\left\langle
C_{i},D_{i}\right\rangle $, which are compared with given level $\varepsilon
_{i},i=1,\ldots ,m$. The necessary and sufficient conditions for solving
this problem are formulated in the following.

\begin{theorem}
Theorem \label{T 2 3 1} The upper bound \textup{(\ref{2 3 3})} is attained
on operators $D_{i}^{\mathrm{o}},i=0,\ldots ,m$ if and only if there is a
trace class operator $L^{\mathrm{o}}\geq R_{i},\;i=0,\ldots ,m$, such that 
\begin{equation}
(L^{\mathrm{o}}-R_{i})D_{i}^{\mathrm{o}}=0,\;i=0,\ldots ,m.  \label{2 3 8}
\end{equation}%
The operator $L^{\mathrm{o}}$ is then the solution to the duality problem 
\begin{equation}
\left\langle R\right\rangle _{+}=\inf_{L}\left\{ \left\langle
L,E\right\rangle |L\geq R_{i},\;i=0,\ldots ,m\right\}  \label{2 3 9}
\end{equation}%
for which conditions \textup{(\ref{2 3 8})} are also necessary and
sufficient for $D_{i}^{\mathrm{o}}\geq 0$, $\sum_{i=0}^{m}D_{i}^{\mathrm{o}%
}=E$, and $\varkappa ^{\mathrm{o}}(R)=\left\langle R\right\rangle _{+}$. The
solution $L^{\mathrm{o}}$ to this problem for operators $R_{i}=R_{i}^{%
\lambda },\;i=0,\ldots ,m$, of the form \textup{(\ref{2 3 1})} represents
the solution to the constraint extremal problem \textup{(\ref{2 3 2})} in
the Lagrange form 
\begin{equation}
\tau ^{\mathrm{o}}(\mathbb{C})=\sum_{i=1}^{m}(\left\langle
C_{i},E\right\rangle +\lambda _{i}^{\mathrm{o}}\varepsilon
_{i})-\left\langle L^{\mathrm{o}},E\right\rangle ,  \label{2 3 10}
\end{equation}%
where the parameters $\lambda _{i}^{\mathrm{o}}\geq 0$ are to be found from 
\begin{equation}
\lambda _{i}^{\mathrm{o}}\left( \varepsilon _{i}-\left\langle C_{i},D_{i}^{%
\mathrm{o}}\right\rangle \right) =0,\;\;\varepsilon _{i}\leq \left\langle
C_{i},D_{i}^{\mathrm{o}}\right\rangle ,i=1,\ldots ,m,  \label{2 3 11}
\end{equation}
\end{theorem}

\begin{proof}
The sufficiency of the optimality conditions (\ref{2 3 8}) for (\ref{2 3 7})
and (\ref{2 3 9}) can be verified directly by employing the property of
monotonicity of the trace,%
\begin{equation*}
L\geq R_{i}\Rightarrow \mathrm{Tr}\;(LD_{i})\geq \mathrm{Tr}\,\;(R_{i}D_{i}),
\end{equation*}%
for $D_{i}\geq 0$. Allowing for the equality $L^{\mathrm{o}%
}E=\sum_{i=0}^{m}R_{i}D_{i}^{\mathrm{o}}$, which is obtained via summation
of (\ref{2 3 8}) over $i=0,\ldots ,m$, for every family $\left( D_{i}\right)
_{i=0}^{m}$ admissible in (\ref{2 3 7}) we have 
\begin{eqnarray*}
\sum_{i=0}^{m}\left\langle R_{i},D_{i}\right\rangle &=&\sum_{i=0}^{m}\mathrm{%
Tr}\;(R_{i}D_{i})\leq \sum_{i=0}^{m}\mathrm{Tr}\;(L^{\mathrm{o}}D_{i}) \\
&=&\mathrm{Tr}\;(L^{\mathrm{o}}E)=\sum_{i=0}^{m}\left\langle R_{i},D_{i}^{%
\mathrm{o}}\right\rangle .
\end{eqnarray*}%
In a similar manner for every operator $L$ admissible in (\ref{2 3 9}) we
have 
\begin{equation*}
\left\langle L,E\right\rangle =\mathrm{Tr}\;(LE)=\sum_{i=0}^{m}\mathrm{Tr}%
\;(LD_{i}^{\mathrm{o}})\geq \sum_{i=0}^{m}R_{i}D_{i}^{\mathrm{o}%
}=\left\langle L^{\mathrm{o}},E\right\rangle .
\end{equation*}%
The necessity of the optimality conditions (\ref{2 3 8}) follows from
Lagrange's duality principle 
\begin{eqnarray*}
&&\sup_{D_{i}\geq 0}\left\{ \sum_{i=0}^{m}\left\langle
R_{i},D_{i}\right\rangle \Big|\sum_{i=0}^{m}D_{i}=E\right\} \\
&=&\sup_{D_{i}\geq 0}\inf_{L}\left\{ \sum_{i=0}^{m}\left\langle
R_{i},D_{i}\right\rangle +\left\langle L,E-\sum_{i=0}^{m}D_{i}\right\rangle
\right\} \\
&=&\inf_{L}\sup_{D_{i}\geq 0}\left\{ \sum_{i=0}^{m}\left\langle
R_{i}-L,D_{i}\right\rangle +\left\langle L,E\right\rangle \right\} \\
&=&\inf_{L}\left\{ \left\langle L,E\right\rangle |L\geq R_{i},i=0,\ldots
,m\right\} ,
\end{eqnarray*}%
according to which $\sum_{i=0}^{m}\left\langle R_{i},D_{i}\right\rangle
=\varkappa ^{\mathrm{o}}(R)=\left\langle R\right\rangle =\left\langle L^{%
\mathrm{o}},E\right\rangle $ and 
\begin{equation*}
\sum_{i=0}^{m}\mathrm{Tr}\;\left[ (L^{\mathrm{o}}-R_{i})D_{i}^{\mathrm{o}}%
\right] =\mathrm{Tr}\;(L^{\mathrm{o}}E)-\sum_{i=0}^{m}\left\langle
R_{i},D_{i}^{\mathrm{o}}\right\rangle .
\end{equation*}%
The necessary and sufficient condition for this sum of traces of products of
positive operators to vanish is, obviously, Equation (\ref{2 3 8}).

Employment of the duality principle in the conditional problem (\ref{2 3 2})
reduces this problem by the elementary Lagrange method to problem (\ref{2 3
4}), for which the necessity and sufficiency of conditions (\ref{2 3 11})
can be verified directly. The proof of the theorem is complete.
\end{proof}

Note that the above proof remains unchanged in the case of an infinite
number of hypotheses, $m=\infty $. From this theory follows, for one thing,
Theorem \ref{T 2 2 1} if we put $R_{0}=0,\;R_{i}=C_{i},\;i=1,\ldots ,m$, and 
$L=B$. The existence of a solution to problem (\ref{2 3 7}) and the
uniqueness on the subspace $\mathcal{E}=E\mathcal{H}$ of the solution to
problem (\ref{2 3 9}) can be obtained from the proof in Section \ref%
{Optimality Cond} of these assertions for problems (\ref{2 2 7}) and (\ref{2
2 9}) to which (\ref{2 3 7}) and (\ref{2 3 9}) are reduced by the
substitutions $C_{i}=R_{i}-R_{0}$ and $B=L-R_{0}$.

In the case of positive $R_{i}$'s the problem of testing the hypotheses $%
H_{i},i=0,\ldots ,m$, can be solved as a problem of separating $m-1$ signals 
$H_{i}=R_{i}^{1/2}$ by applying Theorems \ref{T 2 2 2} and \ref{T 2 2 3}.
For nonpositive $R_{i}$'s it has also proved expedient to go over to the
signal space $\mathcal{K}^{m+1}=\bigoplus_{i=0}^{m}\mathcal{K}_{i}$, $%
\mathcal{K}_{i}=H_{i}\mathcal{H}$, $i=0,\ldots ,m$, via a partially
isometric operator $V:\mathcal{H}\rightarrow \mathcal{K}^{m+1}$ of polar
decomposition $H=\sigma ^{1/2}V,\sigma =HH^{\ast }$ for the operator $%
H:\varphi \in \mathcal{H}\mapsto \lbrack H_{i}\varphi ]_{i=0}^{m}$. As a
result, the optimality conditions for the decision operators $D_{i}^{\mathrm{%
o}}$ can be written in the form of conditions imposed on the decomposition $%
\varepsilon =\sum_{i=0}^{m}\delta _{i}^{\mathrm{o}}$, where $\delta _{i}^{%
\mathrm{o}}=VD_{i}^{\mathrm{o}}V^{\ast }$, of the support $\varepsilon ^{%
\mathrm{o}}=VV^{\ast }$ of the signal correlation matrix $\sigma
_{ik}=H_{i}H_{k}^{\ast }$ with$\;i,k=0,\ldots ,m$: 
\begin{equation*}
\left( \lambda _{i}^{\mathrm{o}}-\rho _{i}\right) \delta _{i}^{\mathrm{o}%
}=0,\;\lambda _{i}^{\mathrm{o}}\geq \rho _{i}=\sum_{k=0}^{m}h_{k}^{\ast
}c_{i}^{k}h_{k},\;i=0,\ldots ,m.
\end{equation*}%
Here $\lambda _{i}^{\mathrm{o}}=VL^{\mathrm{o}}V^{\ast },\;h_{i}=1_{i}h,\;h=%
\sqrt{\sigma }$ and $c_{i}^{k}$ is the quality matrix, which defines the
operators $R_{i}=\sum_{k=0}^{m}H_{k}^{\ast }c_{i}^{k}H_{k}$ and which, for a
fixed $m$, it has proved expedient to consider as being a diagonal operator $%
c_{i}=\bigoplus_{k=0}^{m}c_{i}^{k}1_{k}$ in space $K^{m+1}$ because then the
signal matrices $\rho _{i}=VR_{i}V^{\ast }$ can be represented in the form $%
\rho _{i}=hc_{i}h$.

Even if the amplitudes $H_{i}$ are ordinary, that is, $H_{i}=(\psi _{i}|$
and hence the correlation matrix is a number matrix $\sigma _{ik}=(\psi
_{i}|\psi _{k})$, it is difficult to write conditions of optimality
explicitly for $m>1$ for a nonsingular matrix $\sigma $. Below we will study
this problem for the case where the rank of matrix is equal to $2$ and,
hence, all the operators $R_{i},L$, and $D_{i}$ can be represented by $2$-by-%
$2$ matrices in space $\mathcal{E}^{\mathrm{o}}=\mathbb{C}^{2}$.

\subsection{2-d Wave Pattern Recognition}

To the operators $\left\{ R_{i}\right\} $ in the optimization problem (\ref%
{2 3 8}) we assign Hermitian matrices that can be considered nonnegative
without loss of generality. Any $2$-by-$2$ matrix can be decomposed in Pauli
matrices, which are 
\begin{eqnarray*}
1 &=&\left[ 
\begin{array}{ll}
1 & 0 \\ 
0 & 1%
\end{array}%
\right] ,\;\sigma _{x}=\left[ 
\begin{array}{ll}
0 & 1 \\ 
1 & 0%
\end{array}%
\right] ,\;\sigma _{y}=\left[ 
\begin{array}{ll}
0 & -\mathrm{j} \\ 
\mathrm{j} & 0%
\end{array}%
\right] ,\;\sigma _{z}=\left[ 
\begin{array}{ll}
1 & 0 \\ 
0 & -1%
\end{array}%
\right] , \\
R &=&\left[ 
\begin{array}{ll}
\nu +z & x-\mathrm{j}y \\ 
x+\mathrm{j}y & \nu -z%
\end{array}%
\right] =\nu +x\sigma _{x}+y\sigma _{y}+z\sigma _{z}=\nu +\mathbf{r\cdot
\sigma },\;\mathrm{j}=\sqrt{-1},
\end{eqnarray*}%
where $x,y,z$ and $\nu $ are real if matrix $R$ is Hermitian, and $\mathbf{%
\hat{r}}=x\sigma _{x}+y\sigma _{y}+z\sigma _{z}\equiv \mathbf{r\cdot \sigma }
$ is a vector operator represented by vector $\mathbf{r}=(x,y,z)$ of
three-dimensional real space $\mathbb{R}^{3}$. The product of $\mathbf{\hat{r%
}}$ and $\mathbf{\hat{s}}$, with $\mathbf{r}\in \mathbb{R}^{3}$ and $\mathbf{%
s}\in \mathbb{R}^{3}$, is equal to $\mathbf{\hat{r}\hat{s}}=\mathbf{r}\cdot 
\mathbf{s+}\mathrm{j}(\mathbf{r}\times \mathbf{s}),$ where $\mathbf{r\cdot s}
$ and $\mathbf{r}\times \mathbf{s}$ are the scalar and vector products of $%
\mathbf{r}$ and $\mathbf{s}$. Note that $\mathrm{Tr}\;R=2\nu $ and $\mathrm{%
Det}\;R=\nu ^{2}-|\mathbf{r}|^{2}$ (with $|\mathbf{r|}=\sqrt{\mathbf{r\cdot r%
}}$), and that the nonnegativity condition $R\geq 0$ assumes the form $\nu =|%
\mathbf{r}|\neq 0$, and rank $R=0$ at $\nu =0$.

The operators $R_{i}=(\nu _{i}+\mathbf{\hat{r}}_{i})/2,\;i=0,\ldots ,m$,
where $|\nu _{i}|\geq |\mathbf{r}_{i}|$ and $\sum_{i=0}^{m}\nu _{i}=1$, can
be interpreted as density operators related to the tested wave hypotheses
with prior intensities $\nu _{i}=\mathrm{Tr}\;R_{i}$ and represented by
vectors $\mathbf{r}_{i}\in \mathbb{R}^{3}$, which are known as polarization
vectors. A similar problem arises when we must identify the photon
polarization or the electron spin \cite{4}. Let us assume that polarizations 
$\left\{ \mathbf{r}_{i}\right\} $ satisfy the inequalities 
\begin{equation}
|\mathbf{r}_{k}-\mathbf{r}_{i}|>|\nu _{k}-\nu _{i}|  \label{2 3 12}
\end{equation}%
for all $k\neq i$; in the opposite case, that is, $|\nu _{k}-\nu _{i}|\geq |%
\mathbf{r}_{k}-\mathbf{r}_{i}|$, the $k$-th hypothesis dominates the $i$-th
hypothesis or vice versa: $R_{k}>R_{i}$ or $R_{k}=R_{i}$ or $R_{k}<R_{i}$,
and one of the hypotheses (with the smaller $\nu $) can be ignored.

The decision operators $D_{i}$ in the Pauli representation $D_{i}=\delta
_{i}+\mathbf{\hat{d}}_{i}$ are described by nonnegative numbers $\delta
_{i}\geq 0$ and vectors $\mathbf{d}_{i}\in \mathbb{R}^{3}$ ($|\mathbf{d}%
_{i}|\leq \delta _{i}$), with the decomposition of unity $%
\sum_{i=0}^{m}D_{i}=1$ assuming the form 
\begin{equation*}
\sum_{i=0}^{m}\delta _{i}=1,\;\sum_{i=0}^{m}\mathbf{d}_{i}=0.
\end{equation*}%
The solution of the problem of optimal recognition of polarizations $\mathbf{%
r}_{i}$ with intensities $\nu _{i}$ can be reduced to finding a real number $%
\lambda _{i}^{\mathrm{o}}$ and a vector $\mathbf{l}^{\mathrm{o}}\in \mathbb{R%
}^{3}$ that defines the operator $L^{\mathrm{o}}=(\lambda _{i}^{\mathrm{o}}+%
\mathbf{\hat{l}}^{\mathrm{o}})/2$ that satisfies conditions (\ref{2 3 8})
for a collection of $D_{i}^{\mathrm{o}}=\delta _{i}^{\mathrm{o}}+\mathbf{%
\hat{l}}_{i}^{\mathrm{o}},i=0,\ldots ,m$.

\begin{theorem}
\label{T 2 3 2}The solution to the problem of optimal recognition of
polarizations $\left\{ \mathbf{r}_{i}\right\} $ satisfying together with $%
\left\{ \nu _{i}\right\} $ condition \textup{(\ref{2 3 12}) }can be found if
and only if there is a collection of numbers $\mu _{i}^{\mathrm{o}}\geq
0,\;i=0,\ldots ,m$ such that 
\begin{equation}
\left\vert \sum_{k=0}^{m}(\mathbf{r}_{i}-\mathbf{r}_{k})\mu _{k}^{\mathrm{o}%
}\right\vert +\sum_{k=0}^{m}(\nu _{i}-\nu _{k})\mu _{k}^{\mathrm{o}}\geq
1,\;i=0,\ldots ,m,  \label{2 3 13}
\end{equation}%
where the equality takes place at least for those $i$'s for which $\mu _{i}^{%
\mathrm{o}}\neq 0$. The optimal decision operators have the form $\delta
_{i}^{\mathrm{o}}=|\mathbf{d}_{i}^{\mathrm{o}}|$, $\mathbf{d}_{i}^{\mathrm{o}%
}=\mu _{i}^{\mathrm{o}}(\mathbf{r}_{i}-\mathbf{l}_{i}^{\mathrm{o}})$, where 
\begin{equation}
\mathbf{l}^{\mathrm{o}}=\sum_{i=0}^{m}\mu _{i}^{\mathrm{o}}\mathbf{r}_{i}%
\big/\sum_{i=0}^{m}\mu _{i}^{\mathrm{o}},  \label{2 3 14}
\end{equation}%
and the maximal received intensity is 
\begin{equation}
\varkappa ^{\mathrm{o}}=\left( 1+\sum_{i=0}^{m}\mu _{i}^{\mathrm{o}}\nu
_{i}\right) \Big/\sum_{i=0}^{m}\mu _{i}^{\mathrm{o}}=\lambda _{i}^{\mathrm{o}%
}.  \label{2 3 15}
\end{equation}
\end{theorem}

\begin{proof}
In terms of $\lambda _{i}^{\mathrm{o}},\mathbf{l}^{\mathrm{o}},\nu _{i}$,
and $\mathbf{r}_{i}$, the first optimality condition (\ref{2 3 8}), $L^{%
\mathrm{o}}-R_{i}\geq 0$, has the form 
\begin{equation}
\lambda _{i}^{\mathrm{o}}\geq |\mathbf{r}_{i}-\mathbf{l}^{\mathrm{o}}|+\nu .
\label{2 3 16}
\end{equation}%
The equation $(L^{\mathrm{o}}-R_{i})D_{i}^{\mathrm{o}}=0$ can be written in
another form if we nullify the scalar and real vector parts of the product $%
(\lambda _{i}^{\mathrm{o}}+\nu _{i}-\mathbf{\hat{l}}^{\mathrm{o}}-\mathbf{%
\hat{r}}_{i})(\delta _{i}^{\mathrm{o}}+\mathbf{\hat{d}}_{i})$: 
\begin{equation}
(\lambda _{i}^{\mathrm{o}}-\nu _{i})d_{i}^{\mathrm{o}}+(\mathbf{l}^{\mathrm{o%
}}-\mathbf{r}_{i})\delta _{i}^{\mathrm{o}}=0,\;(\lambda _{i}^{\mathrm{o}%
}-\nu _{i})\delta _{i}^{\mathrm{o}}+(\mathbf{l}^{\mathrm{o}}-\mathbf{r}%
_{i})\cdot \mathbf{d}_{i}^{\mathrm{o}}=0.  \label{2 3 17}
\end{equation}%
The imaginary vector equation $\mathrm{j}(\mathbf{l}^{\mathrm{o}}-\mathbf{r}%
_{i})\times \mathbf{d}_{i}^{\mathrm{o}}=0,\;\mathrm{j}=\sqrt{-1}$, follows
from the real vector equation in (\ref{2 3 17}) are equivalent to 
\begin{equation}
\mathbf{d}_{i}^{\mathrm{o}}=\delta _{i}^{\mathrm{o}}(\mathbf{r}_{i}-\mathbf{l%
}^{\mathrm{o}})/(\lambda _{i}^{\mathrm{o}}-\nu _{i}),\;((\lambda _{i}^{%
\mathrm{o}}-\nu _{i})^{2}-|\mathbf{r}_{i}-\mathbf{l}^{\mathrm{o}%
}|)^{2}\delta _{i}^{\mathrm{o}}=0,  \label{2 3 18}
\end{equation}%
in the opposite case ($\lambda _{i}^{\mathrm{o}}=\nu _{i}$ for a certain $i$
in(\ref{2 3 16})) for $i=k$ we obtain $\mathbf{l}^{\mathrm{o}}=\mathbf{r}%
_{i} $, with the result that inequalities (\ref{2 3 18}) and (\ref{2 3 12})
become incompatible. The optimal decision vector can be written in the form (%
\ref{2 3 14}), where $\mu _{i}^{\mathrm{o}}=\delta _{i}^{\mathrm{o}%
}/(\lambda _{i}^{\mathrm{o}}-\nu _{i})$ is nonnegative in accordance with
the inequalities $\delta _{i}^{\mathrm{o}}\geq 0,\;\lambda _{i}^{\mathrm{o}%
}>\nu _{i}$, and (\ref{2 3 16}), while $\mathbf{l}^{\mathrm{o}}$ is
determined by the set $\left\{ \mu _{i}^{\mathrm{o}}\right\} $ in accordance
with the fact that $\sum_{i=0}^{m}\mathbf{d}_{i}^{\mathrm{o}}=0$. The second
equation in (\ref{2 3 18}) implies that inequalities (\ref{2 3 16}) become
amplitudes for the values of $i$ for which $\delta _{i}^{\mathrm{o}}=\mu
_{i}(\lambda -\nu _{i})\neq 0$. Multiplying (\ref{2 3 16}) by $%
\sum_{i=0}^{m}\mu _{i}^{\mathrm{o}}$ and finding $\lambda $ in the form 
\begin{equation*}
\lambda _{i}^{\mathrm{o}}=\left( 1+\sum_{i=0}^{m}\mu _{i}^{\mathrm{o}}\nu
_{i}\right) \Big/\sum_{i=0}^{m}\mu _{i}^{\mathrm{o}}
\end{equation*}%
from the condition that $\sum_{i=0}^{m}\delta _{i}^{\mathrm{o}}=1$ for 
\begin{equation*}
\delta _{i}^{\mathrm{o}}=\mu _{i}^{\mathrm{o}}(\lambda -\nu _{i})=\mu _{i}^{%
\mathrm{o}}|\mathbf{r}_{i}-\mathbf{l}^{\mathrm{o}}|=|\mathbf{d}_{i}^{\mathrm{%
o}}|
\end{equation*}%
we get condition (\ref{2 3 13}) for determining $\left\{ \mu _{i}^{\mathrm{o}%
}\right\} $. Since $\mathrm{Tr}\;L^{\mathrm{o}}=\lambda _{i}^{\mathrm{o}}$,
the maximal intensity of (\ref{2 3 9}) is equal to (\ref{2 3 15}). The proof
of the theorem is complete.
\end{proof}

Note that the equalities in (\ref{2 3 13}) must be true for at least two
indices $i$ and $k$, since there is no such set $\left\{ \mu _{i}\right\}
,\;\mu _{i}\neq 0$, for only one subscript $i$ that satisfies the $i$-th
inequality. For every pair $\mathbf{r}_{i}$ and $\mathbf{r}_{k}$ for which (%
\ref{2 3 12}) is valid there is a unique solution of the $i$-th and $k$-th
equalities in (\ref{2 3 13}) with $\mu _{j}=0$ for all $j\neq i,\;k$ with$%
\;\mu _{i}>0,\;\mu _{k}>0$: 
\begin{equation*}
\mu _{i}=(|\mathbf{r}_{k}-\mathbf{r}_{i}|+\nu _{k}-\nu _{i})^{-1},\;\mu
_{k}=(|\mathbf{r}_{i}-\mathbf{r}_{k}|+\nu _{i}-\nu _{k})^{-1},
\end{equation*}%
but such a set $\left\{ \mu _{i}\right\} $ may not satisfy the other
inequalities in (\ref{2 3 13}) for $j\neq i,\;k$. If there exists a pair $%
\mathbf{r}_{i},\mathbf{r}_{k}$ for which all the inequalities in (\ref{2 3
13}) are valid at $\mu _{j}\neq 0$ only when $j=i,k$, then the optimal
decision vectors $\mathbf{d}_{j}^{\mathrm{o}}$ are zero for $j\neq i,k$ (see
(\ref{2 3 14})) and 
\begin{equation*}
\mathbf{d}_{i}^{\mathrm{o}}=(\mathbf{r}_{i}-\mathbf{r}_{k})/2|\mathbf{r}_{i}-%
\mathbf{r}_{k}|,\;\mathbf{d}_{k}^{\mathrm{o}}=(\mathbf{r}_{k}-\mathbf{r}%
_{i})/2|\mathbf{r}_{k}-\mathbf{r}_{i}|.
\end{equation*}%
Here the optimal decision operators $D_{i}^{\mathrm{o}}=|\mathbf{d}_{i}^{%
\mathrm{o}}|+\mathbf{\hat{d}}_{i}^{\mathrm{o}}$ are orthogonal and
correspond to an error intensity 
\begin{equation*}
\varkappa ^{\mathrm{o}}=\frac{1}{2}(\nu _{i}+\nu _{k})+\frac{1}{2}|\mathbf{r}%
_{i}-\mathbf{r}_{k}|.
\end{equation*}

In the case where the optimal operators $D_{i}^{\mathrm{o}}$ are nonzero for
more than two $i$'s, they define a nonorthogonal decomposition of unity in
the two-dimensional space $\mathcal{E}=\mathbb{C}^{2}$. We will not try to
find a general analytical solution to the system of equation (\ref{2 3 13})
with $\mu _{i}^{\mathrm{o}}\neq 0$ for more than two $i$'s; rather, we will
give a geometric interpretation of such a solution.

\subsection{Geometric Representation of 2-d Patterns}

Let us represent the Hermitian operators (\ref{2 3 10}) by points $r=(\nu
,x,y,z)=(\nu ,\mathbf{r})$ in the four-dimensional Minkowski space $\mathbb{R%
}^{1+3}$. To every nonnegative operator there corresponds a point inside the
light cone $\nu =|\mathbf{r}|$. In these terms \textit{a priori} neither the 
$k$-th nor the $i$-th hypothesis is dominant at $R_{i}=(\nu _{i}+\mathbf{%
\hat{r}}_{i})/2$ and $R_{k}=(\nu _{k}+\mathbf{\hat{r}}_{k})/2$ if and only
if the interval $r_{i}-r_{k}=(\nu _{i}-\nu _{k},\mathbf{r}_{i}-\mathbf{r}%
_{k})$ is spacelike. In accordance with (\ref{2 3 16}), point $l^{\mathrm{o}%
}=(\lambda _{i}^{\mathrm{o}},\mathbf{l}^{\mathrm{o}})$, which represents the
operator $L^{\mathrm{o}}=(\lambda ^{\mathrm{o}}+\mathbf{\hat{l}}^{\mathrm{o}%
})$, is the apex of the four-dimensional cone 
\begin{equation}
\mathcal{C}(l)=\left\{ r=(\nu ,\mathbf{r}):\nu -\lambda _{i}^{\mathrm{o}}+|%
\mathbf{r-l}^{\mathrm{o}}|=0\right\}  \label{2 3 19}
\end{equation}%
covering all the points $r_{i}=(\nu _{i},\mathbf{r}_{i})$ and containing the
subset $\left\{ r_{j_{\alpha }}\right\} \subset \left\{ r_{i}\right\} $ of
the boundary points $r_{j_{\alpha }}$ satisfying (\ref{2 3 16}). On the
other hand, the optimal points $l$ are only those whose projections $\mathbf{%
l}$ belong to the convex hull of the boundary subset of the spatial
projections $\mathbf{r}_{j_{\alpha }},\alpha =0,\ldots ,s,\;s\leq m$: 
\begin{equation}
\sum_{\alpha =0}^{s}\mathbf{r}_{j_{\alpha }}\pi _{j_{\alpha }}=\mathbf{l,\;}%
\sum_{\alpha =0}^{s}\pi _{j_{\alpha }}=1,  \label{2 3 20}
\end{equation}%
where, in accordance with (\ref{2 3 14}), $\pi _{j_{\alpha }}=\mu
_{j_{\alpha }}/\sum_{\alpha =0}^{s}\mu _{j_{\alpha }}\geq 0$ ($\mu _{i}=0$
if $r_{i}$ is covered by cone (\ref{2 3 19}): $\nu _{i}+|\mathbf{r}_{i}-%
\mathbf{l}|\leq \lambda $). We will say that the subset $\left\{
r_{j_{\alpha }}\right\} $ has an apex if the points $r_{j_{\alpha }}$ lie on
the cone: $r_{j_{\alpha }}\subset \mathcal{C}(l)$ with an apex $l$ whose
spatial projection $\mathbf{l}$ is a point on the convex hull $\left\{ 
\mathbf{r}_{j_{\alpha }}\right\} $. In these terms Theorem \ref{T 2 3 2} can
be formulated as follows:

\begin{theorem}
\label{T 2 3 3}To solve the problem of optimal recognition of points $%
r_{i}=(\nu _{i},\mathbf{r}_{i}),\;i=0,\ldots ,m$, separated by spacelike
intervals \textup{(\ref{2 3 12})}, it is necessary and sufficient to find a
subset $\left\{ \mathbf{r}_{j_{\alpha }}\right\} $ with an apex $l^{\mathrm{o%
}}$ belonging to a cone that covers all other points of set $\left\{
r_{i}\right\} $, that is, to specify a subset of vectors $\mathbf{r}%
_{j_{\alpha }}\subset \left\{ \mathbf{r}_{i}\right\} ,\;\alpha =0,\ldots ,s$
whose convex hull contains vector $\mathbf{l}^{\mathrm{o}}$ with respect to
which the sum $|\mathbf{r}_{j_{\alpha }}-\mathbf{l}^{\mathrm{o}}|+\nu
_{j_{\alpha }}$ is the constant $\lambda _{i}^{\mathrm{o}}$: 
\begin{equation}
|\mathbf{r}_{j_{\alpha }}-\mathbf{l}^{\mathrm{o}}|+\nu _{j_{\alpha
}}=\lambda _{i}^{\mathrm{o}},\;\alpha =0,\ldots ,s,  \label{2 3 21}
\end{equation}%
while $|\mathbf{r}_{i}-\mathbf{l}^{\mathrm{o}}|+\nu _{i}\leq \lambda _{i}^{%
\mathrm{o}}$ for all other indices $i\in \left\{ j_{\alpha }\right\} $. The
optimal decision operators are represented by points on the cone%
\begin{equation*}
d_{i}^{\mathrm{o}}=(\delta _{i}^{\mathrm{o}},\mathbf{d}_{i}^{\mathrm{o}%
}),\;\;\delta _{i}^{\mathrm{o}}=|\mathbf{d}_{i}^{\mathrm{o}}|,
\end{equation*}%
with spatial vectors 
\begin{equation}
\mathbf{d}_{i}^{\mathrm{o}}=\pi _{i}^{\mathrm{o}}(\mathbf{r}_{i}-\mathbf{l}^{%
\mathrm{o}})\Big/\sum_{i=0}^{m}\pi _{i}\mathbf{r}_{i},  \label{2 3 22}
\end{equation}%
where $\pi _{i}^{\mathrm{o}}=0$ for $i\notin \left\{ j_{\alpha }\right\} $,
and $\left\{ \pi _{j_{\alpha }}^{\mathrm{o}},\alpha =0,\ldots ,s\right\} $
is any nonnegative solution to the system of equations \textup{(\ref{2 3 20})%
}. The minimal intensity in this case is 
\begin{equation}
\varkappa ^{\mathrm{o}}=\sum_{i=0}^{m}(\nu _{i}+|\mathbf{r}_{i}-\mathbf{l}^{%
\mathrm{o}}|)\pi _{i}.  \label{2 3 23}
\end{equation}
\end{theorem}

Note that every pair of point $r_{i},r_{k}$ separated by a spacelike
interval defines, via two equations from (\ref{2 3 21}), $j_{\alpha }=i,k$,
a set of point $\mathbf{l}\in \mathbb{R}^{1+3}$ whose difference of
distances to the points $\mathbf{r}_{i}$ and $\mathbf{r}_{k}$ is constant: 
\begin{equation}
|\mathbf{l}^{\mathrm{o}}-\mathbf{r}_{k}|-|\mathbf{l}^{\mathrm{o}}-\mathbf{r}%
_{i}|=\nu _{i}-\nu _{k}.  \label{2 3 24}
\end{equation}%
These points lie on one of the two sheets of the hyperboloid of revolution
with foci at $\mathbf{r}_{i}$ and $\mathbf{r}_{k}$ and eccentricity 
\begin{equation*}
\varepsilon =\frac{|\mathbf{r}_{i}-\mathbf{r}_{k}|\,}{\,|\nu _{i}-\nu _{k}|}%
>1.
\end{equation*}%
Here, if $\nu _{i}=\nu _{k}$, the hyperboloid (\ref{2 3 24}) becomes a plane
normal to the segment $\mathbf{r}_{i}\pi _{i}+\mathbf{r}_{k}\pi _{k}$ ($\pi
>0,\;\pi _{i}+\pi _{k}=1$) at point $(\mathbf{r}_{i}+\mathbf{r}_{k})2$,
while if $\nu _{i}\neq \nu _{k}$, we select the sheet in whose plane lies
the focus with the higher intensity, $\nu _{i}$ or $\nu _{k}$. Obviously, if
the subset $\left\{ r_{j_{\alpha }}\right\} $ has an apex $l$, the spatial
projection $\mathbf{l}$ is the common point of all the hyperboloids (\ref{2
3 24}) corresponding to all the pairs of the set $\left\{ r_{j_{\alpha
}}\right\} $ that belong to the convex hull $\left\{ \mathbf{r}_{j_{\alpha
}}\right\} $. We will call this point $\mathbf{l}$ the center of $\left\{ 
\mathbf{r}_{j_{\alpha }}\right\} $ representing $L^{\mathrm{o}}$ is unique,
which means that the center of $\left\{ \mathbf{r}_{j_{\alpha }}\right\} $
is unique, too. It can easily be shown that for every vector $\mathbf{l}$ of
the convex hull $\left\{ \mathbf{r}_{j_{\alpha }}\right\} $ the system of
linear equations (\ref{2 3 20}) has a unique solution $\left\{ \pi
_{j_{\alpha }}^{\mathrm{o}}\right\} $ if and only if vectors $\mathbf{r}%
_{j_{\alpha }}-\mathbf{r}_{j_{0}},\;\alpha =1,\ldots ,s$, are linearly
independent.

\subsection{Optimal and Simplex Solutions}

The reader will recall that a convex hull of a set $\left\{ \mathbf{r}%
_{j_{\alpha }}\right\} $ of points $\mathbf{r}_{j_{\alpha }},\;\alpha
=0,1,2,3$, is called an $s$-simplex (a segment if $s=1$, a triangle if $s=2$%
, a tetrahedron if $s=3$, and so on) if the vectors $\mathbf{r}_{j_{0}},%
\mathbf{r}_{j_{\alpha }},\;a=1,\ldots ,s$, are linearly independent. It is
well-known that each $s$-dimensional face (an $s$-face) of an $n$-simplex $%
(n\geq s)$ is a simplex, too. We will call a subset that generates a simplex
convex hull a simplex subset.

\begin{theorem}
\label{T 2 3 4}The problem of optimal recognition of polarizations $\left\{ 
\mathbf{r}_{i},i=0,\ldots ,m\right\} $ always has a solution that can be
described by the simplex set $\left\{ \mathbf{d}_{i_{\alpha }}^{\mathrm{o}%
},\alpha =0,\ldots ,s\right\} ,s\leq m$, of the nonzero vectors \textup{(\ref%
{2 3 22})} corresponding to the simplex subset $\left\{ \mathbf{r}%
_{j_{\alpha }}\right\} \subseteq \left\{ \mathbf{r}_{i}\right\} $ with a
center at $\mathbf{l}^{\mathrm{o}}$ and a maximal sum 
\begin{equation*}
\nu _{j_{\alpha }}+|\mathbf{r}_{j_{\alpha }}-\mathbf{l}^{\mathrm{o}%
}|=\max_{i=0,\ldots ,m}\left\{ \nu _{i}+|\mathbf{r}_{i}-\mathbf{l}^{\mathrm{o%
}}|\right\} .
\end{equation*}%
This solution is unique if and only if the $s$-simplex generated by subset $%
\left\{ \mathbf{r}_{j_{\alpha }}\right\} $ is an $s$-face of the convex hull
of all vectors $\mathbf{r}_{j_{0}},\ldots ,\mathbf{r}_{j_{m}}$ with a common
center $\mathbf{l}^{\mathrm{o}}$.
\end{theorem}

\begin{proof}
By Theorem \ref{T 2 3 3}, the solution to the problem considered here is
reduced to finding the cone (\ref{2 3 19}) that covers all points $\left\{ 
\mathbf{r}_{i}\right\} $ and has an apex $\mathbf{l}^{\mathrm{o}}$ with a
projection lying inside the convex hull of projections $\left\{ \mathbf{r}%
_{j_{\alpha }}\right\} $ of the tangency points $r_{j_{\alpha }}$.
Obviously, there is always such a cone. Let $n\leq m$ be the number of
tangency points $r_{j_{\alpha }},\alpha =0,\ldots ,n$. If the subset $%
\left\{ \mathbf{r}_{j_{\alpha }},\alpha =0,\ldots ,s\right\} $ ($s\leq m$)
is a simplex set, the validity of Theorem \ref{2 3 4} is obvious. If this
subset is not a simplex, then the convex hull $\left\{ \mathbf{r}_{j_{\alpha
}}\right\} $ can be partitioned into several simplexes with a common vertex $%
\mathbf{r}_{j_{0}}$ via diagonal planes $(\mathbf{r}_{j_{0}},\mathbf{r}%
_{j_{\alpha }},\mathbf{r}_{j_{\beta }})$ or diagonal lines $(\mathbf{r}%
_{j_{0}},\mathbf{r}_{j_{\alpha }})$ when all the vectors $\mathbf{r}%
_{j_{\alpha }}$ are coplanar. Hence, the center $\mathbf{l}^{\mathrm{o}}$ is
an interior point of one of the $s$-simplexes ($s<n\leq m$) with apexes $%
\mathbf{r}_{j_{\alpha }},\alpha =0,\ldots ,s$, which are the projections of
the tangency points and determine the unique positive solution $\left\{ \pi
_{j_{\alpha }}^{\mathrm{o}}\right\} $ of system (\ref{2 3 20}). The set $%
\left\{ \mathbf{d}_{j_{\alpha }}^{\mathrm{o}}\right\} $ of nonzero vectors (%
\ref{2 3 22}) is a simplex set if and only if the set $\left\{ \mathbf{r}%
_{j_{\alpha }}\right\} $ is a simplex and determines the optimal solution
with maximal quality (\ref{2 3 21}) and minimal error intensity (\ref{2 3 23}%
). When center $\mathbf{l}^{\mathrm{o}}$ is an interior point of a
nonsimplex convex hull of the projections of tangency points, the optimal
simplex solution is not unique (the partition into simplexes is note unique)
and there are also optimal nonsimplex solutions. But if point $\mathbf{l}^{%
\mathrm{o}}$ is a boundary point of the convex hull, that is, an interior
point of an $s$-face, the optimal solution is unique if the face is an $s$%
-simplex.
\end{proof}

\begin{corollary}
To solve the problem of optimal testing of several hypotheses in the
two-dimensional space $\mathcal{E}=\mathbb{C}^{2}$, it is sufficient to
limit oneself to $s+1\leq 4$ solutions $j_{0},\ldots ,j_{s}$ corresponding
to a simplex subset of hypotheses $\mathbf{r}_{j_{0}},\ldots ,\mathbf{r}%
_{j_{s}}$. Each such solution procedure can be realized in an indirect
measurement described by an orthogonal decomposition in the observation
space $\mathcal{H}=\mathbb{C}^{2}\otimes \mathbb{C}^{2}$.
\end{corollary}

Indeed, in the three-dimensional space $\mathbb{R}^{3}$ there is not a
single simplex subset $\mathbf{r}_{j_{0}},\ldots ,\mathbf{r}_{j_{s}}$ for $%
s>3$ and, therefore, for every $m$ there always exists an optimal
decomposition in the two-dimensional space $\mathcal{E}$ consisting of $%
s+1\leq 4$ nonzero decision operators $D_{j_{\alpha }}^{\mathrm{o}}=\delta
_{j_{\alpha }}^{\mathrm{o}}+\mathbf{\hat{d}}_{j_{\alpha }}^{\mathrm{o}}$ of
rank one. It is well known that each nonorthogonal decomposition of unity in
operators $D_{j_{0}},\ldots ,D_{j_{s}}$ of rank one can be extended to an
orthogonal decomposition in an $(s+1)$-dimensional space $\mathcal{E}\subset 
\mathcal{H}$. Hence, we can limit ourselves to the four-dimensional
measurement space $\mathcal{H}$, which can always be represented as the
tensor product of two-dimensional spaces $\mathcal{E}:\mathcal{H}=\mathcal{E}%
\otimes \mathcal{E}$, corresponding to the composition of two identical
systems.

Note that the optimal solution may be degenerate (in the sense that a
hypothesis $\mathbf{r}_{i}$ may correspond to $D_{i}=0$) even if the set $%
\mathbf{r}_{0},\ldots ,\mathbf{r}_{m}$ is a simplex set $(m\leq 3)$, for
example, at $m=2$, $\nu _{0}=\nu _{1}=\nu _{2}$, if the vectors $\mathbf{r}%
_{0},\mathbf{r}_{1}$ and $\mathbf{r}_{2}$ form an obtuse triangle.

In conclusion of this section we will consider two particular cases.

(1) \textit{Optimal recognition of pure polarization}. Here the
polarizations are normalized to prior intensities: $|\mathbf{r}_{i}|=\nu
_{i} $, with the representative points $r_{i}=(\nu _{i},r_{i})$ belonging to
the cone $\nu =|\mathbf{r}|$. Expression (\ref{2 3 21}), which determines
the subset of points $\mathbf{r}_{j_{\alpha }}$ of tangency of this cone and
the covering cone (\ref{2 3 19}), has the form 
\begin{equation}
|\mathbf{r}_{j_{\alpha }}-\mathbf{l}^{\mathrm{o}}|+|\mathbf{r}_{j_{\alpha
}}|=\lambda _{i}^{\mathrm{o}}.  \label{2 3 25}
\end{equation}%
In relation to $\mathbf{r}_{j_{\alpha }}$, this is the equation of an
ellipsoid of revolution with foci at $0$ and $\mathbf{l}^{\mathrm{o}}$ and
eccentricity $\varepsilon =|\mathbf{l}^{\mathrm{o}}|/\lambda _{i}^{\mathrm{o}%
}<1$. In accordance with (\ref{2 3 16}), all the other point $\mathbf{r}%
_{i}\notin \left\{ \mathbf{r}_{j_{\alpha }}\right\} $ must lie inside the
ellipsoid. Hence, the problem of optimal recognition of pure polarization is
reduced to finding the ellipsoid described about points $\left\{ \mathbf{r}%
_{i}\right\} $ with foci at $0$ and $\mathbf{l}^{\mathrm{o}}$, where $%
\mathbf{l}^{\mathrm{o}}$ is an interior point of the convex hull of the
points of tangency $\left\{ \mathbf{r}_{j_{\alpha }}\right\} $. The quality $%
\varkappa ^{\mathrm{o}}$ of the optimal solution is equal to the length of
the major axis of the ellipsoid, $\lambda _{i}^{\mathrm{o}}$.

(2) \textit{Optimal recognition of equi-intensity polarizations}. The priori
intensities $\nu _{i}=\nu _{0},\;i=1,\ldots ,m$, and the corresponding
points are points of the hyperplane $\nu =\nu _{0}$. The density operators $%
R_{i}=(\nu +\mathbf{r}_{i})/2$, all having the same trace $\nu _{0}$, are
represented by normalized vectors, $|\mathbf{r}_{i}|\leq \nu _{0}$. The
intersection of the covering cone (\ref{2 3 19}) and the hyperplane $\nu
=\nu _{0}$ is a sphere $|\mathbf{r-l}^{\mathrm{o}}|=\rho $ of radius $\rho
=\lambda _{i}^{\mathrm{o}}-\nu _{0}$. Hence, the problem of optimal
recognition of equiprobable polarizations is reduced to finding a sphere
described about all points $\mathbf{r}_{i}:|\mathbf{r}-\mathbf{l}^{\mathrm{o}%
}|\leq \rho $ with radius $\rho $ and centered at $\mathbf{l}^{\mathrm{o}}$,
the center belonging to the convex hull of the tangency points $\mathbf{r}%
_{j_{\alpha }}:|\mathbf{r}_{j_{\alpha }}-\mathbf{l}^{\mathrm{o}}|=\rho $.
The radius $\rho =\lambda _{i}^{\mathrm{o}}-\nu _{0}$ determines the maximal
intensity (\ref{2 3 15}): 
\begin{equation}
\varkappa ^{\mathrm{o}}=\rho +\nu _{0}=\lambda _{i}^{\mathrm{o}}
\label{2 3 26}
\end{equation}%
($\rho \leq \nu _{0}$ since $|\mathbf{r}_{i}|\leq \nu _{0}$ for all $i$'s).
The minimum of intensity (\ref{2 3 26}) is obtained at $\rho =\nu
_{0}:\varkappa ^{\mathrm{o}}=2\nu _{0}$. This corresponds to the typical
equiprobable case $|\mathbf{r}_{i}|=1$, when there is at least one simplex
subset $\left\{ \mathbf{r}_{j_{\alpha }}\right\} $ for which the center $%
\mathbf{l}^{\mathrm{o}}=0$ is an interior point of the simplex.

\chapter{Efficient Estimation of Wave Patterns}

In this section we develop the noncommutative theory of efficient
measurements and optimal estimation of unknown parameters of wave patterns
as applied to problems of sound and visual pattern recognition. We consider
two variants of the lower bounds for the variance of the measured
parameters, the variants being based on noncommutative generalizations \cite%
{14}, \cite{31}, \cite{49} of the Rao-Cram\'{e}r inequality \cite{50}, and
introduce the notion of canonical states \cite{lenin76}, for which we derive
generalized uncertainty relations similar to the quantum mechanical
uncertainty relations \cite{11}, \cite{12}, \cite{31}. We then establish the
necessary and sufficient conditions for efficient measurements, conditions
that extend the conditions of efficiency of quantum mechanical measurements
obtained in \cite{lenin76}, \cite{31} to the case of classical wave signals
and fields. We formulate the necessary and sufficient conditions for the
optimality of generalized measurements of wave patterns, conditions that
generalize the respective conditions for quantum systems, see for example 
\cite{32}. Finally, we investigate the structure of optimal covariant
measurements for symmetric wave patterns, which in the case of quantum
symmetric fields has been studied in \cite{12}, \cite{29}. The exposition is
based largely on the works of Belavkin \cite{diss72}, \cite{29}, \cite{30}, 
\cite{31}.

\section{Wave Patterns Variances and Invariant Bounds}

We will consider two variants of the lower bound for the variance of
parameters of wave patterns, both based on noncommutative generalizations 
\cite{14}, \cite{31}, \cite{49} of the Rao-Cram\'{e}r inequality. In
contrast to \cite{14}, \cite{49}, these bounds are represented in a form
invariant under diffeomorphisms, the form will be used to obtain generalized
uncertainty relations and efficient measurements of canonical parameters.

\subsection{Classical Variance Bound}

\label{Classical}In Section \ref{Multi} we considered the problem of
recognizing pure or mixed wave patterns taken from a given finite or
denumerable standard family. Generally, sound and visual patterns may
contain unknown parameters that run through an infinite set of continuous
values $\theta \in \Theta $ of finite or denumerable dimensionality. For
example, we may not know the mean frequency and the moment when the sound
signal appears or the mean position and the wave number of the visual
pattern, or we may \textit{a priori} have no information on the expected
amplitudes of the oscillations in the given finite or denumerable family of
standard modes.

It is natural to estimate the unknown parameters by the intensity
distributions in the representations in which the wave packets with distinct
values of $\theta $ are clearly separated; for example, the frequency and
position can be calculated as the mean values on the coordinate
representation, while the mean time of arrival of a signal and the wave
number of a wave packet can be calculated in the momentum representation
(but not vice versa).

We will call a family of wave packets described in a Hilbert space $\mathcal{%
H}$ by amplitudes $\{\psi _{\theta }\}$ resolvable in a representation
defined by the decomposition of unity $I=\int M(dx)$ on a given Borel space $%
X$ if there exists a measurable map $\hat{\theta}:X\rightarrow \Theta $
satisfying the condition 
\begin{equation}
\int \hat{\theta}(x)\mu _{\theta }(dx)=\theta \int \mu _{\theta }(dx)\quad
\forall \theta \in \Theta ,  \label{3 1 1}
\end{equation}%
where $\mu _{\theta }(dx)=(\psi _{\theta }\mid M(dx)\psi _{\theta })$ is the
respective intensity distribution on $X$. Thus, the resolvability of the
family $\{\psi _{\theta }\}$ means that it is possible to calculate the
unknown $\theta \in \Theta $ in a given representation given the observed
distribution of the $\mu _{\theta }$ as the mean values of a function $\hat{%
\theta}(x)$ known as the unbiased estimator of parameters $\theta $.

It is natural to define the quality of the resolvability of family $\{\psi
_{\theta }\}$ by the size of the variance of the unbiased estimator $\hat{%
\theta}$, assuming that the quality for a given $\theta $ is all the higher
the smaller the standard deviation from $\theta $ in the distribution
induced on $\Theta $ by the measure $\mu _{\theta }$ of the wave packet $%
\psi _{\theta }$. To find the lower bound for this variance, we can use the
classical Rao-Cram\'{e}r inequality \cite{50} if the measure $\mu _{\theta }$
possesses the appropriate differentiability properties in $\theta $. For the
sake of simplicity we take the case of one parameter $\theta \in \mathbb{R}$%
. If we assume that there exists a second moment for the logarithmic
derivative $\hat{\gamma}_{\theta }=\partial \ln \mu _{\theta }/\partial
\theta $, which is the Radon-Nikodym derivative of measure $\mu _{\theta
}^{\prime }=\partial \mu _{\theta }/\partial \theta $ defined by 
\begin{equation}
\mu _{\theta }(dx)\hat{\gamma}_{\theta }(x)=\partial \mu _{\theta
}(dx)/\partial \theta ,  \label{3 1 2}
\end{equation}%
we can easily obtain the inequality 
\begin{equation}
\int (\hat{\theta}(x)-\theta )^{2}\mu _{\theta }(dx)\cdot \int (\hat{\gamma}%
_{\theta }(x)-\gamma _{\theta })^{2}\mu _{\theta }(dx)\geq J_{\theta }^{2},
\label{3 1 3}
\end{equation}%
where $J_{\theta }=\int \mu _{\theta }(dx)$ is the total intensity of the
wave packet $\psi _{\theta }$, and $\gamma _{\theta }$ is the mean value of $%
\hat{\gamma}_{\theta }$: 
\begin{equation*}
J_{\theta }\gamma _{\theta }=\int \hat{\gamma}_{\theta }(x)\mu _{\theta
}(dx)=J_{\theta }^{\prime }=2\func{Re}(\psi _{\theta }\mid \psi _{\theta
}^{\prime }).
\end{equation*}%
Inequality (\ref{3 1 3}), which implies the inverse proportionality of the
standard deviation $\sigma _{\hat{\theta}}\geq 1/\sigma _{\hat{\gamma}}$ or
variances 
\begin{equation}
\sigma _{\hat{\theta}}^{2}=\int (\hat{\theta}(x)-\theta )^{2}\mu _{\theta
}(dx)/J_{\theta },\;\sigma _{\hat{\gamma}}^{2}=\int (\hat{\gamma}_{\theta
}(x)-\theta )^{2}\mu _{\theta }(dx)/J_{\theta }  \label{3 1 4}
\end{equation}%
follows in an obvious manner from the Schwarz inequality if we allow for the
fact that the right-hand side can be represented, in accordance with (\ref{3
1 1}), in the form of the square of the scalar product 
\begin{equation*}
J_{\theta }=\int \hat{\theta}(x)\mu _{\theta }^{\prime }(dx)-\theta
J_{\theta }^{\prime }=\int (\hat{\theta}(x)-\theta )(\hat{\gamma}_{\theta
}(x)-\gamma _{\theta })\mu _{\theta }(dx).
\end{equation*}%
In a more general situation, where the estimated parameters $\theta =[\theta
^{i}]_{i=1}^{m}$ are differentiable functions $\theta (\alpha )$ of unknown
parameters $\alpha =[\alpha ^{k}]_{k=1}^{n}$ of the density operators of
mixed wave patterns $S(\alpha )$, we can easily obtain a matrix Rao-Cram\'{e}%
r inequality that is invariant with respect to the choice of the state
parameters: 
\begin{equation}
R\geq D\sigma _{\hat{\gamma}\hat{\gamma}}^{-1}D^{\intercal },  \label{3 1 5}
\end{equation}%
where $D=\left[ \partial \theta ^{i}/\partial \alpha ^{k}\right] $ is the
Jacobian matrix of the $\alpha \mapsto \theta $ transformation, $R=\sigma _{%
\hat{\theta}\hat{\theta}}$ is the covariance matrix 
\begin{equation}
R^{ik}(\alpha )=\int (\hat{\theta}^{i}(x)-\theta ^{i}(\alpha ))(\hat{\theta}%
^{k}(x)-\theta ^{k}(\alpha ))\mu (\alpha ,\;dx)  \label{3 1 6}
\end{equation}%
of unbiased estimators $\hat{\theta}^{i}(x)$ with respect to $\mu (\alpha
,\,dx)=\mathrm{Tr}\;S(\alpha )M(dx)$, 
\begin{equation}
\int \hat{\theta}^{i}(x)\mu (\alpha ,\;dx)=\theta ^{i}(\alpha )J(\alpha
),\;J(\alpha )=\mathrm{Tr}\,S(\alpha ),  \label{3 1 7}
\end{equation}%
and $\sigma _{\hat{\gamma}\hat{\gamma}}$ is a similar covariance matrix for
the logarithmic derivatives $\hat{\gamma}_{k}=\partial \ln \mu (\alpha
)/\partial \alpha ^{k}$, $k=1,\ldots ,n$. We will derive the inequality for
the general noncommutative case.

\subsection{Symmetric Nonclassical Bound}

The lower bound of inequality (\ref{3 1 5}) depends, naturally, on the
choice of the representation determined by the method of measurement. by
using the noncommutative analog of the logarithmic derivative introduced by
Helstrom, we can obtain a more exact bound for the variances of the unbiased
estimators that does not depend on the choice of representation.

If we assume that the family $\left\{ S_{\theta }\right\} $ of the trace
class density operators is strongly differentiable in $\theta $ in a certain
region $\Theta $, we can define a symmetric logarithmic derivative by the
following equation: 
\begin{equation}
\hat{g}_{\theta }S_{\theta }+S_{\theta }\hat{g}_{\theta }=2S_{\theta
}^{\prime }.  \label{3 1 8}
\end{equation}
It is easy to show (see \cite{13}) that if $\left| \mathrm{Tr}\;(S_{\theta
}^{\prime }\hat{x})\right| ^{2}\leq c\mathrm{Tr}\,(S_{\theta }\hat{x}^{2})$
for every Hermitian operator $\hat{x}$, the solution to this equation exists
and is unique, with $\mathrm{Tr}\;\left( S_{\theta }\hat{g}_{\theta
}^{2}\right) <\infty $.

Let us consider the operator 
\begin{equation*}
\hat{x}=\int \hat{\theta}(x)M(dx),\quad \mathrm{Tr}\;(S_{\theta }\hat{x}%
)=\theta J_{\theta },
\end{equation*}%
determined by the unbiased estimator $\hat{\theta}$ for a fixed measurement $%
M$. Since 
\begin{gather*}
\int (\hat{\theta}(x)-\theta )^{2}\mu _{\theta }(dx)=\mathrm{Tr}\left(
S_{\theta }\int (\hat{\theta}(x)-\theta )^{2}M(dx)\right) , \\
\mathrm{Tr}\;\left( S_{\theta }\int (\hat{\theta}(x)-\hat{x})M(dx)(\hat{%
\theta}(x)-\hat{x})+(\hat{x}-\theta )^{2}\right) \geq \mathrm{Tr}%
\;[S_{\theta }(\hat{x}-\theta )^{2}],
\end{gather*}%
it is sufficient to find the lower bound of the variance $\sigma _{\hat{x}%
}^{2}$ of operator $\hat{x}$. By analogy with the commutative case we have 
\begin{eqnarray*}
J_{\theta } &=&\mathrm{Tr}\;(\hat{x}S_{\theta }^{\prime }-\theta J_{\theta
}^{\prime })=\mathrm{Tr}\;[(\hat{x}-\theta )S_{\theta }^{\prime }]=\tfrac{1}{%
2}\mathrm{tr}\;[(\hat{x}-\theta )(\hat{g}_{\theta }S_{\theta }+S_{\theta }%
\hat{g}_{\theta })] \\
&=&\tfrac{1}{2}\mathrm{Tr}\;[S_{\theta }((\hat{x}-\theta )(\hat{g}_{\theta
}-\gamma _{\theta })+(\hat{g}_{\theta }-\gamma _{\theta })(\hat{x}-\theta ))]
\\
&=&\left\langle \hat{x}-\theta |\hat{g}_{\theta }-\gamma _{\theta
}\right\rangle _{+}.
\end{eqnarray*}%
Thus, the total intensity $J_{\theta }$ is equal to the symmetrized scalar
product with respect to $S_{\theta }$ of the operators $\hat{x}-\theta $ and 
$\hat{g}_{\theta }-\gamma _{\theta }$, where $\gamma _{\theta }=J_{\theta
}^{\prime }/J_{\theta }$ is the mean value of the operator $\hat{g}_{\theta
} $ of the logarithmic derivative, and $J_{\theta }^{\prime }=\mathrm{Tr}%
\;(S_{\theta }\hat{g}_{\theta })=\mathrm{Tr}\;S_{\theta }^{\prime }$.
Applying the Schwarz inequality 
\begin{equation*}
\left\vert \left\langle \hat{x}-\theta \mid \hat{g}_{\theta }-\gamma
_{\theta }\right\rangle _{+}\right\vert ^{2}\leq \left\langle \hat{x}-\theta
\mid \hat{x}-\theta \right\rangle _{+}\left\langle \hat{g}_{\theta }-\gamma
_{\theta }\mid \hat{g}_{\theta }-\gamma _{\theta }\right\rangle _{+},
\end{equation*}%
we arrive at the sought inequality: 
\begin{equation}
\sigma _{\theta }^{2}\geq \mathrm{Tr}\;[S_{\theta }(\hat{x}-\theta
)^{2}]/J_{\theta }\equiv \sigma _{x}^{2}\geq J_{\theta }/\mathrm{Tr}%
\;[S_{\theta }(\hat{g}_{\theta }-\gamma _{\theta })^{2}]\equiv 1/\sigma _{%
\hat{g}\theta }^{2}.  \label{3 1 9}
\end{equation}%
Thus, the variance of any unbiased estimator cannot be smaller than the
inverse variance of the operator of the logarithmic derivative (\ref{3 1 8}%
): 
\begin{equation}
\sigma _{\hat{g}\theta }^{2}=\mathrm{Tr\;}[S_{\theta }(\hat{g}_{\theta
}-\gamma _{\theta })^{2}]/J_{\theta }.  \label{3 1 10}
\end{equation}

A similar result can be obtained in the case where there are several
parameters $\theta =[\theta ^{i}]_{i=1}^{m}$ for the estimator $\hat{\theta}%
(x)=[\hat{\theta}^{i}(x)]_{i=1}^{m}$ satisfying the unbiasness conditions (%
\ref{3 1 1}), which when met make matrix (\ref{3 1 5}) the covariance matrix
of estimators $\hat{\theta}^{i}$, and the mean square error at a fixed $%
R_{\theta }$ assumes the minimal value.

For the covariance matrix $R_{\theta }$, Helstrom \cite{14} has established
the lower bound by assuming that the operator function $S_{\theta }=S(\theta
)$ is differentiable and using the concept of the operators $\hat{g}_{i}$ of
partial symmetrized logarithmic derivatives of the functions $S(\theta )$ in 
$\theta ^{i}$. He defined these operators by the following equations: 
\begin{equation}
\hat{g}_{i}S_{\theta }+S_{\theta }\hat{g}_{i}=2(\partial S_{\theta
}/\partial \theta ^{i}).  \label{3 1 11}
\end{equation}%
As in the classical case \cite{50}, this bound is defined by the matrix $%
G_{\theta }=\left[ G_{ik}(\theta )\right] $ of the covariances of the
solutions $\hat{g}_{i}=\hat{g}_{i}(\theta )$ of (\ref{3 1 1}). This matrix
for noncommutative $\hat{g}_{i}$ is taken in symmetrized form 
\begin{equation}
G_{ik}(\theta )=\frac{1}{2}\mathrm{Tr}S_{\theta }(\hat{g}_{i}\hat{g}_{k}+%
\hat{g}_{k}\hat{g}_{i})  \label{3 1 12}
\end{equation}%
(the mathematical expectations of $\mathrm{Tr}\;(S_{\theta }\hat{g}%
_{i}(\theta ))$ are equal to zero). The corresponding inequality has the
form 
\begin{equation}
R_{\theta }\geq G_{\theta }^{-1},\quad \theta \in \Theta  \label{3 1 13}
\end{equation}%
and is understood as the nonnegative definiteness of matrix $[R^{ik}(\theta
)-G^{ik}(\theta )]$, where $G^{ik}(\theta )$ are the elements of the inverse
matrix $G_{\theta }^{-1}:G^{ij}\left( \theta \right) G_{jk}(\theta )=\delta
_{k}^{i}$. Inequality (\ref{3 1 13}) is the noncommutative analog of the
Rao-Cram\'{e}r inequality \cite{50}. The matrix $G_{\theta }$ plays the role
of a metric tensor locally defining the distance $\sigma (\theta ,\theta
+d\theta )=G_{ik}(\theta )\;d\theta ^{i}\;d\theta ^{k}$ in the parameters
space $\Theta $, similar to the Fisher information distance in classical
statistics.

We now turn to a more general situation where the state parameters are not
the measured parameters $\theta ^{i}$ but other parameters $\alpha =\{\alpha
^{k},k=1,\ldots ,n\}$, $S=S(\alpha )$. The parameters $\theta ^{i}$ are
differentiable functions $\theta ^{i}=\theta ^{i}(\alpha )$ of the unknown
parameters. The respective generalized Helstrom inequality (\ref{3 1 13})
represents a bound for the covariance matrix $R=R(\alpha )$ of the
estimators $\hat{\theta}^{i}$ in a form invariant with respect to the choice
of the variables of states $S(\alpha )$, 
\begin{equation}
R\geq DG^{-1}D^{\intercal },  \label{3 1 14}
\end{equation}%
where $D=\left[ \partial \theta ^{i}/\partial \alpha ^{k}\right] $, and $%
G=G(\alpha )$ is the covariance matrix (\ref{3 1 12}) of the operators $\hat{%
g}_{k}=\hat{g}_{k}(\alpha )$ of symmetrized logarithmic derivatives of the
operator function $S(\alpha )$ in $\alpha ^{k}$.

Inequality (\ref{3 1 14}), which is equivalent to inequality (\ref{3 1 13})
only if $m=n$ and matrix $D=D(\alpha )$ is nonsingular, can be verified by a
line of reasoning similar to the one that will lead us to inequality (\ref{3
1 17}) (see Section \ref{proof of inequality}).

Inequality (\ref{3 2 5}) can be reduced to the classical Rao-Cram\'{e}r
inequality only where the family $\{S(\alpha )\}$ is commutative. For
noncommutative families other generalizations \cite{22} of the Rao-Cram\'{e}%
r inequality are possible, generalizations that are based on other
definitions of logarithmic derivatives and that lead to other lower bounds
for $R$ differing from the invariant Helstrom bound $DG^{-1}D^{\intercal }$.
For real-valued parameters $\alpha $ these generalizations may serve equally
well as analogs of the Rao-Cram\'{e}r inequality and coincide only if $%
\{S(\alpha )\}$ constitutes a commutative family, in which case they are
reduced to the classical Rao-Cram\'{e}r inequality. However, in the event of
complex-valued parameters $\alpha $ a special invariant generalization of
the Rao-Cram\'{e}r inequality becomes especially important. It is based on
the notions of right and left logarithmic derivatives and was suggested
independently by Belavkin \cite{31} and Yuen and Lax \cite{49}.

From now on we shall consider complex parameters $\alpha ^{k}=\alpha
_{1}^{k}+\mathrm{j}\alpha _{2}^{k}$, $\alpha =\{\alpha ^{k}\}\in \mathbb{C}%
^{n}$. The estimated parameters $\theta ^{i}=\theta ^{i}(\alpha ,\alpha )$, $%
i=1,\ldots ,m$, are assumed to be functions independently differentiable in $%
\alpha $ and $\bar{\alpha}$.\footnote{%
The derivatives $\partial /\partial \bar{\alpha}$ and $\partial /\partial
\alpha $ are defined in terms of the partial derivatives $\partial /\partial
\alpha _{1}$ and $\partial /\partial \alpha _{2}$ in the common manner: 
\begin{eqnarray*}
\partial /\partial \alpha &=&\frac{1}{2}(\partial /\partial \alpha
_{1}+j\partial /\partial \alpha _{2}), \\
\partial /\partial \bar{\alpha} &=&\frac{1}{2}(\partial /\partial \alpha
_{1}+j\partial /\partial \alpha _{2}).
\end{eqnarray*}%
} Let us define the non-Hermitian logarithmic derivatives of $S=S(\alpha
,\alpha )$ thus: 
\begin{equation}
S\hat{h}_{k}=\partial S/\partial \bar{\alpha}^{k},\quad \hat{h}_{k}^{\ast
}S=\partial S/\partial \alpha ^{k},\quad k=1,\ldots ,n.  \label{3 1 15}
\end{equation}%
The operators $\hat{h}_{k}=\hat{h}_{k}\left( \alpha ,\bar{\alpha}\right) $
are called the right derivatives with respect to $\bar{\alpha}^{k}$ (and the
operators $\hat{h}_{k}^{\ast }$ the left derivatives with respect to $\alpha 
$) and have zero mathematical expectations. The covariance matrix $H=\left[
H_{ik}\right] $ of 
\begin{equation}
H_{ik}(\alpha ,\bar{\alpha})=\mathrm{Tr}\;[S(\alpha ,\bar{\alpha})\hat{h}_{i}%
\hat{h}_{k}^{\ast }]  \label{3 1 16}
\end{equation}%
is Hermitian and, assuming it is nonsingular, defines a positive definite
metric $\mathrm{d}s^{2}+H_{ik}\;\mathrm{d}\bar{\alpha}^{i}\mathrm{d}\alpha
^{k}$ in a complex domain $\mathcal{O}\subset \mathbb{C}^{n}$ of the
unknowns $\alpha \in \mathcal{O}$.

Suppose that a joint measurement of the parameters $\theta ^{i}$ is
described by a decomposition of unity that defines the estimator $\hat{\theta%
}$. This estimator is represented by a vector quantity that, in general,
assumes complex values $x=\{x^{i}\}\in \mathbb{C}^{m}$, is represented by a
conditional distribution $\mu (dx\mid \alpha ,\alpha )=\mathrm{Tr}%
\;[M(dx)S(\alpha ,\bar{\alpha})]$, and satisfies the unbiasness condition $%
\left\langle \hat{\theta}^{i}\right\rangle =\theta ^{i}(\alpha ,\bar{\alpha})
$. Then the mean square error of measurement is determined by the matrix $%
R=R(\alpha ,\bar{\alpha})$ of covariances $R^{ik}=\left\langle (\hat{\theta}%
^{i}-\theta ^{i})(\hat{\theta}^{k}-\theta ^{k})^{\ast }\right\rangle $, for
which the following inequality holds true: 
\begin{equation}
R\geq DH^{-1}D^{\dagger },  \label{3 1 17}
\end{equation}%
where $D=D(\alpha ,\bar{\alpha})$, as in (\ref{3 1 14}), is the matrix of
the derivatives $\partial \theta ^{i}/\partial \alpha ^{k}$, and $D^{\dagger
}$ is the respective Hermitian conjugate matrix. Even in the real case, that
is $\hat{\theta}^{i}=\bar{\theta}^{i}$, inequality (\ref{3 1 17}) leads to a
lower bound that differs from the Helstrom bound (\ref{3 1 14}). We will
call the lower bound in (\ref{3 1 17}) the right bound. Other bounds can
also be considered, say, the left bound, which is based on the left
logarithmic derivatives with respect to $\bar{\alpha}$. The proof of all
such inequalities is similar to that of inequality (\ref{3 1 17}), which is
given in Section \ref{proof of inequality}. The right bound in (\ref{3 1 17}%
) is invariant under replacement of derivatives with respect to $\alpha ^{k}$%
, by derivatives with respect to new variables $\beta ^{k}=\beta ^{k}(\alpha
)$ only if the functions $\beta ^{k}\left( \alpha \right) $ are analytic,
that is, $\partial \beta ^{k}/\partial \bar{\alpha}^{i}=0$, and the matrix
of derivatives $\partial \beta ^{k}/\partial \alpha ^{i}$ is nonsingular.
Hence, the use of inequality (\ref{3 1 17}) in invariant form $R\geq H^{-1}$%
, where, as in (\ref{3 1 13}), we employ derivatives with respect to the
estimated parameters $\theta ^{i}$ (but, in contrast to (\ref{3 1 13}),
right derivatives rather than symmetrized are used), is inexpedient since
the condition for the equivalence of these inequalities includes not only
the condition that matrix $D$ be nonsingular but the analyticity condition $%
\partial \theta ^{i}/\partial \bar{\alpha}^{k}=0$ as well (that is, the
independence of functions $\theta ^{i}(\alpha ,\bar{\alpha})$ on $\bar{\alpha%
}$), which is not out initial assumption. A similar situation for
complex-valued parameters $\alpha $ exists in the classical case.

\section{Uncertainty Relations and Efficient Measurements}

In this section we will introduce the notion of canonical families of wave
patterns for whose canonical parameters we will establish uncertainty
relations that generalize the quantum mechanical uncertainty relations
obtained in the one-dimensional case by Helstrom \cite{13} and in the case
of multidimensional Lie algebra by Belavkin \cite{31}. We will then find the
limit of accuracy in estimating the canonical Lie parameters of wave
patterns and prove that such limits are exact only for canonical signals for
which there are efficient measurement or quasimeasurement procedures. The
discourse will follow the scheme suggested in \cite{31}; for examples of
uncertainty relations for quantum systems the readers is advised to turn to 
\cite{12},\cite{13}.

\subsection{Canonical Families and Uncertainty Relations}

\label{Uncertainty Relations}In classical mathematical statistics an
important role is played by canonical, or exponential, families of
probability distributions, for which a special selection of parameters $%
\theta $ and $\alpha $ makes the Rao-Cram\'{e}r bound exact. In Section \ref%
{Canonical Form} we will prove that in the noncommutative case a similar
role is played by density operators of the form 
\begin{equation}
S(\beta ,\bar{\beta})=\chi ^{-1}\mathrm{e}^{\beta ^{k}\hat{x}_{k}^{\ast
}}S_{0}\mathrm{e}^{\bar{\beta}^{k}\hat{x}_{k}^{\ast }},  \label{3 2 1}
\end{equation}%
where the $\hat{x}_{k},k=1,\ldots ,n$, are linearly independent operators in 
$\mathcal{H}$, which may be non-Hermitian $(\hat{x}_{k}^{\ast }\neq \hat{x}%
_{k})$ and may not commute with the conjugate operators $(\hat{x}_{i}\hat{x}%
_{k}^{\ast }\neq \hat{x}_{k}^{\ast }\hat{x}_{i})$, and $\chi =\chi (\beta ,%
\bar{\beta})$ is the generating function of the moments of these operators
in state $S_{0}$: 
\begin{equation}
\chi (\beta ,\bar{\beta})=\mathrm{Tr}\;S_{0}\mathrm{e}^{\bar{\beta}^{k}\hat{x%
}_{k}}\mathrm{e}^{\beta ^{k}\hat{x}_{k}^{\ast }},  \label{3 2 2}
\end{equation}%
which is finite ($\chi <\infty $) in a neighborhood of zero $\beta =0$ of
the complex space $\mathbb{C}^{n}$. The family of density operators (\ref{3
2 1}) will be said to be canonical and the parameters $\beta ^{k}$,
canonically conjugate to the $\hat{x}_{k}$. In contrast to the commutative
case, even for Hermitian operators $\hat{x}_{k}$ it is meaningful to assume
that the conjugate parameters $\beta ^{k}$ may have complex values.

Of special interest is the case, which has no classical analog, of canonical
states (\ref{3 2 1}) where the $\beta ^{k}$ are imaginary and the $\hat{x}%
_{k}$ are Hermitian. The parameters $\theta ^{k}=\func{Im}\beta ^{k}/(2\pi )$
acquire a dimensionality and meaning of quantities that are dynamically
conjugate to the $\hat{x}_{k}$; for instance, if $\hat{x}$ is frequency, $%
\theta $ is time, if $x$ is momentum, $\theta $ is displacement, if $\hat{x}$
is angular momentum, $\theta $ is the angle of rotation. The canonical
states (\ref{3 2 1}) at $\beta ^{k}=2\pi \mathrm{j}\theta ^{k}$ assume the
form 
\begin{equation}
S_{\theta }=\mathrm{e}^{2\pi \mathrm{j}\theta ^{k}\hat{x}_{k}}S_{0}\mathrm{e}%
^{-2\pi \mathrm{j}\theta ^{k}\hat{x}_{k}}  \label{3 2 3}
\end{equation}%
and are unitary equivalent to state $S_{0}$, which corresponds to a zero
value of $\theta $. It has been established that if we put $\alpha =\beta $
and apply inequality (\ref{3 1 17}) to the canonical family (\ref{3 2 3}),
we arrive at the exact formulation of the generalized uncertainty principle
for any pair of dynamically conjugate quantities $\hat{\theta}^{k}$ and $%
\hat{x}^{k}$, where the first quantity in the pair may not correspond to the
Hermitian operator that meaningfully describes in $\mathcal{H}$ the
measurement of this quantity.\footnote{%
The Heisenberg uncertainty principle is usually proved only for such
dynamically conjugate quantities described by noncommutative operatiors $%
\hat{p}$ and $\hat{q}$ that satisfy, say, the commutation relations $[\hat{p}%
,\hat{q}]=1/2\pi j$. The proof employs the well-known scalar inequality $%
\left\langle (\hat{p}-\left\langle \hat{p}\right\rangle )^{2}\right\rangle
\left\langle \left( \hat{q}-\left\langle \hat{q}\right\rangle \right)
^{2}\right\rangle \geq \left\vert \left\langle \lbrack \hat{p},\hat{q}%
]\right\rangle \right\vert ^{2}/4$, which is valid for any pair of operators 
$\hat{p}$ and $\hat{q}$. Strenthening and matrix multidimensional
generalization of this inequality in terms of the covariance estimations of
an arbitrary family of noncommutative operators are suggested in \cite{21}.}

Differentiation (\ref{3 2 1}) with respect to $\bar{\beta}^{k}$ and
comparing the result with (\ref{3 1 1}), we get 
\begin{equation}
\hat{h}_{i}=\mathrm{e}^{-\bar{\beta}^{k}\hat{x}_{k}}\chi \tfrac{\partial }{%
\partial \bar{\beta}^{i}}(\chi ^{-1}\mathrm{e}^{\bar{\beta}^{k}\hat{x}_{k}})=%
\hat{x}_{i}(\bar{\beta})-\theta _{i},  \label{3 2 4}
\end{equation}%
where $\hat{x}_{i}(\bar{\beta})=\mathrm{e}^{-\beta ^{k}\hat{x}_{k}}\frac{%
\partial }{\partial \bar{\beta}^{i}}\mathrm{e}^{\bar{\beta}^{k}\hat{x}_{k}}$%
, and $\theta _{i}=\frac{\partial }{\partial \bar{\beta}^{i}}\ln \chi =%
\mathrm{Tr}\;[S\hat{x}_{i}(\bar{\beta})]$. Matrix (\ref{3 1 16}), therefore,
is the covariance matrix. 
\begin{equation}
H_{ik}=\mathrm{Tr}\;[S(\hat{x}_{i}(\bar{\beta})-\theta _{i})(\hat{x}_{k}(%
\bar{\beta})-\theta _{k})^{\ast }]=\frac{\partial ^{2}\ln \chi }{\partial 
\bar{\beta}^{i}\partial \beta ^{k}}  \label{3 2 5}
\end{equation}%
of the operators $\hat{x}_{i}(\bar{\beta})$ analytic in $\bar{\beta}$ and
coinciding with $\hat{x}_{i}$ at $\beta =0$. The inequality (\ref{3 1 17})
in the neighborhood of point $\beta =0$, therefore, can be written in the
form of the uncertainty relation 
\begin{equation}
R\gtrsim DS^{-1}D^{\dagger },  \label{3 2 6}
\end{equation}%
which establishes the inverse proportionality between he matrix $S=\left[
S_{ik}\right] $ of the covariances 
\begin{equation}
S_{ik}=\mathrm{Tr}\;[S(\beta ,\bar{\beta})(\hat{x}_{i}-\mu _{i})(\hat{x}%
_{k}-\mu _{k})^{\ast }]  \label{3 2 7}
\end{equation}%
of the operators $\hat{x}_{i},\;\mathrm{Tr}\;[S(\beta ,\bar{\beta})\hat{x}%
_{i}]=\mu _{i}$, and the covariance matrix $R$ of the estimators $\hat{\theta%
}^{i}$ of the functions $\theta ^{i}(\beta ,\bar{\beta})$ of the conjugate
parameters $\beta $ and $\bar{\beta}$. At $\theta =\beta $, (\ref{3 2 6})
transforms into strict inequality in the entire domain $\mathcal{O}\ni \beta 
$. Putting $\theta =\func{Im}\beta /(2\pi )$ and allowing for the fact that $%
\partial \theta /\partial \beta =1/4\pi \mathrm{j}$, we obtain $\hat{x}^{k}=%
\hat{x}$ the generalized uncertainty relation 
\begin{equation}
(2\pi )^{2}R_{\theta }\geq (1/4)S^{-1}  \label{3 2 8}
\end{equation}%
in terms of the variances $R_{\theta }=\left\langle (\hat{\theta}-\theta
)^{2}\right\rangle _{\theta }$, with $S=\mathrm{Tr}\;[S_{0}(\hat{x}-\mu
)^{2}]$, valid for any pair of dynamically conjugate quantities $\hat{\theta}
$ and $\hat{x}$ defining the canonical family (\ref{3 2 3}). For the quantum
case of pure states $S_{0}=|\psi _{0})(\psi _{0}|$, the scalar inequality (%
\ref{3 1 13}) by Helstrom \cite{11} by a complicated procedure for
calculating the matrix elements of the operators of symmetrized logarithmic
derivatives.

In the multidimensional case, when the operators $\hat{x}_{k}$ are pairwise
commutative (but not necessarily with $\hat{x}_{k}^{\ast }$ and $S_{0}$),
the situation is the same: $\hat{x}_{k}(\bar{\beta})=\hat{x}_{k}$ for every $%
\beta \in \mathcal{O}$ and inequality (\ref{3 2 6}) is strict. The averages $%
\mu _{k}$ and the covariances (\ref{3 2 7}) at $\hat{x}_{k}=\hat{x}%
_{k}^{\ast }$ and $\beta ^{k}=2\pi \mathrm{j}\theta ^{k}$ are independent of 
$\theta $ and, therefore, coincide with the respective values at $\theta
=\theta $: $\mu _{k}=\mathrm{Tr}\;(S_{0}\hat{x}_{k})$ and 
\begin{equation}
S_{ik}=\mathrm{Tr}\;[S_{0}(\hat{x}_{i}-\mu _{i})(\hat{x}_{k}-\mu _{k})].
\label{3 2 9}
\end{equation}%
The uncertainty relation (\ref{3 2 8}) in this case acquires a matrix
meaning: $R_{\theta }$ is the covariance matrix of estimators $\hat{\theta}%
^{i}$ of the canonical parameters belonging to a translation group in state $%
S_{\theta }$, and $S$ is the covariance matrix (\ref{3 2 9}) of the
generators $\hat{x}_{k}$ of the group, defining the lower bound $%
S^{-1}/16\pi ^{2}$ for $R_{\theta }$ uniformly in every $\theta \in \mathbb{R%
}^{n}$.

We now take up the case of noncommutative $\{\hat{x}_{k}\}$. Suppose that
the operators $\hat{x}_{k}$ form a Lie algebra: 
\begin{equation}
\hat{x}_{i}\hat{x}_{k}-\hat{x}_{k}\hat{x}_{i}=C_{ik}^{j}\hat{x}_{j},
\label{3 2 10}
\end{equation}%
where $C_{ik}^{j}$ are structure constants. Here the operators $\hat{x}_{i}(%
\bar{\beta})$ in (\ref{3 2 4}) are linear combinations of the generators $%
\hat{x}_{i}$: 
\begin{equation}
\hat{x}_{i}(\bar{\beta})=L^{-1}(-\bar{\beta})_{i}^{j}\hat{x}_{j},
\label{3 2 11}
\end{equation}%
with $L(\xi )=\xi ^{k}C_{k}(I-\mathrm{e}^{-\xi C})^{-1}$ an $n$-by-$n$
matrix that exists in a neighborhood $\mathcal{O}\in \mathbb{C}^{n}$ of zero 
$\xi =0$, and the $C_{k}=\left[ C_{ik}^{j}\right] $ are the generators of
the adjoint representation of the commutation relations (\ref{3 2 10}).
Expressing the covariance matrix $H$ of the operators (\ref{3 2 11}) in
terms of the covariances (\ref{3 2 7}) of the generators $\hat{x}_{i}$, we
get instead of (\ref{3 2 6}) the inequality 
\begin{equation}
R\geq DL^{\dagger }S^{-1}LD^{\dagger },  \label{3 2 12}
\end{equation}%
where $L=L(-\beta )$. In the case of (\ref{3 2 3}), the family of $S_{\theta
}$ is unitary homogeneous with respect to the Lie group with Hermitian
operators $\hat{x}_{k}$ and canonical parameters $\theta ^{k}$. Similarly to
(\ref{3 2 8}), we obtain a more general relationship 
\begin{equation}
(2\pi )^{2}R_{\theta }\geq \frac{1}{4}L_{\theta }^{\intercal
}S^{-1}L_{\theta },  \label{3 2 13}
\end{equation}%
where $L_{\theta }=\theta ^{i}G_{i}(I-\mathrm{e}^{-\mathrm{j}\theta
^{k}G_{k}})^{-1}$, $G_{k}=2\pi \mathrm{j}C_{k}$. Inequality (\ref{3 2 13})
determines in the domain $\mathcal{O}\subset \mathbb{R}^{n}$ of the
convergence of the series expansion 
\begin{equation*}
(I-\mathrm{e}^{-\theta ^{k}G_{k}})^{-1}=\sum_{m=0}^{\infty }\mathrm{e}%
^{-m\theta ^{k}G_{k}},\quad \theta =\{\theta ^{i}\}\in \mathcal{O},
\end{equation*}%
the lower bound of the mean square error in estimating the canonical
parameters of the unitary representation $\mathrm{e}^{2\pi \mathrm{j}\theta
^{k}\hat{x}_{k}}$ of a Lie group.

\subsection{Efficient Measurements and Quasimeasurements}

\label{Quasimeasurements}In classical statistics, estimations whose
covariance matrix assumes the minimal value and thus transforms, locally or
globally, the Rao-Cram\'{e}r inequality into an equality are known as
efficient (locally or globally, respectively). In the noncommutative case,
the concept of efficiency introduced by analogy with the classical concept
looses its universality because the generalization of the Rao-Cram\'{e}r
inequality is not unique and the definitions of locally efficient estimates 
\cite{14}, \cite{22}, \cite{49} based on different variants of this
generalization are not equivalent. For this reason we distinguish between
the efficient measurements (or estimates) for which the invariant Helstrom
bound (\ref{3 1 14}) is attained and those for which the right bound (\ref{3
1 17}) is attained, with the former called Helstrom efficient and the
latter, right efficient. As we show below, the notion of right efficiency is
more universal: measurements that are Helstrom efficient are right
efficient, but not vice versa. Let us first prove that Helstrom efficient
estimates exist globally for canonical families of density operators (\ref{3
2 1}) if the operators $\hat{x}_{k}$ are Hermitian and pairwise commutative
and if for the estimated parameters $\theta $ we take the derivatives $%
\theta _{k}=\partial \ln \chi /\partial \varkappa ^{k}$ of the generating
function $\chi (\varkappa )=\mathrm{Tr}(S_{0}\mathrm{e}^{\varkappa ^{k}\hat{x%
}_{k}})$, where $\varkappa =\beta +\bar{\beta}$. The parameters $\theta _{k}$
selected in this manner coincide with the averages defined by the canonical
subfamilies of the density operators, 
\begin{equation}
S(\varkappa )=\chi ^{-1}(\varkappa )\mathrm{e}^{\varkappa ^{k}\hat{x}%
_{k}/2}S_{0}\mathrm{e}^{\varkappa ^{k}\hat{x}_{k}/2},  \label{3 2 14}
\end{equation}%
with $\func{Im}\beta ^{k}=0$: 
\begin{equation}
\theta _{k}(\varkappa )=\mathrm{Tr}\;[S(\varkappa )\hat{x}_{k}]=\partial \ln
\chi /\partial \varkappa ^{k}.  \label{3 2 15}
\end{equation}%
Taking for parameters $\alpha ^{k}$ the canonical parameters $\varkappa ^{k}$
and differentiating the operator functions (\ref{3 2 14}), we obtain the
symmetrized logarithmic derivatives in $\varkappa ^{k}:\hat{g}_{k}=\hat{x}%
_{k}-\theta _{k}$. Thus, the covariances (\ref{3 1 12}) coincide with the
covariances of operators $\hat{x}_{k}$, 
\begin{equation}
G_{ik}=\mathrm{Tr}\;[S(\varkappa )(\hat{x}_{i}-\theta _{i})(\hat{x}%
_{k}-\theta _{k})]=\frac{\partial ^{2}\ln \chi }{\partial \varkappa
^{i}\partial \varkappa ^{k}},  \label{3 2 16}
\end{equation}%
which are equal to the derivatives $D_{ik}=\partial \theta _{i}/\partial
\varkappa ^{k}$ defining matrix $D$ in (\ref{3 1 14}) for $\alpha =\varkappa 
$. Therefore, inequality (\ref{3 1 14}) assumes the form $R\geq G$, or $%
\left[ R_{ik}-G_{ik}\right] \geq 0$, where $R_{ik}=\left\langle (\hat{\theta}%
_{i}-\theta _{i})(\hat{\theta}_{k}-\theta _{k})\right\rangle $ are the
covariance of the unbiased estimators $\hat{\theta}_{k}:\left\langle \hat{%
\theta}_{k}\right\rangle =\theta _{k}$. If for these estimators we take the
results $x_{k}$ of measurements of the observables $\hat{x}_{k}$ (which are
compatible), then matrix $R$ assumes the minimal value $R=G$. Thus, for the
canonical families (\ref{3 2 14}) with commutative operators $\hat{x}_{k}$
there exists a Helstrom efficient measurement of the functions (\ref{3 2 15}%
) of the canonical parameters $\varkappa _{k}$, which is the usual
compatible measurement of the observables $x_{k}$. The domain of this
efficiency, obviously, coincides with the domain $\mathcal{O}\subset \mathbb{%
R}^{n}$ for which $\chi (\varkappa )<\infty ,\varkappa \in 0$. It has been
established that the converse is true in the following sense.

Let the estimators $\hat{\theta}_{k}$ (i.e. the results of a measurement)
have averages $\theta _{k}(\alpha )$ and covariances $R_{ik}(\alpha )$ that
are differentiable in a certain domain, and let the matrices $%
R=[R_{ik}\left( \alpha \right) ],D=[\partial \theta /\partial \alpha ^{k}]$
satisfy the conditions 
\begin{equation}
\partial (R^{-1}D)_{k}^{j}/\partial \alpha ^{i}=\partial
(R^{-1}D_{i}^{j})/\partial \alpha ^{k}  \label{3 2 17}
\end{equation}%
(the regularity conditions). We can then introduce the canonical parameters $%
\varkappa ^{k}=\varkappa ^{k}(\alpha )$ defined uniquely by the derivatives $%
\partial \varkappa ^{i}/\partial \alpha ^{k}=(R^{-1}D)_{k}^{i}$ if we put $%
\varkappa ^{k}(\alpha _{0})=0$ for a fixed $\alpha _{0}$. It can easily be
verified that for a family of density operators $S(\alpha )$ of canonical
form (\ref{3 2 14}), with $\varkappa ^{k}=\varkappa ^{k}(\alpha )$
differentiable functions possessing a nonzero Jacobian, the regularity
conditions are met in an efficient measurement at $\theta _{k}\left( \alpha
\right) =\partial \ln (\chi (\varkappa (\alpha ))/\partial \varkappa ^{k}$
such that $R_{ik}=G_{ik}(\varkappa (\alpha ))$ and $(R^{-1}G)_{k}^{i}=%
\partial \varkappa ^{i}/\partial \alpha ^{k}$. Proof of the converse
assertion that under the regularity conditions the global Helstrom
efficiency comes into play only for canonical families (\ref{3 2 14}) is
given in Section \ref{proof of inequality} for a more general situation
involving complex variables.

Hence we have proved the following

\begin{theorem}
\label{T 3 2 1}Under appropriate regularity conditions, inequality \textup{(%
\ref{3 1 14})} \textup{is} transformed into an equality in a certain domain $%
\mathcal{O}\subset \mathbb{R}^{n}$ if and only if the family of density
operators $S(\alpha )$ has the canonical form \textup{(\ref{3 1 14})}, where 
$\hat{x}_{k},\;k=1,\ldots ,n$, are commutative Hermitian operators in $%
\mathcal{H}$, and the canonical parameters $\varkappa ^{k},\;k=1,\ldots ,n$,
are functions of parameters $\alpha $ defined by the equations 
\begin{equation*}
\partial \ln \chi /\partial \varkappa ^{k}=\theta _{k}(\alpha ),\quad
k=1,\ldots ,n.
\end{equation*}
\end{theorem}

\subsection{The Theorem Regarding the Canonical Density Operators}

\label{Canonical Form}Suppose that in a certain domain $\mathcal{O}\subset 
\mathbb{C}^{n}$ the unbiased estimators $\hat{\theta}_{k}$ possess averages $%
\theta _{k}(\alpha ,\bar{\alpha})$ and covariances $R_{ik}(\alpha ,\bar{%
\alpha})$ that satisfy the regularity conditions (\ref{3 2 17}), to which we
adjoin the analyticity condition 
\begin{equation}
\tfrac{\partial }{\partial \bar{\alpha}^{k}}R^{-1}D=0.  \label{3 2 18}
\end{equation}
Here we can introduce, as we did in Section \ref{Uncertainty Relations},
canonically conjugate parameters $\beta ^{k}=\beta ^{k}(\alpha )$ via the
equations $\partial \beta ^{i}/\partial \alpha ^{k}=(R^{-1}D)_{k}^{i}$ and
conditions $\beta ^{k}(\alpha _{0})=0$ for a fixed $\alpha _{0}\in \mathcal{O%
}$ with the functions $\beta ^{k}\left( \alpha \right) $ being analytic in
view of conditions (\ref{3 2 18}).

\begin{theorem}
\label{T 3 2 2}Under the formulated regularity conditions, inequality 
\textup{(\ref{3 1 17}) }is transformed into an equality in a certain domain $%
\mathcal{O}\subset \mathbb{C}^{n}$ if and only if the family $\{S(\alpha
,\alpha ),\alpha \in \mathcal{O}\}$ has the canonical form \textup{(\ref{3 2
1})}, with $S_{0}=S(\alpha _{0},\bar{\alpha}_{0})$ for an $\alpha _{0}\in 
\mathcal{O}$, the operators $\hat{x}_{k},\;k=1,\ldots ,n$, \ simultaneously
possess in $\mathcal{H}$ the property of the right proper decomposition of
unity 
\begin{equation}
I=\int M(dx),\quad \hat{x}_{k}M(dx)=x_{k}M(dx),\quad x=\{x_{k}\}\in \mathbb{C%
}^{n},  \label{3 2 19}
\end{equation}
and the parameters $\beta ^{k},\;k=1,\ldots ,n$, are analytic functions $%
\beta ^{k}(\alpha )$ determined by the equations 
\begin{equation*}
\partial \ln \chi /\partial \bar{\beta}^{k}=\theta _{k}(\alpha ,\bar{\alpha}%
),\quad \alpha \in \mathcal{O}.
\end{equation*}
Optimal estimation is then reduced to a quasimeasurement of the
non-Hermitian operators $\hat{x}_{k}$ described by the decomposition of
unity \textup{(\ref{3 2 19})}, while the minimal mean square error is
determined by the covariance matrix 
\begin{equation}
R_{ik}=\mathrm{Tr}\;[S(\hat{x}_{i}-\theta _{i})(\hat{x}_{k}-\theta
_{k})^{\ast }].  \label{3 2 20}
\end{equation}
\end{theorem}

\begin{proof}
Sufficiency is proved in the same way as in Section \ref{Uncertainty
Relations}. Employing the fact of invariance oft he right bound (\ref{3 1 17}%
) under the analytic transformations $\alpha \rightarrow \beta $, we select
for the variables $\alpha ^{k}$ determining this bound the parameters $\beta
^{k}$ of the family of density operators (\ref{3 2 1}). The elements $%
\partial \theta _{i}/\partial \beta ^{k}$ of matrix $D$ then coincide with
the elements (\ref{3 2 5}) of matrix $H$ if we allow for the fact that $%
\theta _{i}=\partial \ln \chi /\partial \bar{\beta}^{i}$. Since according to
(\ref{3 2 19}) the operators $\hat{x}_{k}$ are commutative, $\hat{x}_{i}\hat{%
x}_{k}=\int x_{i}x_{k}M(dx)=\hat{x}_{k}\hat{x}_{i}$, we have $\theta
_{k}=\mu _{k}$ and $H_{ik}=S_{ik}$, where the $\mu _{k}$ are the averages of
the $\hat{x}_{k}$, and the $S_{ik}$ are the covariances (\ref{3 2 7}) of
these operators. Hence, inequality (\ref{3 1 17}) assumes the form $R\geq S$%
. What remains to be proved is that the measurement described by the
decomposition of unity (\ref{3 2 19}) leads to an estimation for which $R=S$
even when the operators are not commutative with the respective conjugates: $%
\hat{x}_{i}\hat{x}_{k}^{\ast }\neq \hat{x}_{k}^{\ast }\hat{x}_{i}$ (which
occurs when decomposition (\ref{3 2 19}) is nonorthogonal). To do this, it
is sufficient to allow for the representation 
\begin{equation}
\hat{x}_{i}=\int xM(dx),\quad \hat{x}_{i}\hat{x}_{k}^{\ast }=\int x_{i}\bar{x%
}_{k}M(dx),\quad x\in \mathbb{C}^{n},  \label{3 2 21}
\end{equation}%
which is obtained by integrating the equations in (\ref{3 2 19}) and the
adjoint equation $M(dx)\hat{x}_{k}^{\ast }=\bar{x}_{k}M(dx)$. Thanks to (\ref%
{3 2 21}) the covariances 
\begin{equation}
R_{ik}=\int (x_{i}-\theta _{i})(\bar{x}_{k}-\theta _{k})\mathrm{Tr}\;[SM(dx)]
\label{3 2 22}
\end{equation}%
of the estimators $\hat{\theta}_{k}$ obtained as a result of a
quasimeasurement of operators $\hat{x}_{k}$ coincide with the covariances $%
S_{ik}$ of these operators, which proves the efficiency of this
quasimeasurement for the density operators (\ref{3 2 1}). The proof of the
converse of Theorem \ref{T 3 2 2} follows from the derivation of inequality (%
\ref{3 1 17}) and will be discussed in Section \ref{proof of inequality}.
\end{proof}

\subsection{Discussion and an Example}

Thus, the condition of (right) efficiency requires the existence of
commutative operators possessing a joint right spectral decomposition and
playing the role of sufficient statistics, which it is natural to call right
efficient. Here it is sufficient to restrict the discussion to the operators
in the minimal subspace generated by the regions $S(\beta ,\bar{\beta})%
\mathcal{H}$ with the density operators $S(\beta ,\bar{\beta})$ for all $%
\beta (\alpha )\in \mathbb{C}^{n}$ for which $\alpha \in \mathcal{O}$. Even
if we consider only the real values of parameters $\theta (\alpha ,\bar{%
\alpha})$, optimal estimation can be described by non-Hermitian and
noncommutative (with the conjugate) operators of the right-efficient
statistics and, therefore, may not be Helstrom efficient. However, estimates
that are Helstrom efficient correspond, according to Theorem \ref{T 3 2 1},
to the particular case of right efficiency where the $\hat{x}_{k}$ are
Hermitian. If the operators $\hat{x}_{k}$ in (\ref{3 2 1}) are non-Hermitian
but commutative with the conjugate operators, the right-efficient estimates
also coincide with complexified estimates, which are Helstrom efficient.
However, commutativity $\hat{x}_{k}\hat{x}_{i}^{\ast }=\hat{x}_{i}^{\ast }%
\hat{x}_{k}$, may not occur either.

\begin{example}
Let $\hat{x}_{k}=\varphi _{k}(\hat{a})$, where the $\varphi _{k}$ are entire
functions $\mathbb{C}^{r}\rightarrow \mathbb{C}$, and let $\hat{a}=\{\hat{a}%
_{i},\;i=1,\ldots ,r\}$ be the annihilation operators satisfying the
commutation relations%
\begin{equation*}
\hat{a}_{i}\hat{a}_{k}-\hat{a}_{k}\hat{a}_{i}=\hat{0},\;\;\;\;\hat{a}_{i}%
\hat{a}_{k}^{\ast }-\hat{a}_{k}^{\ast }\hat{a}_{i}=\delta _{i}^{k}\hat{1}.
\end{equation*}%
Show the right efficient measurement for the parameters $\theta
_{k}=\partial \ln \chi /\partial \bar{\beta}^{k}$ of the density operators 
\textup{(\ref{3 2 1})}.
\end{example}

It is well-known that the operators $\hat{a}$ have right eigenvectors $%
|\alpha )\in \mathcal{H},\;\alpha \in \mathbb{C}^{r}$, that define the
nonorthogonal decomposition of unity 
\begin{equation*}
I=\int |\alpha )(\alpha |\prod_{i=1}^{r}\pi ^{-1}\mathrm{d}\func{Re}\alpha
_{i}\mathrm{d}\func{Im}\alpha _{i},\quad \hat{a}_{i}|\alpha )=\alpha
_{i}|\alpha ).
\end{equation*}%
It is obvious then that the operators $\hat{x}=\varphi (\hat{a})$ have a
right proper decomposition of unity \textup{(\ref{3 2 19})}, where 
\begin{equation*}
M(dx)=\mathrm{d}x\int \delta (x-\varphi (\alpha ))\;|\alpha )(\alpha
|\prod_{i=1}^{r}\pi ^{-1}\mathrm{d}\func{Re}\alpha _{i}\mathrm{d}\func{Im}%
\alpha _{i}
\end{equation*}%
\textup{(}$\mathrm{d}x$ is the Lebesgue measure on $\mathbb{C}^{n}$, and $%
\delta (x-\varphi )$ is the Dirac delta function\textup{)}. Hence, optimal
estimation of the parameters $\theta _{k}=\partial \ln \chi /\partial \bar{%
\beta}^{k}$ of the density operators \textup{(\ref{3 2 1})} at $\hat{x}%
=\varphi (\hat{a})$ is right efficient and can be reduced to a coherent
measurement and estimation of $\theta =\varphi (\alpha )$ by the result $%
\alpha $. In the particular case where $\varphi (\alpha )$ is a linear
function and $S_{0}$ is a Gaussian state this fact has been established in 
\textup{\cite{5}}.

Note that along with right and left lower bounds one can consider other
combined bounds via the factorization $\theta =\theta _{+}+\theta _{-}$ by
appropriately defining the right derivatives with respect to $\theta _{+}$
and the left derivatives with respect to $\theta _{-}$. An interesting
question arising in this connection is whether the class of efficient
estimations is exhausted by the estimations for which at least one such
bound is attained.

Let us now consider the (right) efficiency of estimating the parameters $%
\beta ^{k}$ of the canonical families (\ref{3 2 1}). The inequality (\ref{3
1 17}) corresponding to this case with $\theta ^{k}=\beta ^{k}$ has the form 
$R\geq H^{-1}$, where $H$ is the matrix of derivatives (\ref{3 2 5}).
Without loss of generality, we can assume that $\mathrm{Tr}\;(\hat{x}%
_{k}S_{0})=0$.

\begin{theorem}
\label{T 3 2 3}The inequality $R\geq H^{-1}$ transforms into an equality if
and only if the operators $\hat{x}_{k}$ in \textup{(\ref{3 2 1})} possess a
right joint decomposition of unity \textup{(\ref{3 2 19})}, the generating
function of the moments \textup{(\ref{3 2 2}) }of these operators in state $%
S_{0}$ is Gaussian, $\chi (\beta ,\bar{\beta})=\exp \{\bar{\beta}%
^{i}H_{ik}\beta ^{k}\}$, with $H_{ik}$ independent of $\beta $ and $\bar{%
\beta}$ and linear functions $y^{k}=(H^{-1})^{ki}x_{i}$ of the results $%
x_{k} $ of joint quasimeasurement of observables $\hat{x}_{k}$ are selected
for the estimators $\hat{\beta}^{k}$.
\end{theorem}

\begin{proof}
Sufficiency of the above-formulated conditions for the existence of
right-efficient estimation is obvious: the fact that matrix $H$ coincides
with the covariance matrix $S$ of operators $\hat{x}_{k}$ implies that the
covariance matrix $R=H^{-1}SH^{-1}$ is equal to $H^{-1}$. Necessity follows
from the necessary conditions of right efficiency in Theorem \ref{T 3 2 2},
according to which the family $S(\beta ,\bar{\beta})$ must have the form 
\begin{equation}
S(\beta ,\bar{\beta})=\psi ^{-1}\mathrm{e}^{\theta _{k}\hat{y}^{k\ast }}S_{0}%
\mathrm{e}^{\bar{\theta}_{k}\hat{y}^{k}},  \label{3 2 23}
\end{equation}%
where $\psi =\mathrm{Tr}\;[S_{0}\mathrm{e}^{\bar{\theta}_{k}\hat{y}^{k}}%
\mathrm{e}^{\theta _{k}\hat{y}^{k\ast }}]$, $\beta ^{k}=\partial \left(
\theta _{k}\ln \psi \right) /\partial \theta _{k}$, and the operators $\hat{y%
}^{k}$ possess the joint right unity decomposition in the sense: 
\begin{equation*}
I=\int M(dx),\quad \hat{y}^{k}M(dy)=y^{k}M(dy),\quad y=\left( y^{k}\right)
\in \mathbb{C}^{n}.
\end{equation*}%
Comparing (\ref{3 2 14}) with (\ref{3 2 1}), we conclude that $\theta _{k}%
\hat{y}^{k}=\bar{\beta}^{k}\hat{x}_{k}$, whence 
\begin{equation*}
\theta _{k}=H_{ki}\bar{\beta}^{i},\quad \psi (\theta ,\bar{\theta})=\chi
(\beta ,\bar{\beta})=\bar{\beta}^{i}H_{ik}\beta ^{k},\quad \hat{y}%
^{k}=(H^{-1})^{ki}\hat{x}_{i}.
\end{equation*}%
The proof of Theorem \ref{T 3 2 3} is complete.
\end{proof}

\subsubsection{Proof of Inequality (\protect\ref{3 1 17})}

\label{proof of inequality}

\textbf{1. }We start with the one-dimensional case. Let $\hat{x}$ be a
non-Hermitian operator in $\mathcal{H}$ for which 
\begin{equation}
\mathrm{Tr}\;[\hat{x}S(\alpha ,\bar{\alpha})]=\theta (\alpha ,\bar{\alpha}).
\label{3 2 24}
\end{equation}%
Differentiating (\ref{3 2 24}) with respect to $\alpha $ and employing
definition (\ref{3 1 15}) and the normalization condition $\mathrm{Tr}%
\;S(\alpha ,\bar{\alpha})=1$, according to which $\mathrm{Tr}\;(S\hat{h}%
^{\ast })=0$, we obtain 
\begin{equation*}
\partial \theta /\partial \alpha =\mathrm{Tr}\;[S(\hat{x}-\theta )\hat{h}%
^{\ast }].
\end{equation*}%
Since the covariance $\mathrm{Tr}\;[S(\hat{x}-\theta )\hat{h}^{\ast }]$
obeys the Schwarz inequality 
\begin{equation}
|\mathrm{Tr\;}[S(\hat{x}-\theta )\hat{h}^{\ast }]|^{2}\leq \mathrm{Tr}\;[S(%
\hat{x}-\theta )(\hat{x}-\theta )^{\ast }]\mathrm{Tr}\;(S\hat{h}\hat{h}%
^{\ast }),  \label{3 2 25}
\end{equation}%
which reflects the fact that the determinant of the $2$-by-$2$ covariance
matrix $\mathrm{Tr}\;(S\hat{h}_{i}\hat{h}_{k}^{\ast }),\;i=0,1$, with $\hat{h%
}_{0}=(\hat{x}-\theta )$ and $\hat{h}_{1}=\hat{h}$, is nonnegative, we can
write 
\begin{equation}
\mathrm{Tr}\;[S(\hat{x}-\theta )(\hat{x}^{\ast }-\bar{\theta})\geq |\partial
\theta /\partial \alpha |^{2}/\mathrm{Tr}\;(S\hat{h}\hat{h}^{\ast }).
\label{3 2 26}
\end{equation}%
This inequality, obviously, specifies the lower bound on the variance of the
estimation of parameter $\theta =\theta \left( \alpha ,\bar{\alpha}\right) $
in the class of ordinary measurements described by normal operators $\hat{x}$%
. But since the normality condition, $\hat{x}\hat{x}^{\ast }=\hat{x}^{\ast }%
\hat{x}$, was not used in deriving (\ref{3 2 26}), this bound is the lower
one for the variance of any estimators $\hat{\theta}$ obtained as a result
of arbitrary generalized measurements described in $\mathcal{H}$ by
decompositions of unity $I=\int M(dx),\;x\in \mathbb{C}$ that may be
nonorthogonal. Indeed, the nonnegative definiteness 
\begin{equation}
(\hat{x}-x)M(dx)(\hat{x}-x)^{\ast }\geq 0\quad (M\geq 0)  \label{3 2 27}
\end{equation}%
implies 
\begin{equation}
\int |x-\theta |^{2}M(dx)\geq (\hat{x}-\theta )(\hat{x}-\theta )^{\ast },
\label{3 2 28}
\end{equation}%
where $\hat{x}=\int xM(dx)$, and $\theta =\mathrm{Tr}\;(S\hat{x})$. Taking
the mathematical expectations of both sides of (\ref{3 2 27}), allowing for
the fact that the variance $R$ of the estimator $\hat{\theta}=x$ is equal to 
$\mathrm{Tr}\;S\times \int |x-\theta |^{2}M(dx)$, and combining the result
with (\ref{3 2 26}), we find that 
\begin{equation}
R\geq \mathrm{Tr}\;[S(\hat{x}-\theta )(\hat{x}=\theta )^{\ast }]\geq
|D|^{2}/H,  \label{3 2 29}
\end{equation}%
where $D=\partial \theta /\partial \alpha $, and $H=\mathrm{Tr}\;(S\hat{h}%
\hat{h}^{\ast })$. This proves inequality (\ref{3 1 17}) for the
one-dimensional case.

\textbf{2. }The equality in (\ref{3 2 29}) occurs if, first, the averages of
both sides of (\ref{3 2 28}) coincide and if, second, the Schwarz inequality
transforms into an equality. Actually, the first requirement establishes an
equality in (\ref{3 2 27}). Specifically, we have the following.

\begin{lemma}
Suppose that the ranges $S(\alpha ,\bar{\alpha})\mathcal{H}$ of the density
operators from a family $\{S(\alpha ,\alpha ),\;\alpha \in \mathcal{O}\}$
generate the entire space $\mathcal{H}$. Then the fact that $\mathrm{Tr\;}%
(SA)=0$ for every nonnegative definite operator $A$ in $\mathcal{H}$ and all 
$\alpha \in \mathcal{O}$ implies that $A=0$.
\end{lemma}

\begin{proof}
It is sufficient to prove that in $\mathcal{H}$ there is no vector $\chi $
of the form $\chi =S^{1/2}\psi $ for which $(\chi |A|\chi )\neq 0$. But this
follows from the well-known inequality 
\begin{equation*}
\mathrm{Tr}\;(S^{1/2}AS^{1/2})\geq (\psi |S^{1/2}AS^{1/2}|\psi ),
\end{equation*}
which is true for every nonnegative $A$ at $(\psi \mid \psi )=1$.

Applying this result to the operator $A$ that is equal to the difference
between the right- and left-hand sides of (\ref{3 2 28}), we find that under
the lemma's hypothesis the equality in (\ref{3 2 28}) occurs only if 
\begin{equation*}
(\hat{x}-x)M(dx)(\hat{x}-x)^{\ast }=0,\quad \text{i.e.}\quad \hat{x}%
M(dx)=xM(dx).
\end{equation*}%
This proves that right-efficient estimation in a certain region $\mathcal{O}%
\ni \alpha $ exists if there is an operator of minimal sufficient
statistics, $\hat{x}$, possessing a right proper decomposition of unity in
the subspace generated by the subspaces $S(\alpha ,\bar{\alpha})\mathcal{H}$%
. In the real case, $x\in \mathbb{R}$, such an operator $\hat{x}$ is
obviously Hermitian.

The second requirement for equality to occurs in (\ref{3 2 29}) is
equivalent to the condition of linear dependence, $s\hat{h}=\bar{\lambda}%
S(x-\theta )$, where $\lambda =D/R$ if the first condition for equality in (%
\ref{3 2 28}) is met. Extending this condition over the entire region $%
\mathcal{O}\ni \alpha $ in which the analyticity condition (\ref{3 2 8}), $%
\partial \lambda /\partial \bar{\alpha}=0$, is assumed to hold true, we
arrive at the equation%
\begin{equation*}
\partial S/\partial \bar{\alpha}=\bar{\lambda}S(\hat{x}-\theta ),\;\partial
S/\partial \alpha =(\hat{x}-\theta )^{\ast }S\lambda
\end{equation*}
in $S=S(\alpha ,\bar{\alpha})$. Its solution combined with the boundary
condition $S(\alpha _{0},\bar{\alpha}_{0})=S_{0}$ has the canonical form (%
\ref{3 2 1}), where $\beta \left( \alpha \right) =\int_{\alpha _{0}}^{\alpha
}\lambda (\alpha )$ is an analytic function, and $\hat{x}$ is the operator
of right-efficient statistics. This proves that in the one-dimensional case
the existence of right-efficient estimation requires that the density
operators $S(\alpha ,\bar{\alpha})$ be canonical. This condition is
formulated in Theorem \ref{T 3 2 2}. For the real case, $\hat{x}^{\ast }=%
\hat{x}$, this fact also proves the necessity in Theorem \ref{T 3 2 1}.

\textbf{3. }The multidimensional generalization can be carried out if for $%
\hat{x}-\theta $ and $\hat{h}$ we take the sums $(\hat{x}^{i}-\theta ^{i})%
\bar{\eta}_{i}$ and $\hat{h}_{k}\bar{\xi}^{k}$, where $\eta
_{i},\;i=1,\ldots ,m$ and $\xi ^{k},\;k=1,\ldots ,n$, are complex numbers.
If we allow for the fact that here%
\begin{equation*}
\mathrm{Tr\;}[S(\hat{x}-\theta )\hat{h}^{\ast }]=\bar{\eta}_{i}(\partial
\theta ^{i}/\partial \alpha ^{k})\xi ^{k},
\end{equation*}%
then from (\ref{3 2 25}) at $\xi ^{k}=(H^{-1}D^{\dagger })^{ki}\eta _{i}$ we
arrive at the inequality 
\begin{equation*}
R^{ik}\bar{\eta}_{i}\eta _{k}\geq \mathrm{Tr}\;[S(\hat{x}^{i}-\theta ^{i})(%
\hat{x}^{k}-\theta ^{k})^{\ast }\eta _{i}\eta _{k}]\geq (DH^{-1}D^{\dagger
})^{ik}\bar{\eta}_{i}\eta _{k}
\end{equation*}%
valid for an arbitrary $\hat{x}^{i}$ for which $\mathrm{Tr}\;(S\hat{x}%
^{i})=\theta ^{i}$. Putting $\hat{x}^{i}=\int x^{i}M(dx)$, where $\int
M(dx)=I,\,\hat{x}\in \mathbb{C}^{m}$, is the decomposition of unity
describing the estimator $\hat{\theta}^{i}=x^{i}$, and applying inequality (%
\ref{3 2 28}) with $\hat{x}=\hat{x}^{i}\bar{\eta}_{i}$ and $\theta =\theta
^{i}\bar{\eta}_{i}$, we obtain for the matrix $R$ of covariances of $\hat{%
\theta}^{i}$ the first inequality in (\ref{3 2 29}), which in view of the
arbitrariness of $\eta _{i}$ yields (\ref{3 1 17}).

Inequality (\ref{3 2 29}) transforms into an equality at $\alpha \in 
\mathcal{O}$ only when $\hat{x}^{i}M(dx)=x^{i}M(dx)$ and $\partial
S/\partial \bar{\alpha}^{k}=\bar{\lambda}_{ki}S(x^{i}-\theta ^{i})$, where $%
\lambda _{ik}=(R^{-1}D)_{ik}$, whence, if we allow for the regularity
conditions $\lambda _{ik}$, we arrive at (\ref{3 2 1}).
\end{proof}

\section{Optimal and Covariant Estimation of Wave Patterns}

In this section we will consider the necessary and sufficient conditions for
the optimality of measuring sound and visual patterns by the criterion of
mean square error in parameter estimation and by the maximal intensity
criterion. To avoid substantiation of the operator integrals involved in the
discussion this is done in \cite{32}, we interpret them as operator-valued
Radon measures. The solution to the optimal measurement problem will be
found for homogeneous families of wave patterns for which it coincides with
optimal covariant measurements of the corresponding parameters of quantized
fields, with the latter measurements introduced in \cite{29}.

\subsection{Optimal Measurements}

The problems of optimal estimation of continuous wave parameters constitute
essentially multialternative problems with an infinite-dimensional solution
space (or manifold) $X$. without loss of generality, we can assume that the
information parameter space $\Theta $ coincides with $X$ equipped with
measure $\mathrm{d}\lambda $. Let us assume that a wave signal, which in
general is described by a density operator $S$, depends in a continuous
manner on real- or complex-valued random parameters $\theta =(\theta
_{1},\ldots ,\theta _{n}),\;S=S_{\theta }$, having a given \textit{a priori}
distribution $P(d\theta )$. The deviation of the estimate $x\in X$ from $%
\theta $ is penalized by an integrable cost function $c_{x}(\theta )$ of the
form, say, $(x-\theta )^{2}$. On $X$ we must find an optimal
quasimeasurement that (a) is described by an operator-valued measure $M(dx)$%
, (b) determines the decomposition of unity in the Hilbert space $\mathcal{H}
$, and (c) minimizes the mean estimation cost 
\begin{equation*}
\left\langle c\right\rangle =\iint \mu _{\theta }(dx)c_{x}(\theta )P(d\theta
)=\int \mathrm{Tr}\;R_{x}M(dx),
\end{equation*}%
where $\mu _{\theta }(dx)=\mathrm{Tr}\;M(dx)S_{\theta }$ is the observed
intensity distribution on $X$ for a given $\theta $, and $R_{x}=\int
c_{x}(0)\;S_{\theta }P(d\theta )$ is the operator of the mean cost $x\in X$.
We will now formulate the necessary and sufficient conditions for the
optimality of solution $M^{\mathrm{o}}$ to this extremal problem, which in 
\cite{32} were introduced to estimate the parameters of quantum states. This
will be done in a manner similar to that of Theorem \ref{T 2 3 1}:

\begin{theorem}
The lower bound 
\begin{equation}
\inf_{M\geq 0}\left\{ \int \left\langle R_{x},M(dx)\right\rangle \Big|\int
M(dx)=I\right\}  \label{3 3 1}
\end{equation}%
is attained on measure $M^{\mathrm{o}}$ if and only if for almost all $x\in
X $ there exists a minorant operator $\Lambda ^{\mathrm{o}}\leq R_{x}$ such
that 
\begin{equation}
(R_{x}-\Lambda ^{\mathrm{o}})M^{\mathrm{o}}(dx)=0\quad \forall x\in X.
\label{3 3 2}
\end{equation}%
The operator $\Lambda ^{\mathrm{o}}$ is a trace class operator, or $\mathrm{%
Tr}\;\Lambda ^{\mathrm{o}}=\left\langle \Lambda ^{\mathrm{o}},I\right\rangle
<\infty $, that determines the solution to the duality problem 
\begin{equation}
\sup_{\Lambda }\{\left\langle \Lambda ,I\right\rangle |\Lambda \leq
R_{x},\;x\in X\}  \label{3 3 3}
\end{equation}%
for which conditions \textup{(\ref{3 3 2})} are also necessary and
sufficient \textup{(}if we allow for the fact that $M^{\mathrm{o}}\geq 0$
and $\int M^{\mathrm{o}}(dx)=E$\textup{).}
\end{theorem}

\begin{proof}
For a proof of this theorem as well as for the existence conditions for a
solution see \cite{32}.

Allowing for the fact that the operators $M(dx)$ can be decomposed into
operators of the form $|\chi _{x})(\chi _{x}|\;\mathrm{d}\lambda (x)$, where
the $\chi _{x}$ are the generalized elements of space $\mathcal{H}$, we find
that the problem of optimal estimation of wave parameters will be solved if
and only if we can find a family of reference waves, $\{\chi _{x}\}$
satisfying the completeness condition 
\begin{equation}
\int |\chi _{x})(\chi _{x}|\;\mathrm{d}\lambda (x)=I  \label{3 3 4}
\end{equation}%
and a Hermitian operator $\Lambda $ for which 
\begin{equation}
R_{x}=\Lambda \geq 0,\quad (R_{x}-\Lambda )\chi _{x}=0,\quad x\in X.
\label{3 3 5}
\end{equation}

Note that, in contrast to problems of signal discrimination, in problems of
parameter estimation the commutative case $R_{x}R_{x^{\prime }}=R_{x^{\prime
}}R_{x}$, which can be reduced to the classical case, is of no practical
interest and will not be discussed here.

The solution of problem (\ref{3 3 5}) poses no fundamental difficulties in
the case of a single unknown real-valued parameter $\theta (\chi =\mathbb{R}%
^{1})$ and a quadratic penalty function 
\begin{equation*}
C_{x}(\theta )=(x-0)^{2}.
\end{equation*}%
The mean estimation cost operator 
\begin{equation*}
R_{x}=\int (x-\theta )^{2}S_{\theta }P(d\theta )
\end{equation*}%
in the case of (\ref{3 3 5}) can be represented, via three Hermitian
operators 
\begin{equation}
R^{(k)}=\int \theta ^{k}S_{\theta }P(d\theta ),\quad k=0,1,2,  \label{3 3 6}
\end{equation}%
in the form 
\begin{eqnarray*}
R_{x} &=&x^{2}R^{(0)}-2xR^{(1)}+R^{(2)} \\
&=&(\hat{x}-x)R^{\left( 0\right) }(\hat{x}-x)+R^{(2)}-\hat{x}R^{(0)}x,
\end{eqnarray*}%
where $\hat{x}$ is an operator satisfying the equation 
\begin{equation}
\hat{x}R^{(0)}+R^{(0)}\hat{x}=2R^{(1)}.  \label{3 3 7}
\end{equation}%
and for $\chi _{x}$ take the complete orthogonal system of generalized
eigenvectors determining the spectral decomposition of the Hermitian
operator $\hat{x},$%
\begin{equation*}
(\hat{x}-x)\chi _{x}=0,\hat{x}=\int x|\chi _{x})(\chi _{x}|\;\mathrm{d}x,
\end{equation*}%
the conditions (\ref{3 3 5}) are satisfied in an obvious manner: 
\begin{equation*}
(\hat{x}-x)R^{(0)}(\hat{x}-x)\geq 0,\quad (\hat{x}-x)R^{(0)}(\hat{x}-x)\chi
_{x}=0.
\end{equation*}%
Thus, the solution of the parameter estimation problem by criterion (\ref{3
3 5}) is reduced to measuring operator $\hat{x}$ satisfying (\ref{3 3 7}).
The result of such a measurement, $x$, leads to the minimal error $%
\left\langle c\right\rangle =\mathrm{Tr}\;\Lambda ^{\mathrm{o}}$ equal to
the a posteriori variance 
\begin{equation*}
\sigma ^{2}=\mathrm{Tr}\;(R^{(2)}-\hat{x}R^{(0)}x).
\end{equation*}
\end{proof}

As an example, let us consider the estimation of the amplitude of a coherent
signal of known shape received against a background of Gaussian noise. The
density operator of the corresponding mode has the Gaussian form 
\begin{equation}
S(\theta )=\int |\alpha )(\alpha |\bar{n}^{-1}\exp \left\{ -\frac{|\alpha
-\theta |^{2}}{\bar{n}}\right\} \pi ^{-1}\mathrm{d}\func{Re}\alpha \mathrm{d}%
\func{Im}\alpha ,  \label{3 3 8}
\end{equation}%
where $\theta $ is the amplitude, which assumes real values, $\theta \in 
\mathbb{R}^{1}$. We assume that amplitude $\theta $ has a Gaussian prior
density 
\begin{equation*}
p(\theta )=(2\pi \bar{s})^{-1/2}\exp \{-\theta ^{2}/2\bar{s}\},
\end{equation*}%
where $\bar{s}$ is the prior variance, $\left\langle \theta
^{2}\right\rangle =\bar{s}$. It is then easy to find, via the formulas of
Gaussian integration, the operators (\ref{3 3 6}), which define the mean
decision cost operator 
\begin{equation*}
R_{x}=\left( x-\frac{2\bar{s}}{2\bar{s}+\bar{n}+1/2}Q\right) S\left( x-\frac{%
2\bar{s}}{2\bar{s}+\bar{n}+1/2}Q\right) +\frac{2\bar{s}(\bar{n}+1/2)}{2\bar{s%
}+\bar{n}+1/2}S.
\end{equation*}%
Here $Q=\int \func{Re}\alpha |\alpha )(\alpha |\pi ^{-1}\;\mathrm{d}\func{Re}%
\alpha \;\mathrm{d}\func{Im}\alpha $ is the operator of the
\textquotedblleft coordinate\textquotedblright\ of the harmonic oscillator
representing this mode, and 
\begin{equation*}
S=\int |\alpha )(\alpha |\left( (2\bar{s}+\bar{n})\bar{n}\right) ^{-1/2}\exp
\left\{ -\frac{(\func{Re}\alpha )^{2}}{2\bar{s}+\bar{n}}-\frac{(\func{Im}%
\alpha )^{2}}{\bar{n}}\right\} \pi ^{-1}\,\mathrm{d}\func{Re}\alpha \,%
\mathrm{d}\func{Im}\alpha
\end{equation*}%
is the density operator, with $\int S_{\theta }p(\theta )\;\mathrm{d}\theta
=R^{(0)}$. Hence, optimal estimation of the amplitude of a Gaussian signal
is reduced to measuring the coordinate operator $Q$, whose result $q$
determines the optimal estimate 
\begin{equation*}
x=2\bar{s}q/(2\bar{s}+\bar{n}+1/2)
\end{equation*}%
with a minimal mean square error 
\begin{equation*}
\sigma ^{2}=2\bar{s}(\bar{n}+1/2)(2\bar{s}+\bar{n}+1/2).
\end{equation*}

\subsection{Miltidimensional Optimal Estimation Problem}

In the case $(n>1)$ even for the quadratic quality criterion 
\begin{equation*}
C_{x}(\theta )=\sum_{j=1}^{n}(x_{j}-\theta _{j})^{2}
\end{equation*}%
the general solution to problem (\ref{3 3 5}) is unknown. Only in the
particular case where the operators $\hat{x}=\{\hat{x}_{j}\}$ obeying (\ref%
{3 3 7}), with 
\begin{equation*}
R^{k}=R_{j}^{k}\equiv \int \theta _{j}^{k}S_{\theta }p(\theta )\;\mathrm{d}%
\theta _{1}\ldots \mathrm{d}\theta _{n},\quad k=0,1,2,
\end{equation*}%
commute with each other ($x_{j}x_{i}=x_{i}x_{j}$), optimal estimation is
reduced, obviously, to joint measurement of these operators.

In general, a good estimate of parameters $\theta _{j}$ can be obtained by
an indirect measurement of noncommutative operators $\{\hat{x}_{i}\}$ (see
Section \ref{Quasimeasurements}). However, this estimate is not necessarily
optimal, even if the indirect measurement is ideal.

For an example let us take the complex-valued one-dimensional case ($X=%
\mathbb{C}^{1}$), which can also be interpreted as the real-valued
two-dimensional: 
\begin{equation*}
C_{x}(\theta )=|x-\theta |^{2}=(\func{Re}(x-\theta ))^{2}+(\func{Im}%
(x-\theta ))^{2}.
\end{equation*}%
An exact solution to the problem of optimal estimation of a single parameter 
$\theta $ has been obtained in \cite{11} for this case of a quadratic
penalty function, a Gaussian state $S_{\theta }$; of the form (\ref{3 3 8}),
with $\theta \in \mathbb{C}^{1}$, and a Gaussian prior probability density 
\begin{equation}
p(\theta )=\bar{s}^{-1}\exp \{-|\theta |^{2}/\bar{s}\},\quad \bar{s}%
=\left\langle |\theta |^{2}\right\rangle .  \label{3 3 9}
\end{equation}%
The density $p(\theta )$ is normalized with respect to 
\begin{equation*}
\mathrm{d}\lambda (\theta )=\pi ^{-1}\mathrm{d}\func{Re}\theta \mathrm{d}%
\func{Im}\theta .
\end{equation*}%
In this case, by the standard formulas of Gaussian integration, we can
easily find the mean decision cost operator 
\begin{equation*}
R_{x}=\left( x^{\ast }-\frac{\bar{s}}{\bar{n}+\bar{s}+1}A^{\ast }\right)
S\left( x-\frac{|\alpha |^{2}}{\bar{s}+\bar{n}}\right) \;\mathrm{d}\lambda
(\alpha )
\end{equation*}%
is the density operator $S=\int S_{\theta }p(\theta )\;\mathrm{d}\lambda
(\theta )$. Assuming that 
\begin{equation*}
\Lambda =\frac{\bar{s}(\bar{n}+1)}{\bar{s}+\bar{n}+1}S,\quad \chi _{x}=c^{-1}%
\Big|\left( 1+\frac{\bar{n}+1}{\bar{s}}\right) x\Big),
\end{equation*}%
where $|\alpha )$, $\alpha =(1+(\bar{n}+1)/\bar{s})x$, are coherent vectors,
and $c=\bar{s}/(\bar{s}+\bar{n}+1)$ is a coefficient that can be found from
condition (\ref{3 3 4}) if we allow for the completeness of coherent states 
\begin{equation*}
\int |\alpha )(\alpha |\;\mathrm{d}\lambda (\alpha )=I,
\end{equation*}%
and allowing for the equation $A|\alpha )=\alpha |\alpha )$, we find that
conditions (\ref{3 3 5}) are met: 
\begin{equation}
(x^{\ast }-cA^{\ast })S(x-cA)\geq 0,\quad (x^{\ast }-cA^{\ast
})S(x-cA)|c^{-1}x)=0.  \label{3 3 10}
\end{equation}%
Thus, optimal estimation in the one-dimensional complex-valued
quadratic-Gaussian case is reduced to a coherent measurement describing an
ideal proper indirect measurement of the annihilation operator $A$ whose
result $\alpha $ determines the estimate 
\begin{equation*}
x=\frac{\bar{s}}{\bar{s}+\bar{n}+1}\alpha
\end{equation*}%
with a minimal error 
\begin{equation*}
\sigma ^{2}=\frac{\bar{s}(\bar{n}+1)}{\bar{s}+\bar{n}+1}.
\end{equation*}%
This error is equal to the error of the appropriate classical problem of
estimation in a Gaussian linear channel with a noise intensity of $\bar{n}+1$%
. The quantity $\bar{n}$ (the mean number of the noise quanta) is determined
by the noise proper in the wave channel, while the unity corresponds to the
\textquotedblleft effective noise\textquotedblright\ thanks to the
inaccuracy in the ideal indirect measurement. The measurement noise of unit
intensity can be interpreted as the noise produced by an ideal wave
amplifier or as the noise produced by an ideal optical heterodyne.

\subsection{Optimal Measurement of Wave States}

Let $X$ be a set of hypotheses concerning the states of a wave field, and $\{%
\hat{R}_{x},\;x\in X\}$ the respective decomposable family of density
operators $R_{x}=\bigoplus_{n}R_{x}^{(n)}$ in the Hilbert space $\mathcal{H}%
=\bigoplus_{n}\mathcal{H}^{(n)}$. Then the set of measurements described by
operator-valued measures $M$ on $X(M(\cdot )\geq 0,\;\int M(\mathrm{d}x)=I)$
of the decomposable form $M(dx)=\bigoplus_{n}M^{(n)}\;\mathrm{d}x$ is
sufficient. The optimal strategy is described by the family of
operator-valued measures $M^{(n)}$ on $X$, i.e. $M^{(n)}(\cdot )\geq
0,\;\int M^{(n)}(dx)=I^{(n)}$, defined independently in $\mathcal{H}^{(n)}$
for every $n$ by the conditions 
\begin{equation}
(R_{x}^{(n)}=\Lambda ^{(n)})M^{(n)}(dx)=0,\quad R_{x}^{(n)}\leq \Lambda
^{(n)}\quad \forall z\in X  \label{3 3 11}
\end{equation}%
(the maximum intensity criterion). Here $\Lambda ^{(n)}$ are Hermitian trace
class operators in $\mathcal{H}^{(n)}$ that are nonnegative (for $%
R_{x}^{(n)}\geq 0$) and can be represented in the form 
\begin{equation*}
\Lambda ^{2}=\int R_{x}^{(n)}M^{(n)}(dx)R_{x}^{(n)}.
\end{equation*}%
Problem (\ref{3 3 11}) is incomparably simpler than the general problem of
optimal discrimination of a family $\{R_{x}\}$ and for every $n$ has a
finite-dimension of space $\mathcal{H}^{(n)}$ if the signal space $\mathcal{L%
}$ is finite-dimensional. In what follows, the index $n$ will be dropped.

Let $\mathcal{U}_{x}=R_{x}\mathcal{H}$ be the range of values of operators $%
R_{x}$ in $\mathcal{H}$, let $\mathcal{U}(dx)$ be their algebraic sum for
all $x\in dx$, and let $\mathcal{U}=\int \mathcal{U}(dx)$ be the sum of all
the subspace $\mathcal{U}_{x}\subset \mathcal{H}$. Each nonnegative operator 
$R_{x}$ can be represented in the form $R_{x}=\psi _{x}\psi _{x}^{\ast }$,
where $\psi _{x}$ is the operator from $\mathcal{U}_{x}$ into $\mathcal{U}$.
The following conjectures are multidimensional generalizations of the
appropriate assertions of Theorem \ref{T 2 2 2} (for a discrete set $X$).

\begin{theorem}
\textup{(1)} Subspace $\mathcal{U}$ is sufficient for solving problem 
\textup{(\ref{3 3 11})}. Every operator $\Lambda $ satisfying conditions $%
\Lambda \geq R_{x}\;\forall x\in X$ for $R_{x}\geq 0$ has an inverse $%
\Lambda ^{-1}$ in $\mathcal{U}$.\newline
\textup{(2)} The solution to problem \textup{(\ref{3 3 11}) }in the
sufficient space $\mathcal{U}$ has the form 
\begin{equation}
M(dx)=\Lambda ^{-1}\psi _{x}\hat{\mu}(dx)\psi _{x}^{\ast }\Lambda ^{-1},%
\text{ where }\Lambda =\int \Big(\int \psi _{x}\hat{\mu}(dx)\psi _{x}^{\ast }%
\Big)^{1/2}  \label{3 3 12}
\end{equation}%
and $\hat{\mu}$ is a measure on $X$ whose values $\hat{\mu}\;(dx)$ are
nonnegative operators in $\mathcal{U}_{x}$ defined by the conditions 
\begin{equation}
(\psi _{x}^{\ast }\Lambda ^{-1}\psi _{x}-I_{x})\hat{\mu}(dx)=0,\quad \psi
_{x}^{\ast }\Lambda ^{-1}\psi _{x}\geq I_{x}\quad \forall x\in X
\label{3 3 13}
\end{equation}%
\textup{(}$I_{x}$ is the identity element in $\mathcal{U}_{x}$\textup{)}.%
\newline
\textup{(3) }If the subspace $\mathcal{U}(dx)$ does not intersect with the
sum $\mathcal{U}\overline{(dx)}$ of all the remaining subspaces $\mathcal{U}%
_{y},\;y\notin dx$, the operator $\hat{\mu}(dx)$ is strictly positive in $%
\mathcal{U}_{x}$ and is defined by the condition $\psi _{x}^{\ast }\Lambda
^{-1}\psi _{x}=I_{x},\;x\in X$.
\end{theorem}

\subsection{Waves with Group Symmetry}

Equations (\ref{3 3 13}) are considerably simpler than Eqn. (\ref{3 3 11})
since the dimensionality of each operator equation in (\ref{3 3 13}) is
equal to rank $r(R_{x})$. For the case where $r(R_{x})=1$ the solution has
been found \cite{4} under the condition that the square root of the
correlation matrix $[\psi _{x}^{\ast }\psi _{y}]$ has equal diagonal
elements. An analog of this condition in the general case where $%
r(R_{x})\geq 1$ is the condition of group (say, cyclic in \cite{5}) symmetry
of the family $\{R_{x}\}$.

Let $X$ be a homogeneous set with respect to a group $G$, that is, group $G$
acts on $X$ transitively, and let $U(g),\;g\in G$, be a unitary
representation of $G$ in $\mathcal{H}$. The family $\{R_{x},\,x\in X\}$ is
said to be $G$-homogeneous (or $G$-invariant) if $X$ is a homogeneous set
with respect to group $G$ and $U(g)R_{gx}^{-1}U^{\ast }(g)=R_{x}$.

Let $G$ be a finite compact or locally compact group, $\mathrm{d}g$ be the
left Haar measure on $G$, the family $\{R_{x}\}$ be homogeneous and
continuous in $U(g)$. The following conjectures are true:

\begin{theorem}
\textup{(1)} The sufficient space $\mathcal{U}$ is a subspace in $\mathcal{H}
$ cyclically generated by the family $\{U(g),\,g\in G\}$ over $\mathcal{U}%
_{0}=R_{x_{0}}\mathcal{H}$, where $x_{0}$ is any element belonging to $X$.
The operators $R_{x}$ in $\mathcal{U}$ can be represented in the form 
\begin{equation*}
R_{x}=U(g)\psi \psi ^{\ast }U^{\ast }(g)\quad \forall g\in G,
\end{equation*}%
where $U(g)$ is a subrepresentation induced in $\mathcal{U}\subset \mathcal{H%
}$, $\psi $ an operator from $\mathcal{U}_{0}$ into $\mathcal{U}$, and $%
G_{x} $ the left coset $G_{x}=\{g:gx_{0}=x\}$ over the stationary subgroup $%
G_{0}=G_{x_{0}}$ of element $x_{0}$.\newline
\textup{(2)} The optimal strategy \textup{(\ref{3 3 12}) }has the covariant
form 
\begin{equation}
M(dx)=\int_{G(dx)}U(g)\Lambda ^{-1}\psi \hat{\mu}\psi ^{\ast }\Lambda
^{-1}U^{\ast }(g)\;\mathrm{d}g,  \label{3 3 14}
\end{equation}%
where $G(dx)=\bigcup_{x\in dx}G_{x}$ the union of the $G_{x}$ over all $x\in
dx$,%
\begin{equation*}
\Lambda =\Big(\int U(g)\psi \hat{\mu}\psi ^{\ast }U^{\ast }(g)\;\mathrm{d}g%
\Big)^{1/2}
\end{equation*}%
is the $G$-invariant, and $\hat{\mu}$ a non-negative operator in $\mathcal{U}%
_{0}$ satisfying the conditions 
\begin{equation}
\left( \psi ^{\ast }\Lambda ^{-1}\psi -\hat{I}\right) \hat{\mu}=0,\quad \psi
^{\ast }\Lambda ^{-1}\psi \leq \hat{I}.  \label{3 3 15}
\end{equation}%
\textup{(3)} If the representation $U(g)$ in $\mathcal{U}$ is topologically
irreducible, then operator $\Lambda $ is a multiple of the identity element $%
\hat{I}$ of space $\mathcal{U}_{0}:\Lambda =\lambda I$ and operator $\hat{\mu%
}$ is proportional to the proper projector $\hat{\pi}$ of operator $\hat{R}%
=\psi ^{\ast }\psi $ corresponding to its maximal eigenvalue $\lambda :\hat{%
\mu}=\mu \hat{\pi}$, $(\hat{R}-\Lambda )\hat{\pi}=0$.

\textup{The proportionality factor can be made equal to unity by
appropriately renormalizing} $\mathrm{d}g$. \textup{For the particular case
where} $G$ \textup{is finite and its action on} $X$ \textup{is effective
this result was obtained earlier in \cite{24}. If} $G$ \textup{is an Abelian
group, the case is trivial and can be of no interest.} \newline
\textup{(4)} Let $U_{\omega }(g),\;\omega \in \Omega $, be the field of
nonequivalent irreducible representations $U_{\omega }(g)$ in space $%
\mathcal{H}_{\omega }$, and $\mathrm{d}\omega $ the Plancherel measure. Then 
\textup{(\ref{3 3 15})} can be represented in the form 
\begin{equation}
\int_{\Omega }\mathrm{Tr}_{\mathcal{H}_{\omega }}(\hat{R}_{\omega }\hat{\mu}%
)^{1/2}\;\mathrm{d}\omega =\hat{\mu},\quad \int_{\Omega }\mathrm{Tr}_{%
\mathcal{H}_{\omega }}(\hat{R}_{\omega }\hat{\mu})^{-1/2}\hat{R}_{\omega }\;%
\mathrm{d}\omega \leq \hat{I},  \label{3 3 16}
\end{equation}%
where $\hat{R}_{\omega }=\int \hat{r}(g)U_{\omega }(g)\;\mathrm{d}g$ is the
Fourier transform of the operator correlation function $\hat{r}(g)=\psi
^{\ast }U^{\ast }(g)\psi $. If the family $\{U_{\omega }(\cdot )\}$ is
discrete and \textrm{d}$\omega $ is the dimensionality of representations $%
U_{\omega }(g)$ \textup{(}formally, if $\mathcal{H}_{\omega }$ is
infinite-dimensional\textup{),} then conditions \textup{(\ref{3 3 16})}
assume the form 
\begin{equation}
\mathrm{Tr}_{\mathcal{H}_{\omega }}(\hat{R}_{\omega }\hat{\mu})^{1/2}d\omega
=\hat{\mu},\quad \mathrm{Tr}_{\mathcal{H}_{\omega }}(\hat{R}_{\omega }\hat{%
\mu})^{-1/2}\hat{R}_{\omega }\;d\omega \leq \hat{I}.  \label{3 3 17}
\end{equation}

\textup{If} $\mathcal{U}_{0}$ \textup{is one-dimensional, (\ref{3 3 16}) and
(\ref{3 3 17}) lead us to the following solutions:} 
\begin{equation*}
\mu =\Big(\int_{\Omega }\mathrm{Tr}_{\mathcal{H}_{\omega }}\hat{R}_{\omega
}^{1/2}\;\mathrm{d}\omega \Big)^{2},\;\mu =\Big(\sum_{\Omega }\mathrm{Tr}_{%
\mathcal{H}_{\omega }}\hat{R}_{\omega }^{1/2}\;d\omega \Big)^{1/2},
\end{equation*}%
\textup{which were found in \cite{25} (that is, the case of group symmetry
for pure states is \textquotedblleft equidiagonal\textquotedblright ). The
rank of operators} $\hat{R}_{\omega }$ \textup{determines the multiplicity
of representations} $U_{\omega }(g)$ \textup{in the representation} $U(g)$ 
\textup{in} $\mathcal{U}$.\newline
\textup{(5)} Suppose that the multiplicity of representations $U_{\omega
}(g) $ in the representation $U(g)$ is unity. Then the operators $\hat{R}%
_{\omega }$ are one-dimensional, $\hat{R}_{\omega }=\psi _{\omega }^{\ast
}\otimes \psi _{\omega \cdot }$ and Eqs \textup{(\ref{3 3 16})} and \textup{(%
\ref{3 3 17})} assume the form 
\begin{equation*}
\Big(\int \hat{S}_{\omega }\;\mathrm{d}\omega /c_{\omega }-\hat{I}\Big)\hat{%
\mu}=0,\quad \Big(\sum \hat{S}_{\omega }\;d\omega /c_{\omega }-\hat{I}%
_{\omega }\Big)\hat{\mu}=0,
\end{equation*}%
where $\hat{S}_{\omega }=\mathrm{Tr}_{\mathcal{H}_{\omega }}\hat{R}_{\omega
} $, and $c_{\omega }=[\mathrm{Tr}_{\mathcal{U}_{0}}(\hat{S}_{\omega }\hat{%
\mu})]^{1/2}$.

In particular, if all $\hat{S}_{\omega }$ are commutative, then operator $%
\hat{\mu}$ is a multiple of the proper projector $\hat{\pi}$ of operators $%
\hat{S}_{\omega }$, which corresponds to the eigenvalues $\lambda _{\omega }$
with maximal%
\begin{equation*}
\mu =\Big(\int \lambda _{\omega }^{1/2}\;\mathrm{d}\omega \Big)^{2}
\end{equation*}%
(or $\mu =\Big(\sum \lambda _{\omega }^{1/2}\;d\omega \Big)^{2}$): $\hat{\mu}%
=\mu \hat{\pi},(\hat{S}_{\omega }-\lambda \omega )\hat{\pi}=0$.
\end{theorem}

\subsection{Application to a Group Symmetry and Indeterminate Phase}

\label{group symmetry}To apply the above results to the case of a
decomposable $G$-homogeneous family of density operators $S_{x}=\otimes
_{n}R_{x}^{(n)}$ it is sufficient to supply all the spaces and operators in (%
\ref{3 3 11})-(\ref{3 3 16}) with an index $n$ and then sum over $n$. In
particular, if the representations $U^{(n)}(g)$ in $\mathcal{H}^{(n)}$ are $n
$-th tensor powers of the representation of $U(g)$ in $\mathcal{L}$, then
the solution of the problem of optimal recognition of audio and optical
fields is reduced to finding the irreducible representations $U_{\omega }(g)$
contained in $U^{(n)}(g)$. The operators $R_{\omega }^{(n)}$ determining (%
\ref{3 3 16}) and (\ref{3 3 17}) are 
\begin{equation*}
\hat{R}_{\omega }^{(n)}=\int \hat{r}^{(n)}(g)U_{\omega }(g)\;\mathrm{d}g
\end{equation*}%
where 
\begin{equation*}
\hat{r}^{(n)}(g)=\psi ^{(n)\ast }U^{(n)}(g^{-1})\psi ^{(n)}.
\end{equation*}%
For example, if the states $S_{x}$ are Gaussian, the family of signals $%
\{\varphi _{x},x\in X\}$ is $G$-homogeneous: $U(g)\varphi _{gx}=\varphi _{x}$%
, and the correlation noise operator $L$ (or $N$) is $G$-invariant:%
\thinspace $U(g)LU^{\ast }(g)=L$, then the family of the $R^{(n)}$ operators
is also $G$-homogeneous with respect to the appropriate tensor powers $%
U^{(n)}(g)$ of the representation $U(g)$ in the subspace $\mathcal{U}$
generated by vector $\varphi =\varphi _{x_{0}}$ for a certain $x_{0}\in X$.
In this manner we can find the exact solution to the following problems:
resolution of several nonorthogonal partially coherent signals or fields
that form a homogeneous family of permutations with respect to a certain
group (symmetric groups $S(r)$ and their subgroups), estimation of the time
lag of pulsed signals and the carrier frequency in quasiperiodic signals
(cycle $Z$ groups), joint measurement of the duration and the frequency of a
wave packet (the symplectic group), separate or joint measurement of momenta
and position of quantum systems (and ensembles of such systems) with $r$
degrees of freedom (the $Z(r)$ groups), detection of photon polarization and
electron spin (the $SU(2)$ group), detection of complex signals and fields
with equal intensities of rank $r$ against a thermal background (the $SU(r)$
groups and their subgroups), and the like.

\backmatter

\chapter{Afterword}

Optimal wave tomography as mathematical wave-pattern recognition theory,
emerged in the 70's first in connection with the problems\ of optimal
estimation and hypothesis testing in quantum theory, is a new pattern
recognition theory. The key problems in this theory, mathematical design of
the optimal dynamic analyzer discriminating between a given family of pure
or mixed \textit{a priori} unknown wave patterns, is thoroughly studied
here. Like the problem of optimal quantum measurement it cannot be tackled
by the methods of classical mathematical statistics which is not concerned
how these date should be obtained from the physical wave states in an
optimal way. We extended here the results of optimal quantum measurement
theory obtained in \cite{3}--\cite{22} into the direction of wave, rather
than particle statistical estimation and hypothesis testing theory,
naturally including into the wave tomography not only quantum matter waves
but also classical wave patterns like optical and acoustic waves. The
developed methods are applied to the problems of mathematical design of
optimal wave analyzer discriminating the visual and sound patterns. Thus,
Hilbert space and operator methods, developed first in quantum theory of
optimal quantum measurement, are found to be equally useful in the classical
wave theory where the possible observations are restricted to only intensity
distributions of waves, i.e. when the wave states are not the allowed
observables, as they are not the observables for individual particles in the
quantum theory. It has been shown that all the attributes of quantum
measurement theory such as complementarity, entanglements or Heisenberg
uncertainty relations have also an exact reflection in the wave pattern
recognition theory.

\chapter{Acknowledgement}

The first author (VPB) acknowledges EEC support through the ATESIT project
IST-2000-29681 which allowed to cover the expenses for retyping and
typesetting this book in TeX.

\end{document}